\documentclass{ar-1col}
\usepackage{xspace}
\usepackage[authoryear,round]{natbib}
\usepackage{biblio}
\usepackage[total={30.5pc,47pc},centering]{geometry}
\jname{Annu. Rev. Astron. Astrophys.}
\jvol{56}
\jyear{2018}
\newbox\grsign \setbox\grsign=\hbox{$>$} \newdimen\grdimen \grdimen=\ht\grsign
\newbox\simlessbox \newbox\simgreatbox
\setbox\simgreatbox=\hbox{\raise.5ex\hbox{$>$}\llap
     {\lower.5ex\hbox{$\sim$}}}\ht1=\grdimen\dp1=0pt
\setbox\simlessbox=\hbox{\raise.5ex\hbox{$<$}\llap
     {\lower.5ex\hbox{$\sim$}}}\ht2=\grdimen\dp2=0pt
\def\simgreat{\mathrel{\copy\simgreatbox}}
\def\simless{\mathrel{\copy\simlessbox}}
\newbox\simppropto
\setbox\simppropto=\hbox{\raise.5ex\hbox{$\sim$}\llap
     {\lower.5ex\hbox{$\propto$}}}\ht2=\grdimen\dp2=0pt

\newcommand{\narreq}{\!=\!}
\newcommand{\nsim}{\!\sim\!}

\newcommand{\nsimeq}{\!\simeq\!}
\newcommand{\nto}{\!-\!}

\newcommand{\nle}{\!\le\!}
\newcommand{\ngt}{\!\ge\!}
\newcommand{\nlt}{\!\le\!}

\def\gta{\;\lower 0.5ex\hbox{$\buildrel > \over \sim\ $}}
\def\lta{\;\lower 0.5ex\hbox{$\buildrel < \over \sim\ $}}
\newcommand{\los}{\rm{LOS}\xspace}

\def\kpc{\ifmmode {\> {\rm kpc}}\else {kpc}\fi}
\def\kms{\ifmmode {{\rm\ km\ s}^{-1}}\else {km s$^{-1}$}\fi}
\def\perkpc {\ifmmode {{\rm kpc}^{-1}}\else {\ kpc$^{-1}$}\fi}
\def\Msun{\ifmmode {\>M_\odot}\else {${\rm M}_\odot$}\fi}
\newcommand{\dg}{^\circ}
\newcommand{\mum}{\,\mu\rm{m}\xspace}

\newcommand{\pc}{\,\rm{pc}\xspace}

\newcommand{\masyr}{\,\rm{mas\,yr^{-1}}\xspace}
\newcommand{\Gyr}{\,\rm{Gyr}}

\begin{document}

% Page heads
\markboth{Beatriz Barbuy, Cristina Chiappini, Ortwin Gerhard}
{Chemodynamical History of the Galactic Bulge}

% Title
\title{Chemodynamical history of the Galactic Bulge}

% Author/affiliation
\author{Beatriz Barbuy$^1$, Cristina Chiappini$^2$, Ortwin Gerhard$^3$ 
\affil{$^1$Department of Astronomy, Universidade de S\~ao Paulo, S\~ao Paulo
 05508-090, Brazil; e-mail: b.barbuy@iag.usp.br, bbarbuy@gmail.com}
\affil{$^2$Leibniz-Institut f\"ur Astrophysik Potsdam (AIP), 
An der Sternwarte 16, 14482, Potsdam, Germany;
e-mail: cristina.chiappini@aip.de}
\affil{$^3$Max-Planck-Institut f\"ur
extraterrestrische Physik P.O. Box 1312. D-85741 Garching, Germany; email: gerhard@mpe.mpg.de}}

% First page note
%\firstpagenote{This is an example of dummy text used to illustrate an example of first page note.}

% Abstract
\begin{abstract}
  The Galactic Bulge can uniquely be studied from large samples of individual stars, and is
  therefore of prime importance for understanding the stellar population structure of bulges in
  general.  Here the observational evidence on the kinematics, chemical composition, and ages of
  Bulge stellar populations based on photometric and spectroscopic data is reviewed.  The bulk of
  Bulge stars are old and span a metallicity range $-$1.5$\simless$[Fe/H]$\simless$+0.5.  Stellar
  populations and chemical properties suggest a star formation timescale below $\sim$2 Gyr. The
  overall Bulge is barred and follows cylindrical rotation, and the more metal-rich stars trace a
  Box/Peanut (B/P) structure.  Dynamical models demonstrate the different spatial and orbital
  distributions of metal-rich and metal-poor stars.  We discuss current Bulge formation scenarios
  based on dynamical, chemical, chemodynamical and cosmological models.  Despite impressive progress
  we do not yet have a successful fully self-consistent chemodynamical Bulge model in the
  cosmological framework, and we will also need more extensive chrono-chemical-kinematic 3D map of
  stars to better constrain such models.
\end{abstract}

% Keywords
\begin{keywords}
bulge, bar, stellar populations, chemical composition, dynamical evolution,
chemical evolution, chemodynamical evolution
 
\end{keywords}

\maketitle

% to generate article TOC
%\tableofcontents

% Head 1

\section{INTRODUCTION}

The Galactic Bulge was first identified as a distinct component of our Galaxy by
\citet{Baade1946} and \citet{stebbinswhitford47}. In the same work, Baade also
used for the first time the term stellar population of type II for the stars in
the globular cluster NGC 6522 and its surrounding fields. The discovery of M
giants by \citet{nassaublanco58} first showed the presence of metal-rich old
populations.  Early historical work on the Galactic or Milky Way (MW) Bulge,
hereafter the Bulge, is reviewed in \citet{madore16}.

Thorough studies of Bulge populations were carried out by
\citet{whitfordrich83} and \citet{rich88} who presented a metallicity
distribution function (MDF) in the range $-$1$<$[Fe/H]$<$+1\footnote{The bracket
  notation denotes logarithmic abundance ratios relative to Solar values, with
  [X/Y] $=$ log(X/Y)$_{\rm star}$ $-$ log (X/Y)$_{\odot}$}.  A Color-Magnitude
Diagram (CMD) for Bulge field stars was obtained by \citet{terndrup88}, in
parallel to
 the infrared survey by \citet{frogelwhitford87}, the radial velocity study by
\citet{Mould1983},  a velocity dispersion derivation from integrated spectra
by \citet{Freeman1988},
 and the multi-fiber TiO band spectroscopic survey of 300 giants
in Baade's Window (BW) by
 \citet{Sharples1990}.  \citet{mcwilliamrich94} presented the first
high-resolution abundance analysis of 11 Bulge giants that revealed their
[$\alpha$/Fe]-enhancement.

Photometric and spectroscopic surveys 
covering large regions of the Bulge 
together with advanced data
interpretation tools have since led to a new challenge, of
combining the multifaceted observations into a consistent chemodynamical
evolution picture for our Bulge, and placing the Bulge into the more general
context of bulges of spiral galaxies. 
 Many issues are still open and debated. One
of the critical issues concerns the age of the Bulge. Ages derived from CMDs
suggest an uniform old age for the Bulge stellar populations
\citep[e.g.][]{ortolani95,clarkson08,gennaro15}, with little room for
intermediate age stars, in contrast to findings by \citet{bensby17}, based on
spectroscopy, which suggest a fraction of at least 15\% of metal-rich
([Fe/H]$\simgreat$$-$0.5) dwarf stars younger than 5-8 Gyr.
A broad metallicity range of $-$1.5$\simless$[Fe/H]$\simless$+0.5 has been
confirmed with the MDF turning out to be dependent on the field location, which
suggests the coexistence of different Bulge populations.  Most Bulge stars with
[Fe/H]$\simless$$-$0.4 show enhancements in $\alpha$-elements, indicative of a
fast chemical evolution, dominated by core collapse supernovae (CCSNe) chemical
enrichment.

\citet{devaucouleurs64} first suggested a barred Bulge in the MW from the high
gas velocities in the central Galaxy.  The existence of the bar was put on firm
grounds in the 1990's, from modeling the non-circular gas motions in the Bulge
region \citep{Binney+91, Englmaier+Gerhard99, Fux99},the near-infrared
data in particular from the COBE satellite \citep{blitzspergel91,Weiland1994,
  Binney+97}, and red clump giant star counts (RCG) \citep{Stanek1994}.  Modern
data have confirmed the Bulge to have a box/peanut (B/P) shape typical for
barred galaxies.  With star counts at high latitudes it was discovered that the
Bulge has an X-shape component \citep{McWilliam+Zoccali10,Nataf2010}, which is a
manifestation of a B/P bulge as later demonstrated by the reconstruction of the
3D Bulge density \citep{Wegg+Gerhard13} from VVV data (see Table \ref{surveys}).
Because B/P bulges are the inner 3D parts of a longer, flatter bar in the disk
 \citep[e.g.,][]{Bureau1999, Athanassoula2005,erwin13}, 
the definition of `the Bulge' has
become somewhat blurred, affecting also estimates of the Bulge mass (see
Section~\ref{s:dynamics}). General practice is to use a region on the sky with
size $\pm10\dg$ in both longitude and latitude $(l,b)$, which however poses the
problem of separating Bulge stars from foreground and background disk stars.

Bulges of spiral galaxies can be classified into two categories, according to
their observational properties, often designated as classical bulges (ClBs) and
pseudobulges, according to 
\citet{Kormendy2004,Athanassoula2005,fisherdrory16}: classical
bulges have properties similar to elliptical galaxies, with steep surface
brightness profiles with Sersic index n$>$2, pressure supported kinematics,
generally rounder isophotes, steeper dispersion profiles, spheroidal rotation,
anisotropy parameter $({\rm V}_{\rm rot}\!/\!\sigma)_{\star}\nlt 1$
\footnote{V$_{rot}$ stands for rotation velocity, and $\sigma$ for velocity
  dispersion along a line of sight.} \citep{Kormendy1982}, bulge to total ratios
B/T$>$1/3, and they follow elliptical galaxy scaling relations in the
fundamental plane. Pseudobulges, including B/P bulges, have near
exponential (n$<$2) surface brightness profiles, more flattened isophotes,
approximately flat dispersion profiles, { they} are rotationally supported with
near-cylindrical rotation $({\rm V}_{\rm rot}\!/\!\sigma)_{\star}\ngt 1)$, B/T
$<$ 1/3, and are disjoint from the classical bulge scaling relations.  In a
number of galaxies both a classical bulge and a pseudobulge coexist 
\citep{erwin15}.
Classical bulges are found in nearly half of early-type spirals and are absent
in Sc and later-type systems.  Pseudobulges spread over all types of spiral
galaxies and their scalelengths are related to the disk scalelength.

The Milky Way galaxy is considered to be of SBbc type, but could be earlier
(SBb) as indicated by its high bulge mass. It appears to fall into the so-called
green valley region that separates the red and blue sequences in the galaxy
colour-magnitude plot \citep{BHGerhard2016}.  Central velocity dispersions of
$\sigma\nsimeq 140$ km s$^{-1}$ measured for the Bulge \citep{Blum1995,Ness2016,
  Zoccali2017} are intermediate between typical values for pseudobulges (70-150
km s$^{-1}$) and for classical bulges (100-230 km s$^{-1}$) as in
\citet{fabricius12}. Lick indices, e.g. Mgb$\sim$3.0 \citep{puzia02} compare
well with estimated values of Mgb $>$ 2.35 for classical bulges
\citep{fisherdrory16}. The barred 3D density and cylindrical rotation of the
Bulge suggest a box-peanut (B/P) pseudobulge \citep{Shen2010,
  Wegg+Gerhard13}.

The first detailed models focusing on the formation of the Bulge were
chemical evolution models \citep{matteuccibrocato90,Molla1995,ferreras03}, in
which the common hypothesis was that the Bulge would form on short timescales, 
as inspired by early disssipative collapse models \citep[e.g.][]{Larson1974}. In
modern hierarchical $\Lambda$CDM models, galaxies form upon the collapse of the
baryonic matter within dark matter haloes. It was first found that these initial
galaxies would then subsequently merge with other discs, thus spheroids would
grow through early mergers \citep{steinmetz95,springelhernquist05}. However,
when mergers are prevented, galaxies with predominant disks form from the inside
out \citep[e.g.,][]{samland03}.  Nowadays, strong feedback is believed to
decouple the gas from the merging dark matter halos, leading to the formation of
realistic disk galaxies
in cosmological simulations \citep[see, e.g.][]{Stinson2006, Aumer2013, naabostriker17}.
  Models based on observations of nearby galaxies
including the MW suggested another channel for the formation of spiral bulges, { where}
through secular evolution by gas accretion and star formation in the disk
first a bar forms and then the bulge is built by buckling of the bar
\citep{Combes1990,Raha1991,Norman1996}.  Rapid gas accretion in the early phases
of the Universe, forming starbursts \citep{Elmegreen1999} and turbulent
\citep{brook04} clumpy disks have also been shown to produce bulges
\citep{elmegreen08, bournaud09}.

These early different ideas, and sometimes conflicting scenarios for bulge
formation, are discussed in excellent reviews such as
\citet{bouwens99,Kormendy2004,Athanassoula2005}.
IAU Symposia dedicated to galaxy bulges and to the Galactic Bulge were held by
\citet{dejonghehabing93}, and \citet{Bureau2008}.  Annual Review articles
dedicated to bulges and the Galactic Bulge were written by \cite{frogel88},
\cite{wysegilmorefranx97}, and \citet{Kormendy2004}. More recent reviews on
bulges in general were presented by \citet{laurikainen16} and
\citet{naabostriker17}, whereas recent ones more focused on the MW Bulge were
presented by \citet{rich13}, the PASA volume 33 (2016), \citet{nataf17},
and part of the contributions to IAU 334.

In this review, we describe the various tracers of the multiple populations in
the Bulge, as well as their transitions into the present thin disk, thick disk,
and extant early spheroid (old bulge/inner halo), and their chemodynamical
properties. We begin in Section 2 with a review of the structure and dynamics of
the Bulge.  In Section 3 we describe our current knowledge of the Bulge stellar
population properties (ages, metallicity distribution, chemical abundances, and
initial mass function), discuss nucleosynthesis implications, and compare with
halo and thick disk properties.  In Section 4 we present various chemical and
chemodynamical bulge models from the literature and discuss how the different
approaches can lead to a more complete fully self-consistent chemodynamical
bulge model in the cosmological framework. Conclusions and outlook are presented
in Section 5. This review is being written to set the scene for several major
surveys which will provide us with much-improved information and insights about
the MW Bulge, such as the full Gaia astrometric catalog, the photometric surveys
VVV/X, LSST, Plato, and several large spectroscopic surveys such as
APOGEE-South, 4MOST, and MOONS (Table \ref{surveys}).

% Table
\begin{table}
\tabcolsep3.25pt
\caption{Surveys cited in the present review. }
\label{surveys}
\begin{center}
\begin{tabular}{|@{}l|c|c|c|c|c|c|}
\hline
Acronym  & Name & Reference  \\
\hline 
APOGEE & Apache Point Observatory Galactic Evolution Experiment & \citet{Majewski2015} \\
\hline
ARGOS & Abundances and Radial velocity Galactic Origins Survey & 
\citet{Freeman2013,Ness2013a,Ness2013b} \\ 
\hline
BRAVA & The Bulge Radial Velocity Assay & \citet{Rich2007,Kunder2012} \\
\hline
COBE & Cosmic Background Explorer & \citet{Dwek+95} \\
\hline
GES &  Gaia-ESO Survey &\citet{Gilmore2012} \\
\hline
GIBS &  GIRAFFE Inner Bulge Survey &\citet{Zoccali2014,Zoccali2017} \\
\hline
SWEEPS & Sagittarius Window Eclipsing Extrasolar Planet Search & \citet{sahu08}\\ 
\hline
GBTP & Galactic Bulge Treasury Program & \citet{Brown2009} \\
\hline
EMBLA & Extremely Metal-poor BuLge stars with AAOmega survey & \citet{howes16}  \\
\hline
HST & Hubble Space Telescope & http://hubblesite.org  \\
\hline
LIGO & Laser Interferometer Gravitational-Wave Observatory &
 http://www.ligo.caltech.edu  \\
\hline
2MASS & Two Micron All-Sky Survey & \citet{Skrutskie2006} \\
\hline
MACHO & Massive Compact Halo Objects  & \citet{Alcock2000} \\
\hline
OGLE & Optical Gravitational Lensing Experiment  & \citet{Udalski2002} \\
\hline
SDSS & Sloan Digital Sky Survey  & \citet{York2000} \\
\hline
VVV & Vista Variables in the Via Lactea  & \citet{Minniti2010,Saito2012a} \\
\hline
VIRGO & VIRGO Interferometric Gravitational Wave Observatory  & http://www.virgo-gw.eu \\
\hline
WISE & Wide-field infrared Survey Explorer & \citet{wright10} \\
\hline
\hline
LSST & Large Synoptic Survey Telescope & \citet{Tyson2002} \\
\hline
4MOST & 4-metre Multi-Object Spectrograph Telescope  & \citet{deJong2016} \\
\hline
MOONS & Multi-Object Optical and Near-Infrared Spectrograph
 & \citet{Cirasuolo2014} \\
\hline
PLATO & Planetary Transits and Oscillations of Stars
 & \citet{Rauer2014,Miglio2017} \\
\hline
\end{tabular}
\end{center}
\end{table}

%%%%%%%%%%%%%%%%%%%%%%%%%%%%%%%%%%%%%%%%%%%%%%%%%%
\section{Structure and  dynamics}
%%%%%%%%%%%%%%%%%%%%%%%%%%%%%%%%%%%%%%%%%%%%%%%%%%
\label{s:dynamics}

In this section we describe the spatial distribution and kinematics of stars in the Bulge, and how these depend
on the stellar metallicities.  We review the evidence that the bulk of the Bulge stars constitute a barred bulge within
which the metal-rich stellar populations have a pronounced box/peanut (B/P) shape similar as
in other barred galaxies.  Finally, we review constraints on the stellar mass and dark matter in the Bulge.

%%%%%%%%%%%%%%%%%%%%%%%%%

\begin{figure}
\centering
\parbox{5.8cm}{
\includegraphics[width=5.3cm]{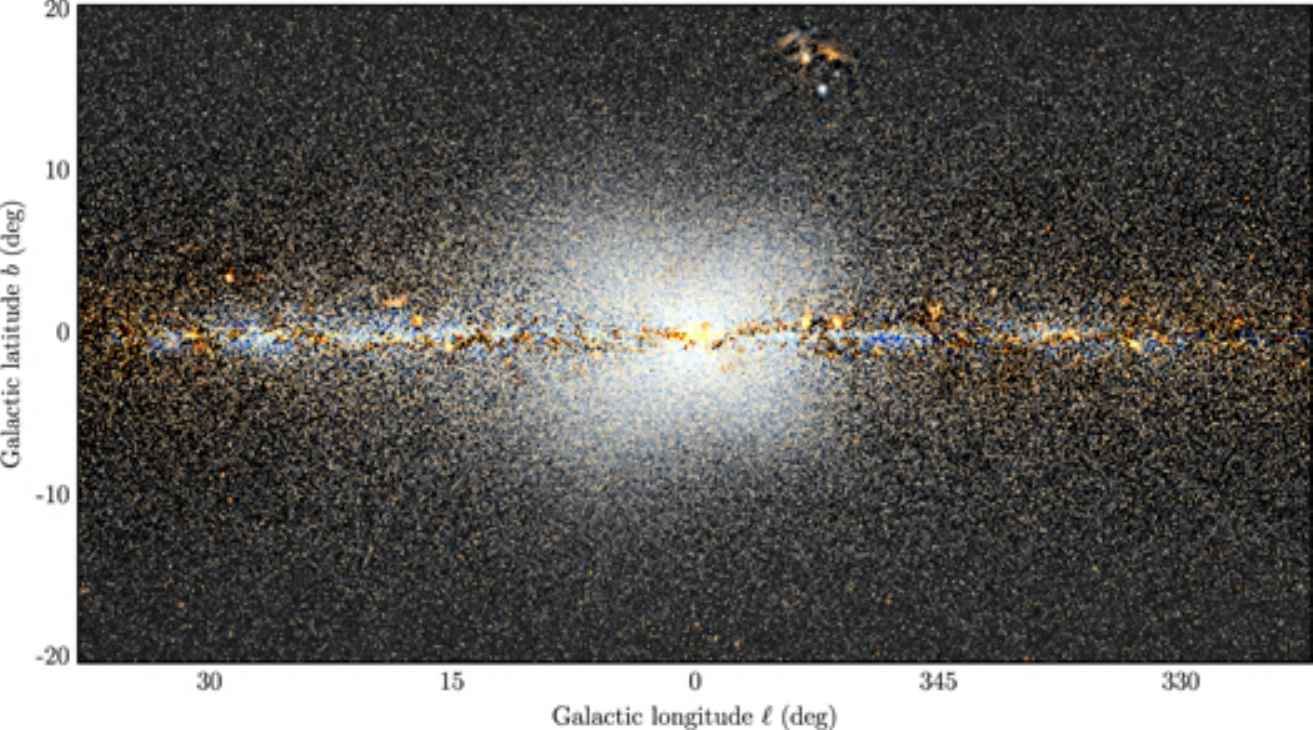}
\includegraphics[width=6.2cm]{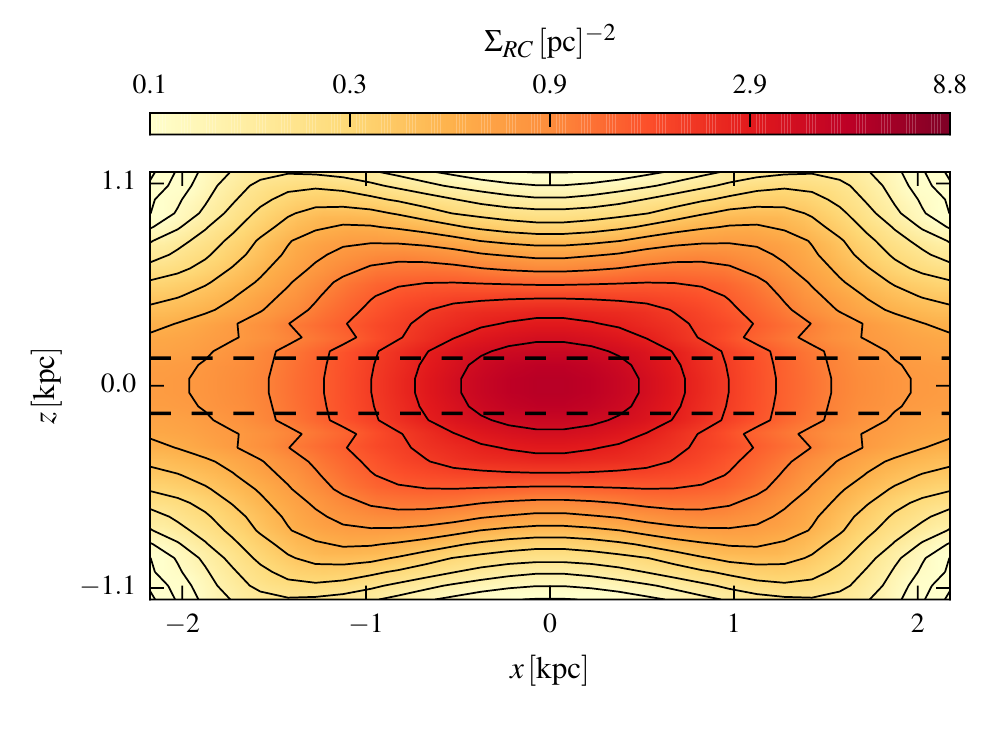}}
\begin{minipage}{6.4cm}
\includegraphics[width=6.4cm]{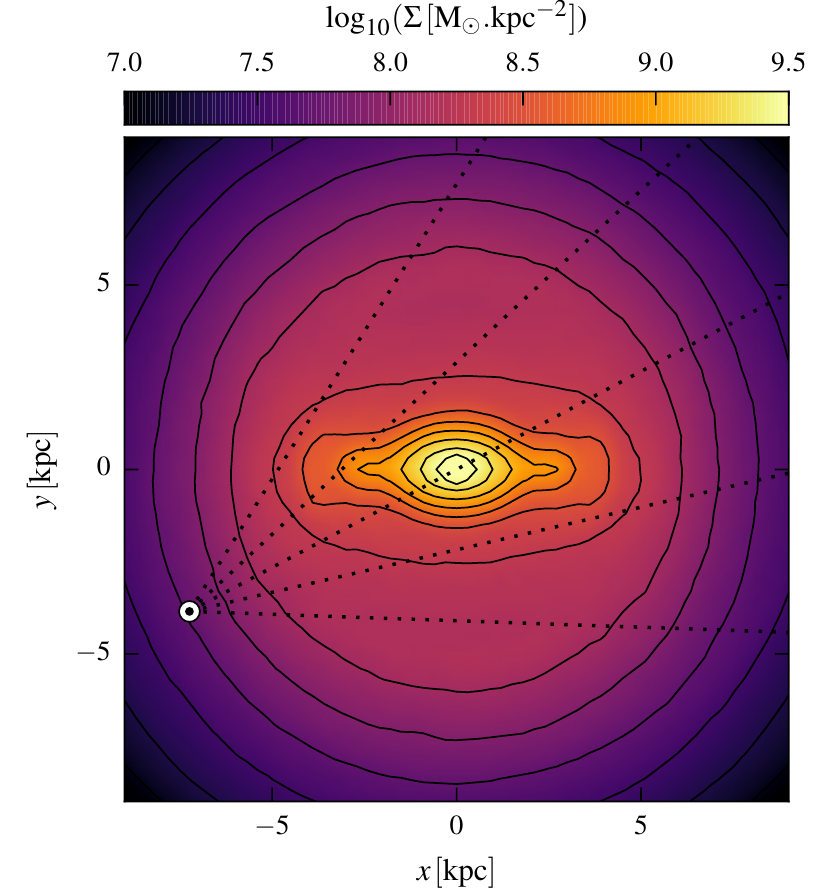}
\end{minipage}
\caption{Top left: WISE 3.4$\mum$/4.6$\mum$ contrast-enhanced image of the Bulge region, showing the X-structure.
  Bottom left: surface density of the Galactic bulge viewed side-on, obtained by projecting the 3D density of red clump
  giant stars measured by \citet{Wegg+Gerhard13} from VVV data. Between the dashed lines ($|z|\narreq160\pc$,
  $|b|\narreq1\dg$), the density has been extrapolated down from higher latitudes, due to crowding uncertainties.
  Contours are plotted every 0.33 mag and show the exponential vertical density profile.  Right: stellar surface density
  for the entire bulge/bar region, according to best-matching model of \citet{Portail2017a}. This shows how the B/P
  bulge transits continuously into the outer planar bar which extends to $\sim\! 5 \kpc$ from the Galactic Center. Also
  shown are lines-of-sight from the Sun towards longitudes $l$=$-$10$^{\circ}$, $-$5$^{\circ}$, 0, +5$^{\circ}$,
  +10$^{\circ}$. Adapted respectively from \citet{Ness2016b}, \citet{Portail2015a}, \citet{Portail2017a}, with
  permission.}
\label{f:bulgex}
\end{figure}

\subsection{Bulge structure and morphology from star counts}
\label{s:structure}

The first secure morphological evidence for a triaxial bulge in our Galaxy came from COBE near-infrared photometry which
unambiguously showed that the Bulge was both larger and brighter at $l>0$.  Several photometric models for the Bulge
modeled this asymmetry \citep{Dwek+95, Binney+97, Bissantz+Gerhard02} and showed that the Bulge had a barred shape,
which extended to $\sim 2\kpc$ from the center and had its near side in the second Galactic quadrant, with a {\sl bar
  angle} of $\phi_{\rm bar}=\sim25\dg\pm10\dg$ between the bar's long axis and the Sun-Galactic center line. Early star
counts using the Red Clump (RC) confirmed that indeed stars on the $l>0$ near side of the Bulge are closer to us
\citep{Stanek1994}.

Red clump giants (RCGs) are very useful tracers for the Bulge. They are He-core burning stars with standard
candle properties except for young (ages $<2\nto 4\,\mathrm{Gyr}$) or metal-poor ([Fe/H]$<$$-$1) populations; see
\citet{Salaris+Girardi02,girardi16}.
 For typical Bulge populations RCGs have a narrow range of absolute
 magnitudes and
colors $\sigma(\mathrm{K}_s)\simeq0.17$ and $\sigma(\mathrm{J-K}_s)\simeq0.05$, and individual distances can be
determined to within $\sim\!10\%$.  Variations in $\mathrm{K}_s$ due to age are $\sim\!0.03/ \mathrm{Gyr}$ at age
$10\mathrm{Gyr}$, and due to metallicity lead to a spread $\sigma_{Ks}(\mathrm{Fe/H})\sim\!0.11$ for the
Bulge MDF. { Relative numbers} of RCGs are $\sim 10^{-3}/\Msun$ and are predicted to vary by less than 10\% for old stars with
metallicities in the range 0.02 to 1.5 solar. Among the 14,150 stars of the ARGOS survey within Galactic radius
$R_G<3.5\kpc$, RC stars are prominent to [Fe/H]$=\!-1.0$, corresponding to $\sim\!95\%$ of the ARGOS sample
\citep{Ness2013a}; i.e., RCGs are representative for most of the Bulge 
stellar mass.

RCGs have therefore played an important role in clarifying the structure of the Bulge \citep{Babusiaux2005,
  Rattenbury2007}.  RCG star counts in the OGLE, 2MASS, and VVV surveys (see Table~\ref{surveys}) showed the presence of
an X-shaped structure \citep{Nataf2010, McWilliam+Zoccali10, Saito2012b} similar to B/P bulges in barred
galaxies \citep{Bureau2006} and N-body simulations \citep{Li2012}.  \citet{Ness2012} showed that only stars with
[Fe/H]$>-0.5$ participate in the split RC (X-shape). 
\citet{Gonzalez2015b} showed that a scenario of one population significantly enhanced
in helium does not explain the observed variation of the RC
across different $(l,b)$.

Using $\sim\!8$ million RCGs from the VVV survey, \citet{Wegg+Gerhard13} obtained distance distributions along
$\sim\!300$ sightlines through the Bulge ($-10\dg\nle l\nle 10\dg$, $-10\dg\nle b\nle 5\dg$). Combining these they
measured the three-dimensional (3D) Bulge density distribution in a box-shaped volume of $\pm 2.2 \times \pm 1.4 \times
\pm 1.2\kpc$ (hereafter, the VVV box), outside the most crowded region $|b|<1\dg$, and with typical (systematic) error
of $\sim 10\%$.  They found a strongly barred and peanut-shaped density with bar angle
$\phi_{\rm bar}\narreq27\dg\pm2\dg$, and face-on projected axis ratio $\simeq(1:2)$. Viewed from the Sun the RCG Bulge
appears boxy, slightly peanut-shaped, consistent with the earlier COBE and 2MASS data and with recent WISE photometry
\citep{Ness2016b}.  Figure~\ref{f:bulgex}  illustrates these results. 
Unsharp masking results in a strong off-centered
X-shape structure \citep{Portail2015a}, similar to some galaxies in the sample of \citet{Bureau2006}. Along the bar
axes, the central ($<\!1\kpc$) density distributions were all found to be nearly exponential. In particular the minor
axis profile is exponential in the range $500\pc < z < 1.2\kpc$, with short scalelength $z_0=180\pc$, and shows no
indication of a central $R^{1/4}$ component.

Subsequently \citet{Wegg2015} showed, using the combined 2MASS, UKIDDS and VVV surveys, that the MW's B/P Bulge
continuously transits outwards into a planar bar with half-length of $5.0\pm0.2\kpc$ and bar angle $\phi_{\rm
  bar}=\sim28\dg-33\dg$ (assuming $R_0=8.3\kpc$).  This work superseded earlier work on the long bar which had suggested
different orientations of the Bulge bar and long bar \citep{Hammersley+00, Benjamin+05, Cabrera-Lavers+08}; see also
\citet{BHGerhard2016}.  The MW is thus similar to external barred galaxies that often contain a B/P bulge embedded in a
longer, thinner bar \citep{Kuijken1995,Bureau1999,erwin13}.  N-body models for bar-unstable disk evolution regularly
find that B/P bulges are the inner 3D parts of a longer, planar bar and form through a buckling instability and/or
through orbits in vertical resonance \citep{Combes1990, Raha1991, Athanassoula2005}.  Fig. \ref{f:bulgex} shows the
surface density in the MW bulge/bar and inner disk region of the best dynamical model of \citet{Portail2017a} fitted,
amongst others, to the RCG density of \citet{Wegg+Gerhard13} and the RC magnitude distributions of \citet{Wegg2015}.

Recently, large samples of Bulge RR Lyrae (RRL) stars have been identified in the OGLE and VVV surveys. RRL are
pulsating, low-metallicity core-helium burning horizontal branch giants \citep{Marconi2015} with ages $>11\Gyr$
\citep{Walker1989} and trace an old, { relatively} metal-poor population in the Bulge. From light curve information
accurate luminosities and distances ($\sim 5\%$ in optical, $\sim 3\%$ in NIR) can be determined, making RRL valuable
tracers of the old Bulge.  The Bulge RRL have metallicities [Fe/H]$=-1.0\pm0.2$ with extended [Fe/H] tails, and a steep
density profile $\sim r^{-3}$ \citep{Pietrukowicz2015}. They do not participate in the B/P-Bulge \citep{Dekany2013,
  Pietrukowicz2015, Gran2016}, consistent with the ARGOS result that only stars with [Fe/H] $\gta\!-0.5$ participate in
the split RC \citep{Ness2012}.  While \citet{Dekany2013} find weak if any triaxiality in the central kpc,
\citet{Pietrukowicz2015} report strong triaxiality out to $\sim 1.5\kpc$, approximately aligned with the main
Bulge/bar. This controversy still needs to be resolved. In the outer Bulge \citep[][$|b|=8\dg\nto 10\dg$]{Gran2016} and
near the Galactic plane for $350\dg>l>295\dg$ \citep{Minniti2017} the RRL distribution do not trace the bar seen in the
RCG. Recent spectroscopy shows that the RRL have no significant rotation for $|l|\lta3.5\dg$, rising to $v_{\rm
  rot}\simeq 50\kms$ at $l\narreq9\dg,b\narreq3\dg$ \citep[][and in preparation]{Kunder2016}, compared to $\simeq
90\kms$ for RGB and RCG stars \citep{Kunder2012}.

These properties suggest that the Bulge RRL trace a different component from the RCG stars.  
From the number ratio of
RRL to RCG this very old population is estimated to contribute of order $1\%$ of the Bulge stellar mass
\citep{Kunder2016}. 
A natural question is whether it is related to the inner stellar halo, which must
invariably be present in the Bulge. There is currently no evidence for a break in the mean metallicity
\citep{Pietrukowicz2016} or density \citep{Perez-Villegas2017} of the RRL from the Bulge into the halo.
\citet{Minniti2017} suggest from the RRL MDF that their sample is dominated by halo RRL. 
More detailed density maps,
kinematics, and metallicity maps must show whether there exists a separate Bulge RRL component.

\citet{Perez-Villegas2017} illustrate that in regions where the main B/P Bulge dominates the gravitational potential,
also the RRL orbits are generally non-axisymmetric.  They show that an initially spheroidal halo-bulge perturbed by the
evolving bar and B/P Bulge would at the end show some triaxiality in the vicinity of the main stellar Bulge, due to a
fraction of orbits trapped by the bar. This process \citep[see also][]{Saha2012} induces rotation velocities consistent
with \citet{Kunder2016} if the initial rotation is small.  At larger distances from the B/P Bulge, the perturbing
effects on the initial halo distribution decrease rapidly, with only mild effects expected in the outer Bulge.

In conclusion, RCG star counts show that the Bulge is a B/P bulge similar to bulges in many barred galaxies.
RCG trace $\nsim95\%$ of the Bulge stars over a wide range of metallicities, and the combined distribution of these
populations is strongly barred with a nearly exponential minor-axis density profile. However, only stars with
[Fe/H]$>-0.5$ participate in the B/P Bulge feature.
 A small fraction of the Bulge traced by very old, metal-poor RRL stars
($\nsim1\%$) appears to have a distinct morphological structure.

\subsection{Kinematics}
\label{s:kinematics}

{ Early kinematic studies of the Bulge concentrated on BW, measuring a line-of-sight (LOS) velocity dispersion
  $\sigma_r\narreq110\kms$ for M and K giants in this low-extinction field \citep[][]{Mould1983, Sharples1990,
    Rich1990}. With the recent large spectroscopic surveys we now have radial velocity information for many 1000s of
  stars spread over most of the Bulge, leading to a greatly improved understanding of Bulge kinematics and, ultimately,
  dynamics.}

\subsubsection{The velocity dispersion profile}

Studying K-giants in two fields, \citet{Minniti1992, Minniti1996a} found that the velocity dispersion 
for Bulge giants steeply decreases outwards.  This was confirmed in
much more detail first using Planetary Nebulas (PNs) by \citet{Beaulieu2000}, and then by the recent large surveys, the
BRAVA survey for M-giants \citep{Rich2007}, and the ARGOS, GES, and GIBs surveys for (mostly) RC giants.
Using NIR spectroscopy of the $2.3\mum$ CO bandhead,
\citet{Blum1995} found that the dispersion increases inwards to $\sigma_r\nsimeq140\kms$ at $1.2\dg$ from the
center. This is consistent with the minor axis profile from BRAVA \citep{Kunder2012} and the $\sigma\narreq140\kms$ peak
at $\Vert b\Vert\narreq1$-$2\dg$ found by GIBS using the infrared Ca triplet \citep{Zoccali2014}.  Combining data from
the ARGOS and APOGEE surveys, \citet{Ness2016} also found that the Bulge dispersion reaches the highest values
($\sigma\nsim130\kms$) within $2$-$3\dg$ of the center. The last two papers give approximate velocity dispersion maps
across the Bulge, which show the overall decrease of the dispersion from the center outwards.

\subsubsection{Cylindrical rotation}

The first Bulge rotation curve was measured by \citet{Minniti1992}, combining K-giant velocities in two outer-Bulge
fields at $l,b=(8\dg,7\dg), (12\dg,3\dg)$ with results from BW. They found an increase of $v_{\rm rot}$ to $80\kms$ in
their field at $l\narreq12\dg$ which was confirmed by \citet{Beaulieu2000}.  Much more detailed studies of the Bulge
rotation field became possible with modern surveys.  The BRAVA survey measured velocities for $\sim$8,600 M giants in
sequences of Bulge fields at different longitudes for $b=-4\dg, -6\dg, -8\dg$. These data showed that the Bulge rotation
curve turns away from solid body rotation at $l\sim5\dg$ and that at any given $l$ the rotation velocity is nearly
independent of $b$ for $b<-4\dg$. This so-called cylindrical rotation was suggested earlier by the PN survey of
\citet{Beaulieu2000} and is confirmed by the ARGOS survey for (mostly RC) stars with [Fe/H]$>-1.0$. The combined ARGOS
and APOGEE surveys show that the Bulge rotates cylindrically across all latitudes up to $\Vert b\Vert\lta10\dg$
\citep{Ness2016}. Both the rotational properties and the velocity dispersion show a smooth transition from the Bulge
into the disk and bar region further out.

These kinematic properties are similar to the cylindrical rotation seen in edge-on barred galaxies \citep{Shaw1993,
  Molaeinezhad2016} and in N-body models of B/P bulges that originated from dynamical instabilities of their initial
disks.  \citet{Shen2010} compared such models to the BRAVA data, and found a very good match to the cylindrical rotation
in the Bulge with a pure B/P bulge model while models containing an initial classical bulge \citep[ClB; see Introduction
  and][]{Kormendy2004} gave less good fits to the data.  \citet{Saha2012} and \citet{Saha2013} studied the spin-up of
ClB by evolving bars and found that in some models the ClB may acquire cylindrical rotation as well, but that the
induced rotation profile rises slowly and approximately linearly with radius. Thus while the cylindrical rotation per se
cannot be used to rule out a ClB component in the MW Bulge, more detailed studies of the rotation properties of
different stellar populations in the Bulge may set more stringent constraints on a possible ClB.

\subsubsection{Streaming velocities along the line-of-sight}

Several studies have obtained distance estimates from the magnitudes of their target stars, thereby measuring variations
of the radial streaming velocities with distance. For a bar at the orientation of the Galactic bar, one expects \los
streaming velocities to be more negative on the near side of the Bulge where stars approach the observer, and more
positive on the far side.  While at low latitudes ($b\lta6\dg$) such streaming velocity differences have been seen
\citep{Rangwala2009, Babusiaux2010, Babusiaux2014, Vasquez2013}, studies at higher latitudes have not found a similar
signal \citep{DePropris2011, Uttenthaler2012, Rojas-Arriagada2014}. The ARGOS data were recently reanalyzed by
\citet{Portail2017a} including the full survey selection function. Their Fig.~17 illustrates the dependence of the
streaming motions on distance modulus for $b=-5\dg$ and $b=-10\dg$, both in the ARGOS data and in their overall best B/P
bulge model. A related variation with distance has been found in some proper motion studies, due to stars streaming
towards positive (negative) $l$ on the near (far) side of the bar \cite[e.g.,][and below]{Kuijken2002,
  Poleski2013}. These streaming velocity patterns provide clear evidence for the barred nature of the Bulge.

\subsubsection{Proper motions, vertex deviation and anisotropy}

In a barred bulge tilted with respect to our \los, the principal directions of the velocity ellipsoid are also tilted.
To see this effect requires measuring line-of-sight velocities and proper motions (PM).  In a pioneering paper,
\citet{Spaenhauer1992} measured PMs for $\sim400$ stars in BW from photographic plates taken in 1950 and 1983. An
analysis of 61 of these stars with 3D kinematics by \citet{Zhao1994} indeed showed velocity ellipsoid tilts suggesting
bar dynamics. Another important application of PM studies is to separate foreground disk stars from Bulge stars
\citep{Kuijken2002, Vieira2007}.  The disk stars have an offset centered around $\mu_l=4 \masyr$ relative to the mean
(used as zero point in the absence of an absolute reference frame). For population studies a {\sl pure Bulge} sample can
then be isolated (e.g., requiring $\mu_l<0$), but this will be kinematically biased.

Selecting instead Bulge stars in the CMD and measuring their PMs and radial velocities enables kinematically unbiased
studies of the 3D velocity distribution.  This can generally be characterized by the velocity dispersions $\sigma_r$,
$\sigma_l$, $\sigma_b$ and the cross terms $\sigma_{rl}$, $\sigma_{rb}$, and $\sigma_{lb}$.  \citet{Soto2012} showed
that the vertex deviation $l_v$, where $\tan 2l_v= 2\sigma^2_{rl}/[\sigma_r^2 - \sigma_l^2]$, decreases in a field at
$b=-6\dg$ relative to BW and the SgrI field at $b=-2.5\dg$. The actual value of the vertex deviation is not easily
interpreted - it depends on the distance distribution of the stellar sample and must be related to 3D information
through a good dynamical model for the Bulge. \citet{Babusiaux2010} and \citet{Soto2012} give measurements for the
anisotropy in their Bulge fields, typically $\sigma_b/\sigma_r\narreq0.8$-$0.9$, and \citet{Kozlowski2006} and
\citet{Rattenbury2007} measure anisotropies $\sigma_l/\sigma_b\narreq1.1$-$1.2$ and a significant negative cross term
$\sigma^2_{lb}/(\sigma_l \sigma_b)\simeq-0.10$.  Altogether these measurements imply a barred potential.

PMs are also very useful for more targeted investigations into the properties of the stars in the Bulge X-shape.
\citet{Poleski2013} and \citet{Vasquez2013} studied kinematics of the nearby and distant arms of the X defined through
RC stars and detected clear differences in the radial velocities and PMs of the two arms. A quantitative
interpretation of these data with Bulge dynamical models is still pending.

%%%%%%%%%%%%%%%%%%%%%%%%%%%%%%%%%%%%%%%%%%%%%%%%%%%%%%%%%%%%%%%%%%%%%%%%%%%%%%%%%%%%
\subsection{Kinematics with Metallicity}
\label{s:kinmet}

First detections of kinematic differences with metallicities go back to at least Sharples et al (1990) and Rich (1990).
Rich found a lower $\sigma_r=92\pm14\kms$ for a metal-rich subsample
 of K giants with [Fe/H]$\geq0.3$, than
$\sigma_r=126\pm22$ for the metal-poor subsample with [Fe/H]$<-0.3$.  Minniti (1996) observed that halo giants
with [Fe/H]$<-1.5$ in a field at $l,b=(8\dg,7\dg)$ have higher dispersion and lower rotation than the Bulge giants
([Fe/H]$>-1.0$), and also that the Bulge giants themselves have kinematics correlated with metallicity. Combining with
BW data he found that for the Bulge giants $\sigma_r$ decreases steeply with distance from the Galactic Center, whereas
for the halo giants it hardly decreases.

\citet{Zhao1994, Soto2007, Babusiaux2010} and \citet{Hill2011} found differences between the vertex deviations of
metal-rich and metal-poor Bulge stars. { Fig.2 gives the vertex deviation as a function of metallicity in BW,
together with the chemodynamical model of Portail et al. (2017b) discussed in Section 4.2.}
The more metal-rich ([Fe/H]$\gta-0.5$) stellar population is found to have a clear
tilt ($\sigma_{rl}\ne0$), providing evidence that its spatial distribution (and the potential) is barred.

Recent spectroscopic surveys agree on distinct kinematic properties between metal-rich and metal-poor Bulge stars. The
ARGOS survey observed the Bulge in multiple fields with $|l|\nle20\dg$, $|b|\narreq5\nto10\dg$, selecting equal numbers
of stars in 3 magnitude bins based on 2MASS photometry. This strategy includes Bulge stars along the entire \los, giving
high weight to the outer Bulge \citep{Freeman2013, Portail2017a}. For \los distances within $\pm3.5\kpc$ around the
MW's center, \citet{Ness2013b} found distinct kinematics in 4 metallicity bins A ([Fe/H]$\ngt 0$), B ($-0.5\nlt$
[Fe/H]$\nle 0$), C ($-1.0\nlt$[Fe/H]$\nle -0.5$), and D ([Fe/H]$\nle-1.0$). Stars in bins A-C show fairly similar,
near-cylindrical rotation, but bin B shows the fastest rotation, about 20\% larger than A and C at $b\narreq-5\dg$.  The
small number of stars in D have significantly slower, non-cylindrical rotation. Stars in A and B have similar dispersion
profile shapes, with A being kinematically colder at all latitudes. C has clearly different, flatter dispersion profiles
with less latitude dependence. \citet{Ness2013b} suggest a common disk origin for stars in A and B which participate in the
B/P Bulge \citep{Ness2012}.  For stars in C which do not participate in the B/P Bulge, they suggest a thick-disk
interpretation, consistent with cylindrical rotation, while stars in D may be part of the stellar halo. Dynamical
models for these different stellar populations are in broad agreement with this interpretation, as discussed in
Section~\ref{s:chemo-dynamics}.

The APOGEE survey has observed a large number of 2MASS giants in the NIR, thereby also covering the low-latitude
inner Galaxy. Combining the APOGEE and ARGOS data, \citet{Ness2016} constructed latitude-combined rotation and
dispersion profiles for the same metallicity bins, showing a smooth transition from the Bulge to the thin bar
and disk. In these average profiles, A and B are very similar, with C showing higher dispersion and lower rotation.

GIBS surveyed fields in the inner Bulge ($|l|\!\lta\!8.5\dg$, $|b|\narreq 1.4\nto8.5\dg$), targeting RC and underlying
RGB stars selected from VVV, which are mostly located in a slice of $\pm600\pc$ width along the \los around the Bulge
center \citep{Zoccali2014}. The GIBS MDF divides into two components, separated approximately at
[Fe/H]$=0$ (see \S \ref{s:populations}), for which
\citet{Zoccali2017} show separate surface density maps and kinematic profiles. While
 for the metal-poor stars $\sigma_r$ varies mildly with latitude, consistent with ARGOS, the metal-rich component has a
steep gradient with $|b|$, so that its $\sigma_r$ becomes roughly equal at $|b|\narreq2\dg$ and higher than that of the
metal-poor component at $|b|\narreq 1.4\dg$.

The GES survey obtained kinematics in 11 Bulge fields with $|l|\!\lta\!10\dg$, $|b|\narreq 4\nto10\dg$, targeting also
RC and RGB stars selected from VVV in a distance range $\simeq\!\pm\!2.3\kpc$ from the center \citep{Rojas-Arriagada2014,
  Rojas-Arriagada2017}. They find a similar bimodal MDF to GIBS and broadly consistent dispersion profiles for the two
components. The overall dispersion versus [Fe/H] profile for the metal-poor component shows a maximum at [Fe/H]$\simeq-0.4$.

In summary, the Bulge kinematics suggest a complex multi-component chemodynamical structure of the Bulge. Differences
between recent surveys are likely to be related, at least in part, to different survey selection functions coupled with
\los metallicity gradients.  The metal-rich Bulge stars are clearly part of the B/P structure, they show cylindrical
rotation, a clear vertex deviation, and their velocity dispersion profile decreases steeply with latitude.  The more
metal-poor stars, which do not participate in the B/P structure, also show near-cylindrical rotation but no vertex
deviation, and their dispersion profile is approximately constant with latitude. The most metal-poor have lower,
non-cylindrical rotation, but these stars are rare in the Bulge and not well-sampled in the current data.

%%%%%%%%%%%%%%%%%%%%%%%%%

\begin{figure}
\includegraphics[width=7cm]{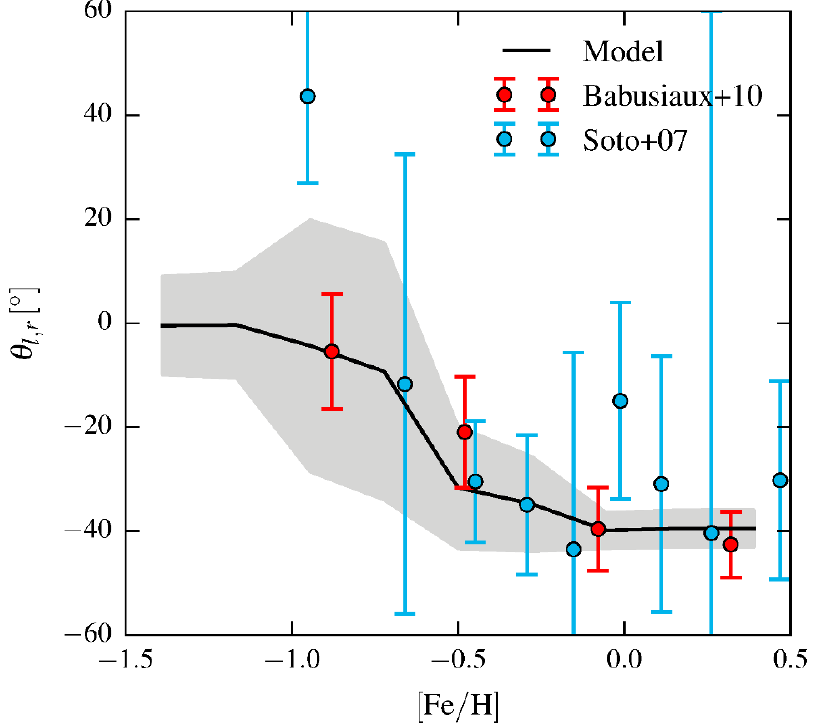}
\caption{Vertex deviation $\theta_{l,r}$ in BW as a function of metallicity. Data are from
  \citet{Babusiaux2010} and \citet{Soto2007}. Overplotted is the fiducial chemod-dynamical model of \citet[][see Section
    4.2]{Portail2017b} including all stellar particles between $6\kpc$ and $10\kpc$ along the line-of-sight from the
  Sun. The shaded area indicates the one-sigma statistical error on the model vertex deviation. The large errors in the
  model predictions at low metallicity arise from the fact that $\sigma_{r}^2$ and $\sigma_{l}^2$ are increasingly
  similar to each other with decreasing [Fe/H], and therefore the angle of the near-circular velocity ellipsoid is poorly
  defined. Adapted from \citet{Portail2017b} with permission. }
\label{vertexdeviation} 
\end{figure}

%%%%%%%%%%%%%%%%%%%%%%%%%

%%%%%%%%%%%%%%%%%%%%%%%%%%%%%%%%%%%%%%%%%%%%%%%%%%%%%%%%%%%
\subsection{Bulge Dynamics and Mass}
\label{globaldynamics}

After the COBE data had established the non-axisymmetric, boxy shape of the Bulge, several barred dynamical models were
built \citep{Zhao1994, Fux1997}.  Little stellar kinematical data was available at the time to constrain these models,
but a barred bulge formed through disk instability appeared to be a promising explanation for the observed kinematics
\citep{Beaulieu2000}. More detailed modeling became possible with the modern survey data.
\citet{Shen2010,Li2012,Ness2013b} showed that the observed X-shape and cylindrical rotation could be 
reproduced with an N-body B/P Bulge model.  \citet{Gardner2014, Qin2015} explained the different kinematics of stars on
the near and far parts of the X in the Bulge.  \citet{Portail2015b, Williams2016, Abbott2017} showed that the B/P Bulge
is maintained by a wide range of orbits, both resonant and non-resonant.

The mass of the Galactic Bulge has been estimated many times; see \citet{Licquia+Newman2015,BHGerhard2016}.  The most
accurate determinations of the dynamical mass are from recent made-to-measure (M2M) dynamical models which combine
well-determined tracer densities with multiple field kinematics \citep{Portail2015a, Portail2017a}. The stellar mass in
the Bulge is best estimated from RCG star counts containing information on the \los density distribution, combined with
an empirical estimate of the RCG abundance per unit mass \citep{Valenti2016, Portail2017b}.  The main systematic
uncertainty is that, because of the continuous transition between the 3D inner B/P Bulge bar and the flatter long bar,
quantities like the {\sl total} Bulge mass (vs.\ the mass within a fixed volume) require a somewhat arbitrary definition
of what stars constitute the Bulge.

\citet{Portail2015a} constructed M2M dynamical models of the Bulge matching the 3D density of RCG stars measured by Wegg
\& Gerhard (2013) from VVV data together with the BRAVA kinematics. They obtained a total dynamical mass of the Bulge in
the 3D VVV box, ($\pm 2.2 \times \pm 1.4 \times \pm 1.2\kpc$), of $1.84\pm0.07\times 10^{10}\Msun$, and comparing to
stellar population models, were able for the first time to estimate the fraction of dark matter mass in the Bulge.
\citet{Portail2017a} subsequently 
  included additional kinematic data from the ARGOS and OGLE surveys,
as well as the planar long bar based on its RCG density from \citet{Wegg2015}.  This leads to better constraints on the
Bulge as well, because long-bar stars moving through the inner Bulge influence the projected velocity dispersions.  With
their new models, \citet{Portail2017a} obtained an improved measurement of the pattern speed of the MW's B/P Bulge and
bar, $\Omega_p=39.0\pm3.5\kms\perkpc$, putting corotation at $6.1\pm0.5\kpc$. They redetermined the total dynamical mass
in the VVV box to $1.85\pm0.05\times 10^{10}\Msun$, in excellent agreement with \citet{Portail2015a}, and found a dark
matter fraction there of $17\pm2\%$. Together with the Galactic rotation curve in $R\narreq 6\nto8\kpc$, this low dark
matter fraction in the Bulge can only be accommodated if the MW's dark matter halo has a core or mild cusp.

\citet{Portail2017a} determined photometric stellar masses for the Bulge, bar, and disk, based on their calibration of
the number of RCG stars per unit stellar mass. Their method is analogous to that of \citet{Valenti2016}, but identifies
RCGs statistically as the excess above the smooth background of stars in the extinction-corrected magnitude
distribution. This results in $1000\pm100\Msun/\rm{RCG}$, in good agreement with the predicted value based on the
initial stellar mass function 
\citep[][see Section~\ref{s:populations}]{Kroupa2001, Calamida2015, Wegg2017}. Then
defining the Bulge and long bar as 'photometric' density excess over the constant-density inner disk,
\citet{Portail2017a} obtain $1.34\pm0.04\times 10^{10}\Msun$,
 $0.54\pm0.04\times 10^{10}\Msun$, and $1.29\pm0.12\times
10^{10}\Msun$ for the mass of the barred Bulge, long bar, and inner disk around the long bar. Added to this is the mass
of the nuclear disk, $2.0\pm0.5\times 10^{9}\Msun$. In total, the region inside radius $5.3\kpc$ thus contains
$\sim65\%$ of the MW's stellar mass \citep[$\sim5\times10^{10}\Msun$,][]{BHGerhard2016}.

\citet{Valenti2016} estimated the stellar mass from counting the number of RCG+RGB stars near the peak of the RCG in the
Bulge fields, and calibrating on a NICMOS field at $(l,b)=(0,-6\dg$) where the stellar mass function was earlier
estimated from HST data \citep{Zoccali2000}.  Scaling to the entire projected $(|l|\nlt10\dg, |b|\nlt9.5\dg)$ Bulge
region, they obtained $2.0\pm0.3\times10^{10}\Msun$ which is higher than the result from \citet{Portail2017a}.  In
determining this number an 18\% correction for disk-star contamination is applied which is determined at
$(l,b)=(0,-6\dg)$ and is taken to be constant in all fields.  This is likely to lead to an overestimate of the Bulge
stellar mass because the fraction of foreground disk RGB stars {\sl increases} towards low latitudes
\citep{Wegg2015}. In addition, the projected mass of \citet{Valenti2016} is likely to include a fraction of the long-bar
mass of \citet{Portail2017a}, reflecting the above-mentioned problem of Bulge definition.

\subsection{Summary}

The evidence from the structural and kinematic properties of the metal-rich stars suggests that
 a major part of the Bulge was built from a thin disk through evolutionary processes leading
 to bar and B/P Bulge formation, as also inferred in many external galaxies 
\citep[so-called {\sl secular evolution},][]{Kormendy2013c}. 
There is still debate about the part of the Bulge in the 
thick-disk metallicity range. As we will
see in Section~\ref{s:chemo-dynamics}, a fraction of these stars is in a barred thick disk in the Bulge, but
there is a central high-density part whose nature is currently open.

The exponential vertical density of the Bulge is typical of pseudobulges in the nomenclature of \citet{Kormendy2004,
  fisherdrory16}, however, a detailed analysis is required to quantify a possible ClB component.  \citet{Howard2008}
estimate a global $v_{\rm max}/\sigma$ for the Bulge and find that in the $v_{\rm max}/\sigma-\epsilon$ diagram of
\citet{Kormendy2004} it is located below the oblate-isotropic line, near ClBs and the B/P bulge of NGC 4565, but less
rotationally supported than other pseudobulges.

\section{STELLAR POPULATIONS}
\label{s:populations}

Stellar populations in the Bulge are characterized by their star formation
history (SFH), initial mass function (IMF), and dynamical history, and can be
studied in terms of their ages, radial velocities and proper motions,
metallicities and chemical abundances. Observational studies use tracers in many
stages of stellar evolution, from dwarf to red giants, red clump, blue
horizontal branch, RR Lyrae, Cepheids, Asymptotic Giant Branch (AGB) stars, and
Planetary Nebulae, besides AGB stages as masers, OH/IR stars among others, and
these stellar populations in the Bulge are described.  The observed chemical
evolution of selected elements in low mass stars is presented \citep[for a
  general view on abundances and chemical evolution see
  e.g.][]{pagel97,Matteucci2001}.

\subsection{Stellar population components and Ages}
\label{s:ages}

\subsubsection{Color-Magnitude Diagrams}

Age derivation from CMDs in the Bulge requires accurate photometric data.
A first modern attempt to dating the Bulge was carried out by 
\citet{ortolani95}, based on a combination of HST and ground-based data, 
indicating  that the main sequence turn-off (MSTO) is faint, 
and therefore older than $\sim$10 Gyr, as later confirmed by several authors
\citep{Kuijken2002,Zoccali2003,Brown2010,clarkson08,valenti13,gennaro15}.

\begin{figure}
\centering
\parbox{0.1cm}{
\includegraphics[width=7cm]{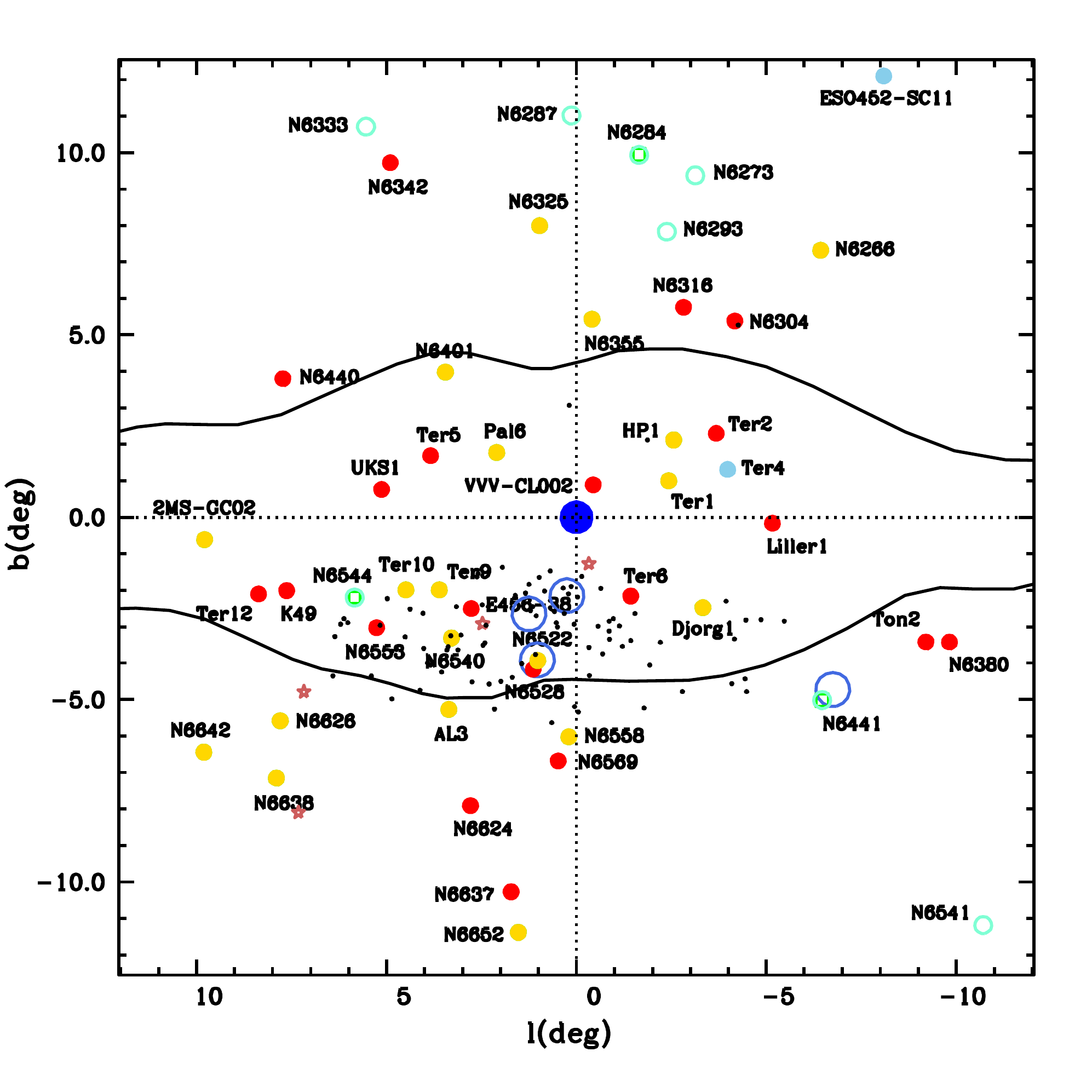}}
\begin{minipage}{3cm}
\includegraphics[width=7cm]{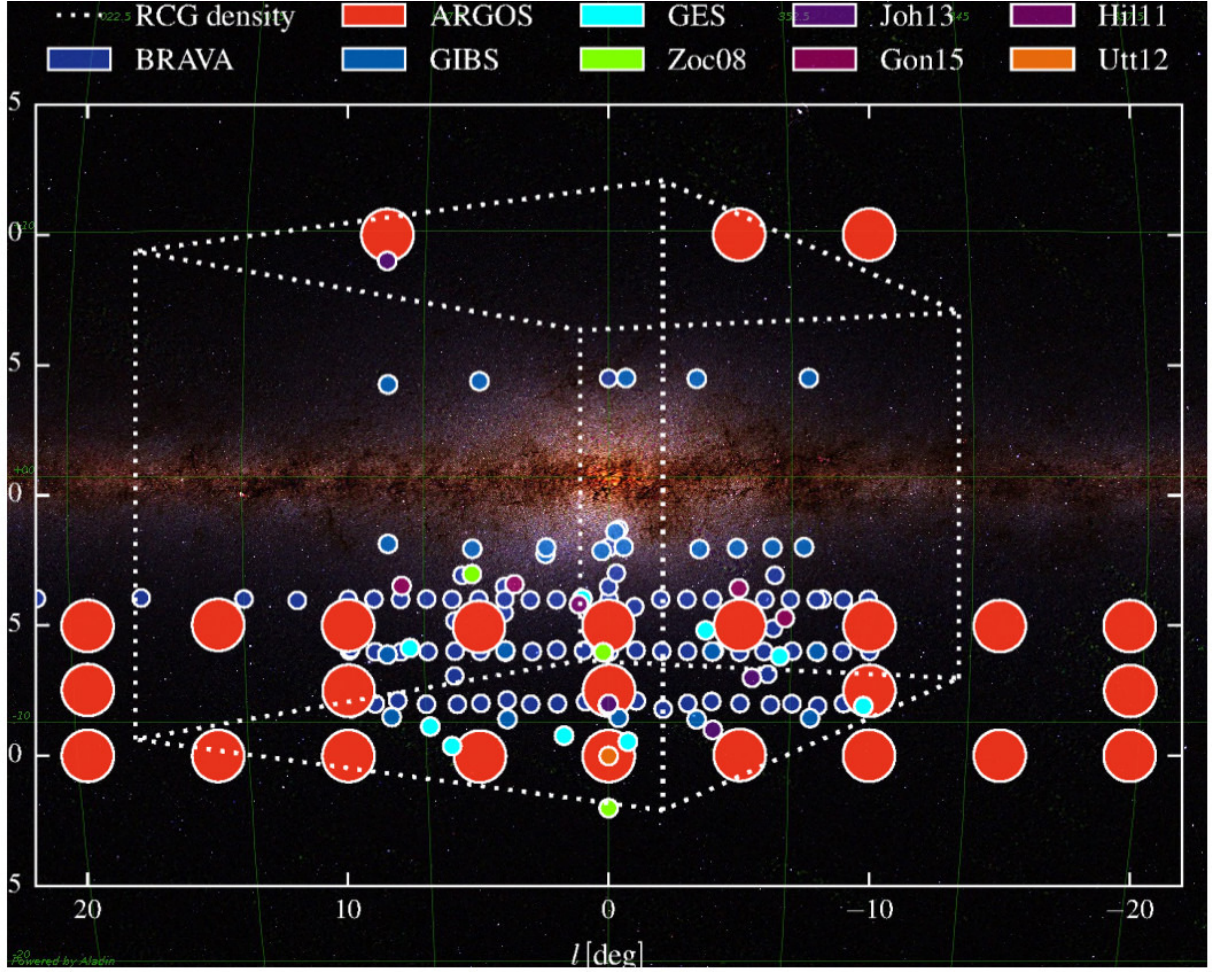}
\end{minipage}
\caption{Left panel: Location in Galactic coordinates of:
 a) Microlensed dwarfs from \citet{bensby17}:
black dots; b)  projected Bulge globular clusters cf.
\citet{bica16}, represented by filled circles, with
[Fe/H]$>$$-$0.8 (red); $-$1.3$<$[Fe/H]$<$$-$0.8 (gold);
[Fe/H]$<$$-$1.3 (blue); halo clusters: acquamarine open circles;
cluster farther than R$_{\rm GC}$$>$4.5 kpc: green open squares. 
The Galactic Center is illustrated by the  blue filled circle.
Contours correspond to COBE/DIRBE outline of the  Bulge 
from \citet{Weiland1994}, and adapted from \citet{jonsson17}. 
Coordinates (l,b) of four low-extinction fields targeted in GBTP:
BW (1.03$^{\circ}$, $-$3.91$^{\circ}$), SWEEPS (1.25$^{\circ}$, $-$2.65$^{\circ}$),
 (0.25$^{\circ}$, $-$2.15$^{\circ}$), and
  OGLE29 ($-$6.75$^{\circ}$, $-$4.72$^{\circ}$) are indicated by
 light-blue open circles. Four C-rich Miras are indicated by
indianred stars.
Right panel: Fields where metallicity distribution functions (MDFs) and 
detailed chemical abundances
were measured (see Table 1). The background is the region as seen by 2MASS. }
\label{plotlb} 
\end{figure}

Even so, one of the still debated issues in the literature is which fraction of
young stars ($\simless$ 5 Gyr) would still be compatible with the observed Bulge
CMDs. In the past, one of the main challenges in obtaining a precise answer to
this question was the CMD contamination from disk stars. More recently, thanks
to proper-motion cleaned CMDs the estimates of the contribution from young stars
became more robust.  \citet{clarkson08} presented a proper-motion cleaned CMD of
field stars at (l,b)=(1.25$^{\circ}$, $-$2.65$^{\circ}$) within SWEEPS (field
indicated in Fig. \ref{plotlb}), based on the Advanced Camera for Surveys -
ACS/HST data.  They found the CMD to be dominated by an old population with an
age of 11$\pm$3 Gyr and a mean metallicity of [Fe/H]=0.0$\pm$0.4.  They estimate
a maximum fraction of 3.5\% of young stars.  \citet{clarkson11} reported the
first detection of Blue Straggler Stars (BSS) in a SWEEPS fields.  These are old
hydrogen-burning stars, that due to accretion, have larger luminosities and
temperatures which are typical of younger stars. Therefore, to better quantify
the fraction of genuine young ($\leq$ 5 Gyr) stars, BSS have to be accounted
for.  \citet{clarkson11} were able to estimate the fraction of young-non BSS
population, made possible with accurate proper-motion cleaning based on
extensive HST data, finding it to be below 3.4\%.

More recently, \citet{gennaro15} presented the first results of the 
multi-epoch, multi-band GBTP Program (fields indicated in 
Fig. \ref{plotlb}). They build proper-motion cleaned CMDs in four
low-extinction Bulge fields, with a set of filters suitable to derive
reliable estimates of metallicities and temperatures of individual
stars \citep{Brown2009,Brown2010}.
 Their preliminary results show that there is no evidence for young
stars (5-10 Gyr), suggesting that most Bulge stars were formed in a short
 burst at early times.
Consistently with the findings of \citet{clarkson08}, the authors claim that
 their observed CMDs of Bulge stars rule out an extended star 
formation period for the Bulge population. 

On the other hand, 
as we will see in \S \ref{microlensing}, Bulge microlensed dwarfs suggest a
 different picture, 
with a much larger fraction of young stars \citep{bensby17}. 
The apparent discrepancy between the results obtained from the CMDs and microlensed dwarfs
 led \citet{haywood16} to suggest that age-metallicity degeneracies 
 could lead to a young population not detectable
 in the observed CMDs of the Bulge.
  One of the focuses of the latter work is to show that an old
 population with a broad metallicity range would not produce the tight TO as observed in the
 SWEEPS HST CMD, and that by assuming a large proportion of younger stars this claimed discrepancy
 disappears. Although the effect is clear when plotting simple isochrones, the simulation of the
 observed CMD is more complex as it depends strongly on the number density of stars corresponding
 to each isochrone. The authors' analysis focused only on colors (horizontally) to extract
 constraints in age. However, in their simulated CMD (their Fig. 8) 
a plume of MS is seen, clearly higher than  the number of RGBs 
at the same magnitude, whereas this is not observed in the data
 (their Fig. 6, specially if one accounts for BSS). If one takes these extra constraints
 into account (not only focusing on the TO)
 it becomes more difficult to believe that a young population
hides in the CMD. Further 
more detailed analysis of the Bulge CMD are needed before firm conclusions can be drawn.

\subsubsection{Microlensed Dwarfs}
\label{microlensing}
Microlensing occurs when a star, or a compact object act as a lens, magnifying a
more distant star behind it, \citet[][and references therein]{bensby17} analysed
90 dwarf stars in the Bulge. Ages were derived based on spectroscopic
determinations of effective temperature (T$_{e}$) and gravity log$~g$: for
[Fe/H]$<-$0.5 stars are 10 Gyr or older, 35\% of metal-rich stars with
[Fe/H]$>$0.0 are younger than 8 Gyr, and 26\% stars of the overall sample
younger than 5 Gyr.  \citet{bensby13,bensby17} analysed the reliability of their
findings, and pointed out biases inherent to their sample, namely, a) the
observing strategy to compose the microlensed dwarf sample tends to be biased
toward metal-rich and younger stars, and b) age derivations from a probability
distribution have a large uncertainty.  \citet{bensby17} propose a conservative
fraction of genuinely young stars of 15\%, when taking into consideration their
observational bias.  As concerns the He abundance effect on ages
\citep[e.g.][]{natafgould12}, \citet{valle15} pointed out that a He enrichment
as high as $\Delta$Y/$\Delta$Z=5 implies an age difference of $\sim$0.6 Gyr.
Even adopting the upper values of age by \citet{bensby13}, there would remain 4
stars younger than 4 Gyr, and 15 younger than 9 Gyr.  The location of
microlensed dwarfs by \citet{bensby17} indicated in Figure \ref{plotlb} are
well-distributed in $(l,b)$, whereas in depth terms, \citet{bensby13} argue that
only $<$3\% of their stars might correspond to foreground/far-side populations.

\subsubsection{Globular clusters and RR Lyrae}
\label{globulars} 
 
Globular clusters located in the Bulge probably have formed in situ, 
therefore their ages should be indicators of an early Bulge formation.

The term Bulge globular cluster comes from the landmark studies
by \citet{minniti95} and \citet{cote99}, based on
metallicity, spatial distribution, rotation with respect to 
field stars, and radial velocities have shown
 that metal-rich globular clusters within 3 kpc from the Galactic center
are associated with the Bulge/bar, and not with the thick disk.
However, \S 2 reports that presently a fraction of the Bulge is dynamically a
thick disk.

The selection from \citet{bica16} lists 43 objects, based on distance to 
the Galactic center within 3 kpc, and metallicity [Fe/H]$>$$-$1.5.
Figure \ref{plotlb}, showing their location, indicates a strikingly large 
number of clusters with metallicities $-$1.3$\simless$[Fe/H]$\simless$$-$0.8,
a metallicity peak similar to that of RR Lyrae (\S \ref{globulars}) -
 see further discussion in \S \ref{mostmp} and \S \ref{inhomogeneous}.

Among Bulge globular clusters, a prototypical moderately
metal-poor case is NGC 6522, estimated to be the oldest globular cluster
 in the Galaxy by \citet{barbuy09}.
\citet{kerber18} found that constraints from RR Lyrae stars require a normal
helium abundance Y$\sim$0.26, therefore taking into account
[Fe/H]$\sim-$1.0, [$\alpha$/Fe]$\sim$+0.3 \citep{barbuy14}, and Y=0.26,
NGC 6522 appears to have an age close to the Universe.
The prototypical metal-rich globular clusters NGC 6528 and NGC 6553
with [Fe/H]$\sim$$-$0.2, were shown to have similar age and
metallicity to BW field stars, and as well being  coeval with the
 inner halo/thick disk reference cluster 47 Tucanae by \citet{ortolani95}.
Therefore, Bulge globular clusters are old and $\alpha$-enhanced 
\citep[e.g.][]{renzini06},
similarly to most Bulge field stars.
Larger samples of Bulge globular clusters with accurate age and metallicity
are needed, in order to further confirm 
the bulge clusters as representative of the old Bulge population, 
being probably  the oldest extant objects in the Galaxy.

Another important indicator of the age and metallicity of stellar populations
 in the Bulge are the RR Lyrae (RRL). 
\citet{walkerterndrup91}
 obtained [Fe/H]$\sim$$-$1.0 with a sample of 59 RRL,
and \citet{lee1992} interpreted this population of
moderately metal-poor RRL as the oldest in the Galaxy.
 \citet{kunderchaboyer08} determined a mean [Fe/H]=$-$1.25,
with a broad metallicity distribution, from 2690 RRL monitored
 with the MACHO survey.
 \citet{Pietrukowicz2012} analysed 16836 RRL from the OGLE-III
survey, and derived [Fe/H]=$-$1.02$\pm$0.18. [Fe/H] is derived
from light curve parameters in all cases.
\citet{Pietrukowicz2015} analysed 27258 RRL from the OGLE-IV survey,
resulting in a metallicity peak at
[Fe/H]=$-$1.025$\pm$0.25. 
In summary, the metallicity peak at [Fe/H]$\sim$$-$1.0 for Bulge RRL
has been found consistently over the years, and this old
stellar population most probably has a spheroidal distribution
\citep[][and references therein]{Minniti2017}.

\subsubsection{Asymptotic Giant Branch Stars}
\label{AGBs}

The Bulge contains a large number of AGB stars, 
 supplying gas to the central parts of the Galaxy \citep{habing96}. 
They are useful tracers of the structure and history of the
 Bulge, as they cover a large age-baseline.
Even if the age-dating of AGBs, Miras and Planetary Nebulae (PNe) is
not well-constrained, as exemplified in \citet{guarnieri97},
it is important to note that the fraction of young microlensed dwarfs
(\S \ref{microlensing}) have to show a counterpart in more advanced 
stages of evolution.

\citet{Uttenthaler2015} report that  among 45 AGBs,  those in the Bulge 
are oxygen-rich, therefore old, and those in 
 the inner Bulge locations are metal-rich 
with kinematics compatible with belonging to the bar.
Outer-bulge ones, located in Plaut's field
at b$\sim$$-$10$^{\circ}$  are metal-poor ([Fe/H]$\sim-$0.6) with radial
velocity dispersion compatible with a metal-poor
spheroidal component of the Bulge. Previously, \citet{Uttenthaler2007} 
identified short-lived technetium in  3 outer-Bulge AGBs.  

Mira variables are AGBs in their last stage of evolution, to end
 later as PNe. \citet*[][and references therein]{catchpole16} 
observed 648 bulge oxygen-rich Mira variables
combined with another 8057 inner-Bulge ones from the OGLE, MACHO and
2MASS surveys. The 5 Gyr and younger long period Miras, including the
 OH/IR stars   show a triaxial bar-like structure. The shorter period
 older ones appear spheroidal or near-spheroidal
 in the outer Bulge, making evident that the Bulge shape depends on
 the tracer used. \citet{matsunaga17} confirm 4 C-rich Miras located in the
 Bulge (indicated in Fig. \ref{plotlb}). They would be
of intermediate age, but could still be old if formed from binary evolution.
Old type II Cepheids are a magnitude more numerous than younger
classical Cepheids in the Bulge.
In a reanalysis of extinction and distances of young classical Cepheids by
 \citet{matsunaga15}, and \citet{Dekany2015},
\citet{matsunaga16}  verified that no classical Cepheids are
found in the Bulge volume within 2.5 kpc of the Galactic center,
except for four of them in the Nuclear Stellar Disk (NSD, \S \ref{NSD}).
Bulge masers, that are single stars of intermediate mass (1-5M$_{\odot}$), and 
ages of 0.8 to 5 Gyr, in a last-episode of AGB stage,
 in a phase slightly less evolved than PNe, are found in the bar
 \citep{habing16}. \citet{gesicki14} derived masses and ages of central stars of
PNe and find ages of 3-10 Gyr.
They derive ages scaling them to those by \citet{bensby13}, therefore
it is not an independent determination, and further investigation is needed.

In conclusion, there might be presence of intermediate-age objects in the Bulge,
although in small numbers.
It is important to point out that the fraction of intermediate age/young dwarf stars
 in the Bulge,  suggested by results from \cite{bensby13,bensby17}, has to show a counterpart
 as C-rich Miras, and PNe. The current data discussed in this section
 shows that there is still no clear evidence for that, and that even in the case of
 the 4 C-rich Miras recently confirmed, there is the possibility that these evolved objects
 are old if belonging to binary systems.

\subsubsection{Detected young populations in the Nuclear Disk}
\label{NSD}

In the innermost parts of the Bulge resides the Nuclear Stellar Disk (NSD)
\citep{launhardt02}, that has a vertical extent of 
b$<$0.4$^{\circ}$ or $\sim$50 pc
 and Galactocentric radius of $\sim$150 pc \citep{schonrich15,BHGerhard2016},
coinciding with the Central Molecular Zone (CMZ) gas motions.
It includes young massive O/B stars in the stellar
 clusters Arches, Quintuplet and the Nuclear Star Cluster (NSC).
 Mass-losing AGBs with SiO masers in the inner bar region were surveyed by 
\citet{fujii06}, where from 291 objects, 163 SiO masers were detected,
with radial velocities compatible with the bar.
 Classical Cepheids from
\citet{matsunaga15}, and \citet{Dekany2015}, reverified by
\citet{matsunaga16}  resulted in four of them in the NSD.
Previously,  \citet{vanloon03} had found  fractions of intermediate
and young M-type RGB and AGB stars in the inner central kpc.

The gas to form these young stars could have been  lost from AGB (and RGB) 
Bulge giants, and/or be due to
  gas flow from the dynamical evolution of the bar, or else
 to accretion of a dwarf galaxy. 

\subsection{Metallicity Distribution Function}
\label{MetallicityDF}

\subsubsection{Metallicity Distribution Function traced by giants and dwarfs}

Metallicity Distribution Function (MDF) surveys of Bulge
stellar populations have used as tracers
 Red Giant Branch (RGB), Red Clump (RC),
and M giant stars. 
RCG tracers (\S 2)  are useful tracers also in terms of spectroscopy
 because their gravity is known, 
with a typical value of log~g=2.5$\pm$0.5 for metal-rich 
and log~g=2.5$\pm$0.2-0.3 for metal-poor stars of
all ages. In the case of a single-aged population, log g is known within 
0.1 dex  \citep{girardi16}.
A limitation of RCGs is that they characterize 
metal-rich populations, with [Fe/H]$\simgreat-$1.5.
 RGB stars have the advantage of including all metallicities, 
but require limits in magnitude range. M giants are bright, 
and as RCGs, they might miss the most
 metal-poor component of the bulge population \citep{Rich2007}.

Early MDF diagrams conclusively indicated that the Bulge contains a metal-rich
population (\S 1).  More recently, \citet{Zoccali2008} presented MDFs for over
500 RGB stars in the Bulge, in three fields along the minor axis, showing that
metal-rich stars were present in inner regions, gradually disappearing with
increasing latitudes.  \citet{Babusiaux2010} further inspected these data in
terms of kinematics, confirming that there are two stellar populations: a
metal-rich one showing a clear vertex deviation compatible with the bar, and a
metal-poor population compatible with an spheroid or a thick disk (see \S 2).
The MDF studies by \citet{Hill2011}, \citet{johnson13},
\citet{bensby13,bensby17}, \citet{rich12}, \citet{Schultheis2017}, and the
larger surveys ARGOS, GES, and GIBS are summarized in Table \ref{MDF}.  In this
table are listed the main MDF studies, indicating the location (l,b), tracer,
number of stars, (J-K) color selection, the metallicity peaks reported, and
corresponding references.  A few more details on the MDF of these authors,
including selection functions, can be found in \S \ref{s:kinmet}.

\begin{figure}
\centering
\parbox{4cm}{
\includegraphics[width=4cm]{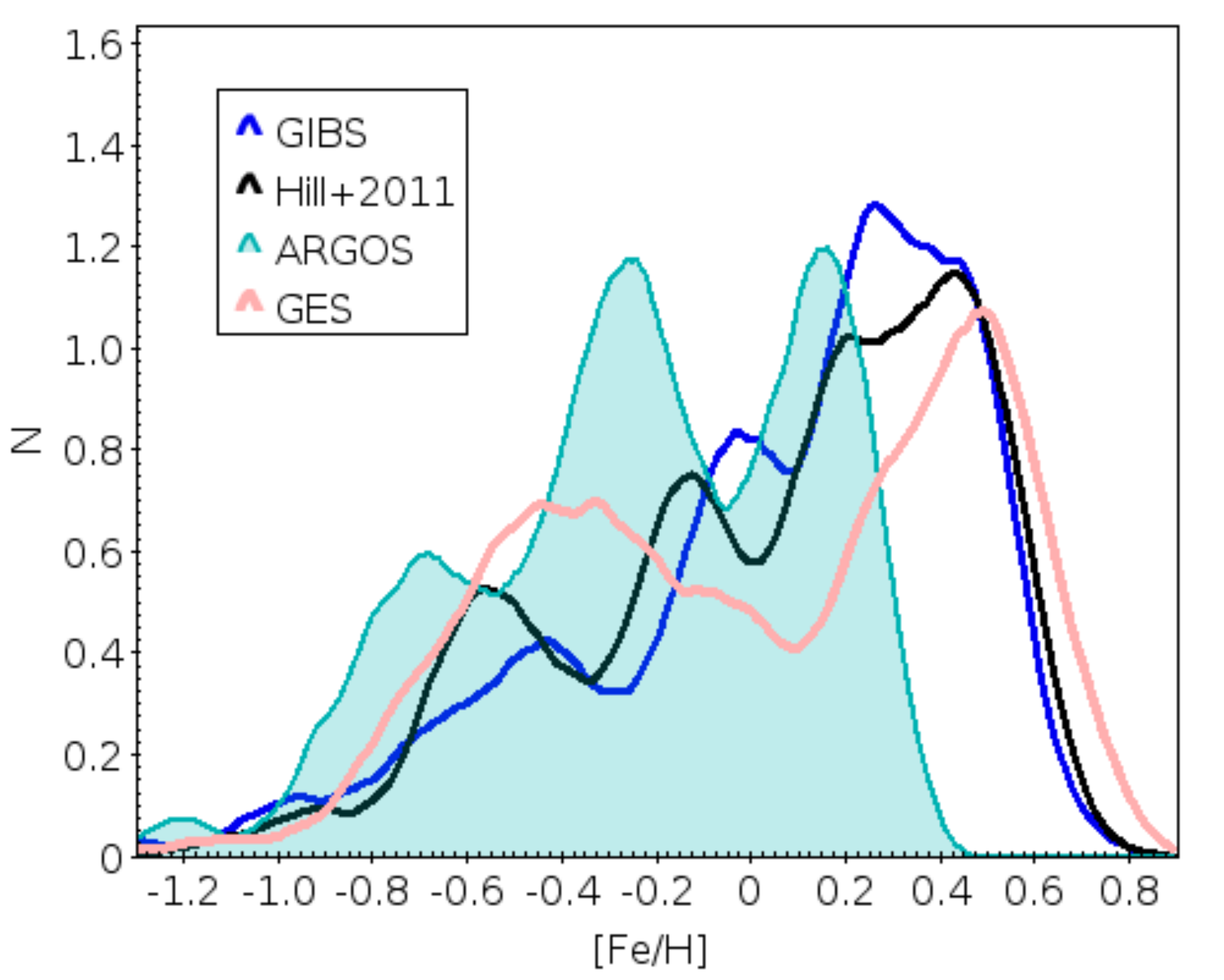}}
\begin{minipage}{4cm}
\includegraphics[width=4cm]{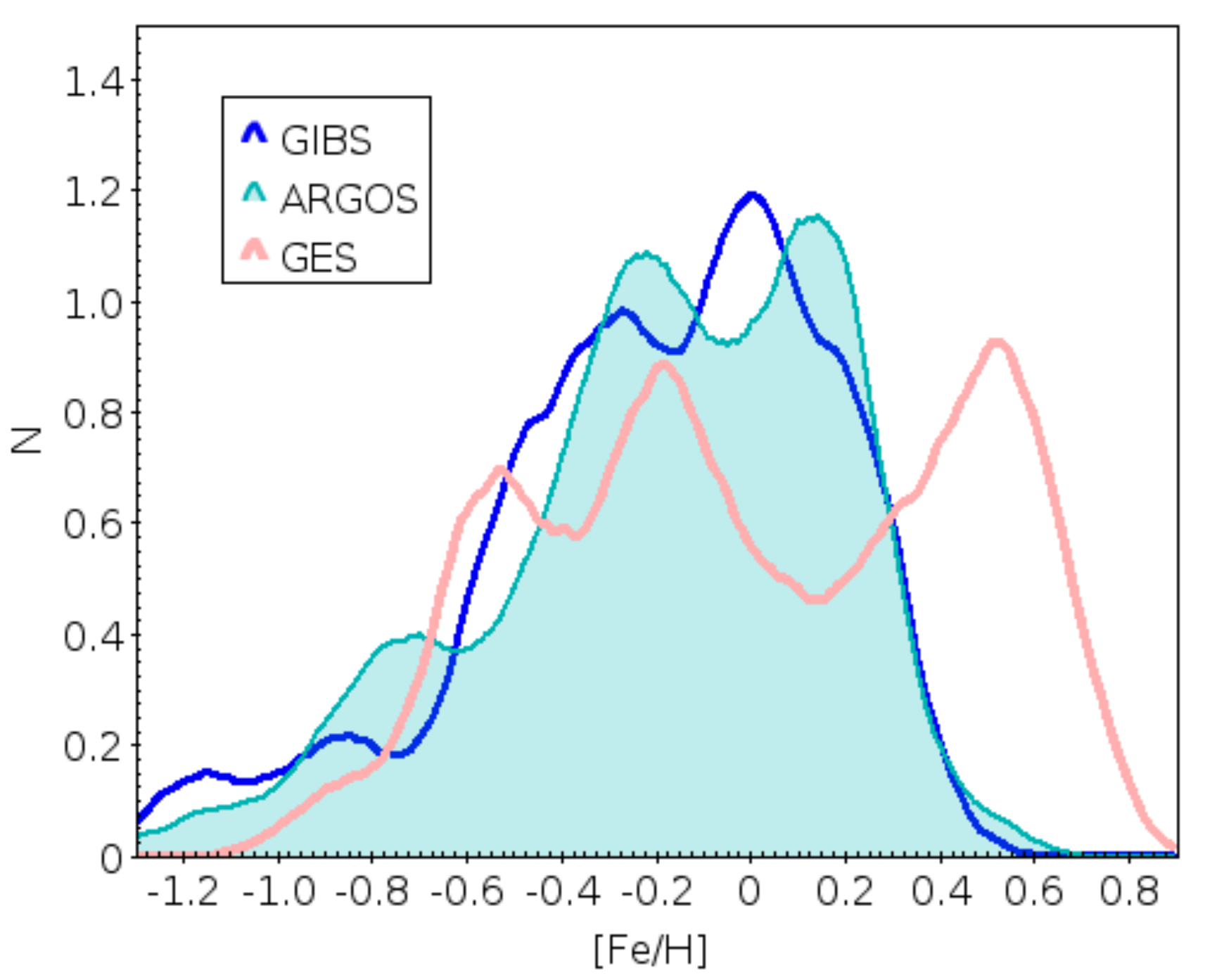}
\end{minipage}\\
\parbox{4cm}{
\includegraphics[width=4cm]{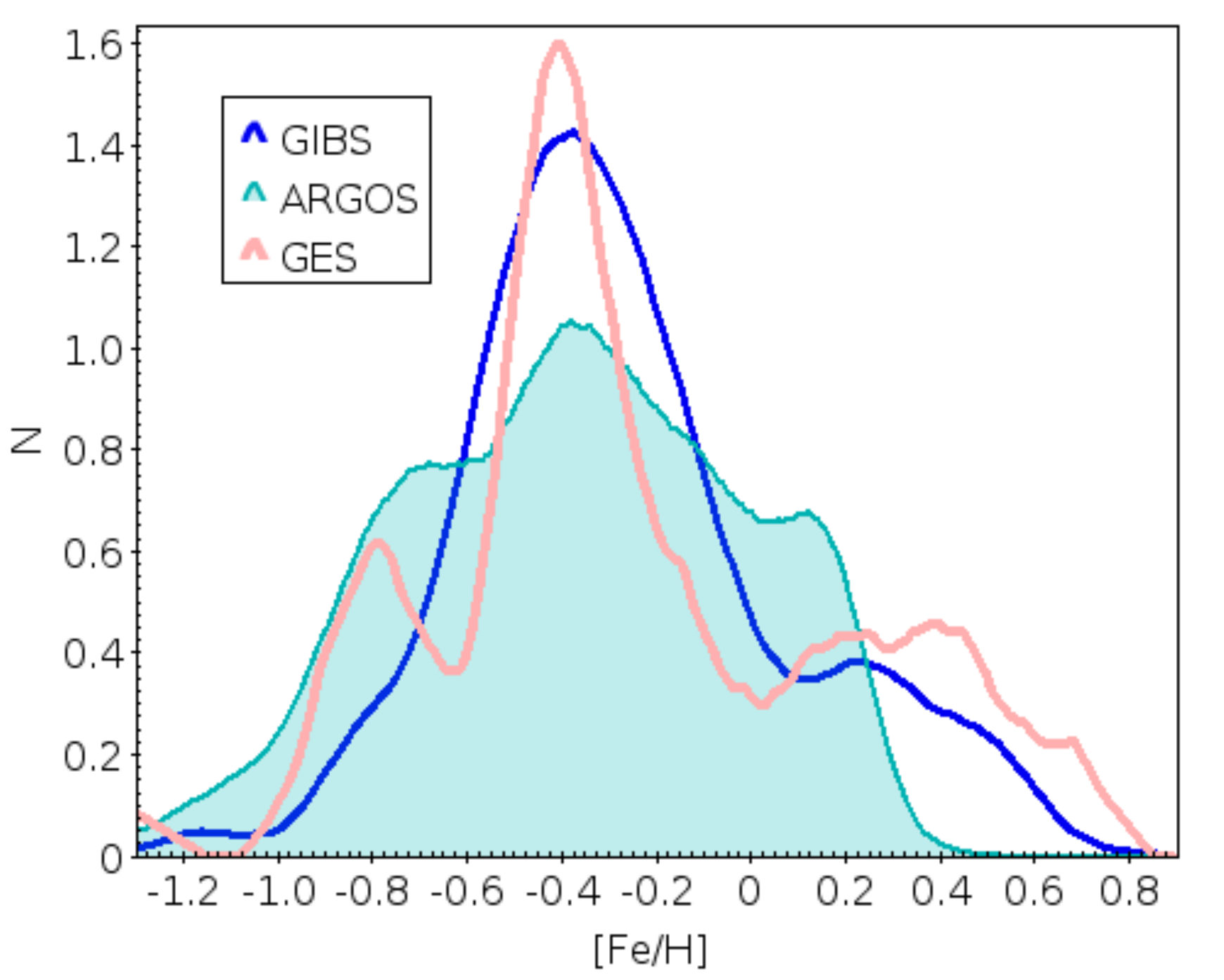}}
\begin{minipage}{4cm}
\includegraphics[width=4cm]{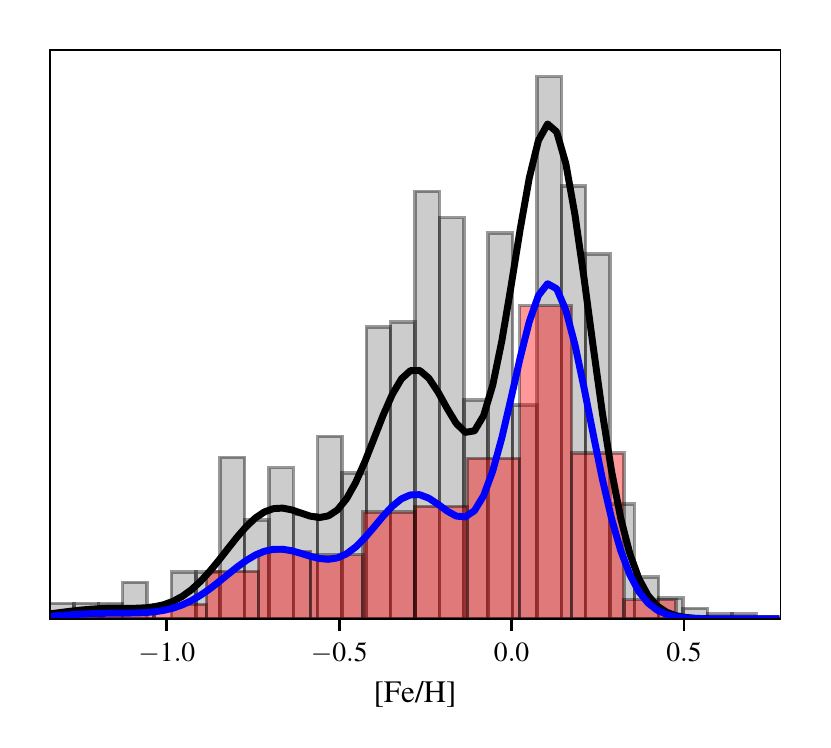}
\end{minipage}
\caption{MDF based on ARGOS, GIBS, GES surveys and Hill et al. (2011),
for fields along the minor axis at: 
(l,b) = (1$^{\circ}$,-3.9$^{\circ}$) -- BW (top left panel), 
 (0.2$^{\circ}$,-6$^{\circ}$) (top right panel), and (359.6$^{\circ}$,-8.5$^{\circ}$)
 (bottom left panel). 
The bottom right panel illustrates the effects of survey selection function and survey statistics on
the measured MDF. Grey histogram: ARGOS MDF for field (l,b) = (0$^\circ$,-5$^\circ$). 'True' bulge MDF
reconstructed  by Portail et al. 2017b from ARGOS and APOGEE data in multiple fields, taking into account
the survey selection functions, is shown in black when integrated over distances $R_0\pm3.5$ kpc, and in blue
when integrated over $R_0\pm1$ kpc. Red histogram: The $R_0\pm1$ kpc distribution binned in a histogram
appropriate for $\sim 200$ stars.} 
\label{mdfok} 
\end{figure}

Figure~\ref{mdfok} shows the MDF from
the ARGOS, GIBS and GES surveys, along the minor axis.
It shows the probability 
density distribution of [Fe/H], using
a kernel function with a smoothing parameter of 0.17 dex, which is 
close to the typical accuracy in metallicity. In the top left panel
corresponding to the BW at (l,b) = (1$^{\circ}$,$-$3.9$^{\circ}$), the data
 from \citet{Hill2011} are also included.
The  top right and bottom left panels show similar comparison for fields located at
 (0.2$^{\circ}$,-6$^{\circ}$),
and (359.6$^{\circ}$,-8.5$^{\circ}$), respectively. The
central field coordinates are from GIBS, and they are compared to GES
and to the several small ARGOS fields
located nearby (within 1.0$^{\circ}$ of the GIBS field centers).
In the case of the GIBS survey, the two first panels have data coming from 
high-resolution spectra, whereas the right panel comes from moderate-resolution spectra 
\citep[see][their Fig. 4]{Zoccali2017}.

One of the caveats of comparing MDFs from different surveys is that their shapes
may be affected by selection effects as illustrated in the last panel of
Fig.~4. Because of the survey selection function designed to give high weight to
the outer bulge (see Section 2), the ARGOS MDF in this
(l,b)=(0$^\circ$,-5$^\circ$) field has a larger number of stars around
[Fe/H]$=-0.3$ than the MDF reconstructed by Portail et al. 2017b when including
the selection function. Their relative number further decreases when only stars
within 1~kpc around $R_0$ are included, which is closer to the selection
function of GIBS. Finally, this panel also illustrates the effect of survey
statistics. In a coarse histogram as would be appropriate for $\sim200$ stars,
the peak around [Fe/H]=-0.3 seen in the original data would be very hard to see.

For this reason, we here do not concentrate on the detailed peak structure
differences among the samples, but on more global aspects such as their
metallicity range. In this framework, the main discrepancy seen between the MDFs
traced by ARGOS with respect to GIBS, GES or \citet{Hill2011} in BW, is that it
finds fewer metal-rich ([Fe/H] $>$ 0.3 dex) stars. Note that the better
agreement between GIBS and Hill et al. (2011) is most probably due to their
similar sample selection, and that ARGOS used a different calibrator (the open
cluster Mellott 66) for the metal-rich end.  Also GES finds a large number of
metal-rich stars in BW.  Another caveat is that the ARGOS fields close to BW are
are not exactly on BW, located around 0.5-1$^{\circ}$ below.  However it is
difficult to explain such a large discrepancy only in terms of field location or
any rigid metallicity shift between the ARGOS and the other metallicity scales.
One explanation could be that the ARGOS field close to BW (but not on BW) goes
less deep into the Bulge, due to the large extinction, and might miss the
metal-rich Bulge population at lower latitudes.  Samples of RCGs in low
extinction regions such as the BW \citep{Hill2011,Schultheis2017} or samples
including more luminous RGB stars, as for instance the one of
\citet{Babusiaux2014}, including high extinction fields, are able to detect the
Bulge metal-rich component.  \citet{Schultheis2017} suggest that the additional
peak found in the ARGOS survey might be an artifact from the derivation of
temperatures from the (J-K)$_{\circ}$ colors, and the adoption of samples over a
large longitudinal area, where small systematic variations are expected. In
fact, differences were found in temperatures derived from J-K, and from
high-resolution spectral analysis of five metal-poor stars from the ARGOS survey
(all of them with [Fe/H]$_{ARGOS}$ $\simeq -$0.8) by \citet{siqueira-mello16},
reaching differences of over 200 K, and [Fe/H]$_{\rm ARGOS}-$[Fe/H]$_{\rm
  UVES}$= $+$0.1, $-$0.2, $+$0.2, $+$0.15 and $-$0.37.  Otherwise, despite the
differences in temperature and metallicity, the [$\alpha$/Fe] ratios from ARGOS
and those from UVES are compatible for 4 out of 5 stars.  Cross-checks of larger
samples with the ARGOS metallicity scale are still missing towards the more
metal-rich end.  Moreover, as discussed before, the ARGOS selection effects can
also play a role on the location of the peaks.  In the second panel the MDF of
GIBS and ARGOS are similar, whereas GES still sees a large number of metal-rich
stars. When combining GIBS fields at different longitudes, but with constant
latitute $|b|=-$6$^{\circ}$, the over-enhancement in metal-rich stars reappears
in GIBS \citep[see][their Fig. 7]{Zoccali2017}, being again compatible with
GES. This indicates that at that latitude, a large number of metal-rich stars
are present, which is again not seen by ARGOS.  In the third pannel, at latitute
$|b|=-$8.5$^{\circ}$, both GIBS and GES give consistent results towards the
metal-rich end.

% Table
\begin{table}
\tabcolsep3.25pt
\caption{Summary of the Bulge Metallicity Distribution Functions properties
 obtained by different authors using different tracers.
Baade's Window (BW) is centered at (l,b)=(1.03$^{\circ}$,-3.91$^{\circ}$), with angular size of $\sim$3.6'x3.6'.}
\label{MDF}
\begin{center}
\begin{tabular}{|l|c|c|c|c|c|c@{}}
\hline
Targets  &[Fe/H] peaks & N &
 J-K & (l,b) & Reference  &\\
\hline
RGB &$\sim$+0.3,$\sim$$-$0.3,$\sim$$-$0.7 &650 & --- & BW;0.21,$-$6.02;0,$-$12 & 1&\\
\hline
RGB & +0.3,$-$0.6 &401 & --- & 0,$-$10 &2 &\\
\hline
RGB & $-$0.08,$-$0.29,-0.44 &264 & --- & $-$5.5,$-$7;$-$4,$-$9;8.5,+9;0,$-$8 & 3 &\\
\hline
RGB & +0.3,$-$0.29 &269 & --- & BW & 4 &\\
\hline
RC &$-$0.30,+0.32 & 219 & (J-K)$_o$$>$0.50 & BW & 5 & \\
\hline
RC & +0.3,$-$0.4,$<$-0.8 & 5500  & (J-K)$_o$$>$0.40 & $-$10$<$l$<$+10,$-$10$<$b$<$+5 & 6 & \\
\hline
RC &
+0.14,$-$0.23,$-$0.60,$-$1.24,$-$1.7
& 14,150  & (J-K)$_o$$>$0.38 & $-$5$<$b$<$$-$10 & 7\\
\hline
RC & +0.1/+0.2,$-$0.37/$-$0.77 & 1,200  & (J-K)$_o$$>$0.38 & $-$10$<$l$<$7,$-$10$<$b$<$$-$4 & 8 &\\
\hline
RGB & $-$0.36$\pm$0.08,+0.40$\pm$0.05 & 1583  & (J-K)$_o$$>$0.38 & $-$10$<$l$<$+10,$-$10$<$b$<$$-$4 & 9 & \\
\hline
M stars & $-$0.20 & 30 &--- &0,$-$1.75, 1,$-$2.65,0,$-$1,BW & 10 &\\
\hline
Dwarfs &$\sim$+0.4,+0.08,$-$0.2,$-$0.68,$-$1.2,$-$1.9 &58 &--- &$-$4.86$<$l$<$6.36,$-$4.51$<$b$<$5.27 &11 &\\
\hline
Dwarfs &$\sim$+0.41,+0.12,$-$0.2,$-$0.63,$-$1.09 &90 &--- &$-$6.00$<$l$<$6.50,$-$6.00$<$b$<$5.27 &12 &\\
\hline
\end{tabular}
\end{center}
\small{Sources: 1. \citet*[][their Fig. 14]{Zoccali2008}; 2. \citet{Uttenthaler2012}; 3. \citet{johnson13}; 
4. \citet{Schultheis2017}; 5. \citet{Hill2011}; 6. \citet{Zoccali2017}; 7. \citet{Ness2013a}; 
8. \citet{Rojas-Arriagada2014}; 9. \citet{Rojas-Arriagada2017}; 10. \citet{rich12}; 
11. \citet[][their Fig. 12]{bensby13}; 12. \citet{bensby17}.}
\end{table}

In summary, this figure illustratres how complex still is the analysis of the
Bulge MDF, and that it is strongly dependent on target selection, tracers,
observed fields, and anaysis methods. This prevents to clearly disentangle the
mixed populations in the Bulge. Therefore, it is still unclear if there is or
not a genuine chemical gradient in the Bulge, or if it is only the result of the
gradient of stellar populations. Because the MDF represents a crucial constraint
to models of bulge formation (\S 4), more complete and homogenous surveys of the
inner Galaxy region are urgently needed.

\subsubsection{Other tracers of the metallicity distribution}
\label{PNe}

 It is important to discuss how other tracers of the Bulge MDF
 compare with those for giant and dwarf stars. 
Planetary nebulae (PNe) are the representatives of the end of AGB
stellar evolution for stars with initial mass between
 $\sim$0.8 and $\sim$8 M$_{\odot}$.
 The currently known number of Bulge PNe is reported to be of
$\sim$800, although estimates as high as 3500 are given in the literature,
 with abundances derived for about 300 of these according to \citet{cavichia17}.
They analysed 17 high-extinction PNe located
in the inner 2$^{\circ}$ of the Bulge, detecting
considerable difference between inner- and outer-Bulge PNe,
with inner-Bulge ones indicating higher progenitor masses, with
enhanced N/O, being probably younger objects, compatible
with their location closer to the bar.
 A large sample of 166 Bulge, and 60 inner-disk PNe, were analysed by
\citet{chiappini09},
 carrying out a detailed comparison between oxygen abundances in these samples and
Bulge field RGBs. PNe have a great advantage with respect to other tracers, namely,
 the fact that in this case it is possible to resort different selection criteria 
for assigning them to the Bulge,
 and hence more easily identify foreground (inner-disk) objects. \citet{chiappini09} adopted 
several independent selection criteria to define their Bulge PNe sample, based on their 
diameter and radio fluxes, as first proposed by 
\citet{stasinska94}, 
and objects not fitting these criteria, but in the Bulge direction, 
were identified as inner-disk PNe. 
They found that abundances of Bulge and inner-disk
PN samples coincide, whereas they show
 a shift of 0.3 dex lower relative to RGB stars,
as recently confirmed by \citet{pottasch15}. To explain this difference, 
apart from several hypotheses on the analyses, a possible one in terms
of stellar evolution is that very metal-rich progenitors would not reach
the PNe stage, or that PNe had slightly more massive progenitors, and
in this case different populations are being compared.

\subsubsection{Vertical (and radial) metallicity distribution gradients}
\label{vertical}

Vertical metallicity gradients measured from
 b $= -$4$^{\circ}$ to $-$12$^{\circ}$ of  $-$0.075 dex/deg.
 $-$0.04 dex/deg, and $-$0.05 dex/deg were estimated by
 \citet{Zoccali2008}, \citet{Gonzalez2013}, and \citet{Rojas-Arriagada2014}
respectively. However,
this apparent vertical metallicity gradient probably mostly comes from the 
changing proportions of stellar population components with latitude,
and not from inherent gradients in each component.
In other words,
\citet{Zoccali2008}, \citet{Hill2011}, \citet{bensby13}, \citet{Ness2013a}, 
\citet{Rojas-Arriagada2014}, \citet{Rojas-Arriagada2017}  have shown that
metal-rich populations are dominant only at low latitudes, i.e.,
there is a variation in the relative weight of components, with a larger
proportion of metal-rich stars in the central regions.
The metal-poor component may show a true gradient with 
latitude \citep{Rojas-Arriagada2014}.

\subsubsection{The most metal-poor stars} 
\label{mostmp}

Why search for the most metal-poor stars in the Bulge? 
It is believed that the first objects could have formed starting from
 around 550 Myr after
 the Big-Bang (Planck collaboration: Ade et al. 2016).  These first stars 
would have formed
 within dark matter minihaloes, and some of them would constitute proto-bulge pieces 
\citep[e.g.][]{whitespringel00,nakasatonomoto03,tumlinson10}.
Some of these models are discussed in more detail in \S \ref{evolutionary}
and \ref{cosmological}.
Recently, evidence that the oldest stars are found towards the Galactic center
has been presented \citep{Carollo2016,Santucci2015},
through a study of Blue HB stars to derive their ages.
Their so-called Ancient Chronographic Sphere extends from the region of
 the Bulge out to 10-15 kpc away, with a slope of the age gradient of $-$25 Myr/kpc.

The search for the earliest metal-poor stars in the Bulge was clearly suggested
in the 2000's \citep[e.g.][]{miralda-escude00}.  A systematic search for very
metal-poor stars in the Bulge is being currently carried out in the large
surveys.  Fractions of stars with [Fe/H]$\leq$$-$1.5 of 1.1\%, 0.2\%, 0.7\%, and
0.2\% only were found respectively by \citet{Rojas-Arriagada2014},
\citet{Rojas-Arriagada2017}, \citet{Ness2013a,Ness2013b}, and
\citet{Zoccali2017}. Note however that these low numbers are partially due to
survey selection.  \citet{garcia-perez13} identified in the APOGEE survey 5
stars with $-$2.1$<$[Fe/H]$<-$1.6, from a sample of 2403 stars.  In EMBLA,
\citet[][and references therein]{howes16} were able to identify over 300
candidate metal-poor stars with [Fe/H]$<-$2.0 in the Bulge, out of 14,000 bulge
stars preselected from photometry, and 500 of these from spectroscopy. Based on
high-resolution spectra of these candidates they analysed and confirmed 37 stars
with [Fe/H]$<$-2.0.  \citet{lamb17} confirmed two candidate RGB stars within
EMBLA, with metallicities [Fe/H] = $-$1.51, $-$2.06.  \citet{schlaufman14} and
\citet{casey15} found 3 stars with $-$3.1$<$[Fe/H]$<$$-$2.7, based on metal-poor
stars being bright in J, WISE W1, W2, whereas W2 is faint in metal-rich stars.
\citet{koch16} analysed 3 bulge stars located around
(0$^{\circ}$,$-$11$^{\circ}$). So far, no more than about 50 stars more
metal-poor than [Fe/H]$\simless$-2.0 were identified in the Bulge.

A dearth of very metal-poor stars in the Bulge
 leads to the question as to whether
 the moderately metal-poor stars are in fact the oldest Bulge stars,
 as a consequence of a fast chemical enrichment in the dense central parts.
The following evidence shows that there may be an old stellar
population peaking at [Fe/H]~$\sim-1.0$ in the Bulge:
a) The metallicity distribution of Bulge  globular clusters
was shown to have two peaks at  
[Fe/H]~$\approx-0.5$ and [Fe/H]~$\approx-1.0$ \citep{bica16},
as had already been pointed out in \citet{barbuy09,rossi15}.
b) Another piece of evidence comes from the old RRL that
have a metallicity peak at [Fe/H]~$\approx-1.0$ (see Sect. \ref{globulars}).
c) \citet{Schultheis2015} found a peak of
$\alpha$-enhanced Bulge-field  stars at [Fe/H]~$\sim-1.0$.
d) \citet{schiavon17} identified a sample of nitrogen-rich
stars that also show metallicities of [Fe/H]~$\sim-1.0$. 
In conclusion, due to the fast chemical enrichment of the central parts of
 the Galaxy, it is not expected 
that the bulk of the  first Bulge stars would be as metal-poor as the
 oldest halo stars. 
Instead, the oldest bulge stars could have metallicities around 
[Fe/H] $\simgreat -$1.5, as suggested by 
\citet{chiappini11,Wise2012}. This is more extensively discussed in
Cescutti et al. (2018, in preparation).

\subsection{Abundances in the Galactic Bulge}
\label{abundances}

Chemical abundances provide signatures of the formation and evolution of
different stellar populations, and their studies are interpreted as near-field
cosmology \citep{freemanBH02}.  This task, which can be complex even in a very
local volume near the Sun \citep{fuhrmann11}, turns out to be extremely
challenging in the Bulge.  As discussed in \S 3.2, the different stellar
populations that co-exist in the Bulge comprise not only the inner-halo, and
traces of younger stars associated with the bar and NSD, but also other two very
important unknowns, namely, the \emph{inner} thick and thin disks.  Therefore
important caveats need to be kept in mind when trying to infer any conclusion
from the chemical abundances of the Bulge and how it compares to that of other
Galactic components.  These are: a) accounting for the fact that different
authors use different selection criteria and definitions for their bulge, thin-
and thick-disk samples, b) Bulge-field samples are most probably contaminated by
foreground or far-side disk stars, particularly in the case of microlensed
dwarfs (due to the large uncertainties in their distances); c) the presence of
halo, thin and thick disk stars in the Bulge.  With these caveats in mind, we
here summarize what at present are the available observations for each of the
chemical elements, with the ultimate goal of being able to extract more solid
constraints for models. Samples with large abundance spread are included in
grey, adopting $\pm$0.18 as a typical uncertainty in metallicities [Fe/H] and
$\pm$0.15 in logarithmic abundance ratios [X/Fe].  Solar abundances adopted can
show small differences among sources of reference.  In the [X/Fe] vs.[Fe/H]
plots presented below, O abundances were scaled to $\epsilon$(O)=8.76
\citep{steffen15} \citep[see][]{friaca17}; the Mg, Mn, Zr, Eu Solar abundances
have negligible differences among authors and were not considered; For La, Ba,
even if differences in scale of 0.1 dex among authors can be present, they were
not scaled, due to difficulties in taking into account differential studies
relative to stars other than the Sun (Arcturus, $\mu$ Leo), and this should be
kept in mind.  Theoretical predictions for the Bulge chemical enrichment are
also overplotted to the data, with the goal to illustrate some of the results of
the different modeling approaches. Details on these models can be found in \S 4.

\subsubsection{$\alpha$-elements}
\label{alphaelements}

The so-called $\alpha$-elements include elements with nuclei multiple of the
alpha ($\alpha$) particle (He nuclei).  The $\alpha$-elements observed in the
Bulge are O, Mg, Si, Ca.  For Ti see \S \ref{iron-peak}.  Among these, O and Mg
are produced during hydrostatic phases of high-mass stars, whereas Si, Ca are
produced mostly by a combination of pre-explosive and explosive burning in
CCSNe, also called supernovae type II (SN II), with smaller contributions from
supernovae of type I (SNIa).

\paragraph{Oxygen}
\label{oxygen}

\begin{figure}
\includegraphics[width=12cm]{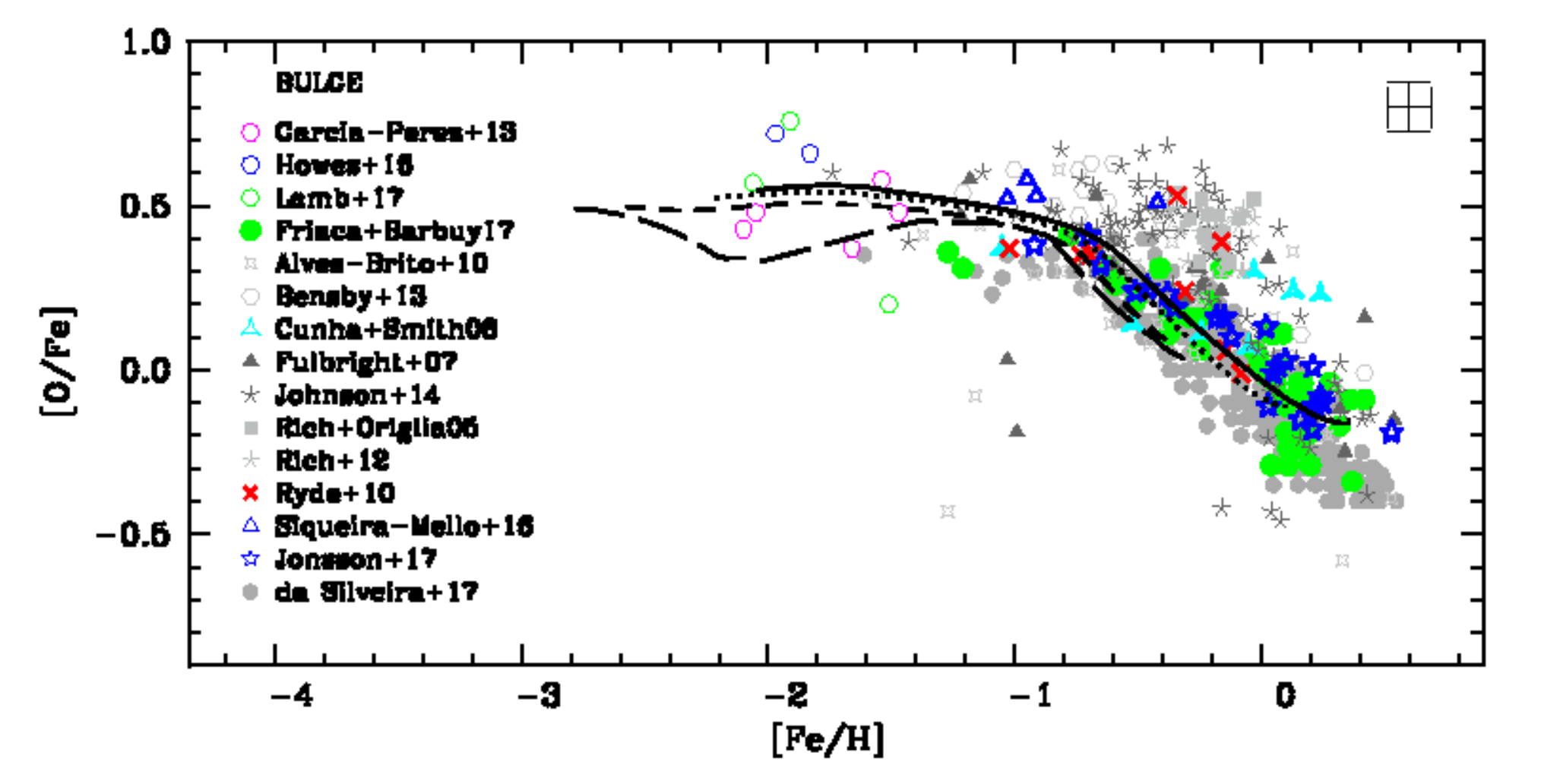}
\caption{ [O/Fe] vs. [Fe/H]: literature abundances for bulge stars. 
magenta open pentagons: \citet{garcia-perez13};
blue open pentagons: \citet{howes16}; 
green open pentagons: \citet{lamb17};
green filled circles: \citet{friaca17};
gray 4-pointed stars: \citet{alves-brito10}
(same stars as \citet{melendez08});
grey open hexagons: bulge dwarfs by \citet{bensby13};
cyan 3-pointed stars: \citet{cunhasmith06};
grey filled triangles: \citet{fulbright07};
grey stars: \citet{johnson14};
grey filled squares: \citet{richoriglia05};
grey stars: \citet{rich12}; 
red crosses: \citet{ryde10};
blue open triangles: \citet{siqueira-mello16};
blue 5-pointed stars: \cite{jonsson17};
grey filled circles: \citet{dasilveira17};
lightgrey filled squares: \citet{Schultheis2017}.
Color-coded choices follow explanation in the text.
Chemodynamical evolution models with formation timescale of 2 Gyr, 
or specific SFR of 0.5 Gyr$^{-1}$ are overplotted.
Solid lines:  r$<$0.5 kpc; dotted lines: 0.5$<$r$<$1 kpc;
dashed lines: 1$<$r$<$2 kpc; long-dashed lines: 2$<$r$<$3 kpc.
}
\label{oo} 
\end{figure}

Oxygen abundances as a function of metallicity are one of the most robust
indicators on the process of Bulge star formation rate (hereafter SFR) 
and chemical evolution,
clearly more so than the other $\alpha$-elements, especially because 
in this case no contribution from SNIa is expected
\citep[e.g.][]{friaca17}. 
 \citet{Woosley2002} described the 
 nucleosynthesis processes in massive stars.

 \cite{Zoccali2006}, \cite{lecureur07}, and \cite{fulbright07}, found Bulge
 oxygen-to-iron abundances slightly higher than those found for thick-disk stars
 by \citet{bensby04}.  This would suggest that the thick disk and Bulge
 components did not share the same chemical-enrichment history (or IMF).  On the
 other hand, \citet{melendez08} and \citet{alves-brito10} examined a sample of
 thick-disk and Bulge giants, covering the same stellar parameter space, and
 found no difference between the two samples.  So far it is not clear if there
 is a systematic shift between the oxygen-to-iron abundances in the Bulge and
 thick disk, as discussed below.

Fig. \ref{oo} shows [O/Fe] vs. [Fe/H]  in Bulge giants.
 The  C,N,O abundances  from \citet{Zoccali2006} and \citet{lecureur07}
 are replaced by revised calculations carried out by \citet{friaca17},
where both the C$_2$(0,1) bandhead at 5635.2 {\rm \AA} and
the CI 5380.3 {\rm \AA} atomic line were taken into account. 
The \citet{ryde10} oxygen abundances,
where a few of the same stars from the \citet{Zoccali2006} sample
 were observed, also take C into account, given that it is derived 
from OH and CN lines in the H-band. \citet{jonsson17} reanalysed
23 stars from \citet{Zoccali2006}, and another 10 red giants.
There is agreement between the results for the larger
samples by  \citet{friaca17}
and \citet{jonsson17} in the optical, and 
\citet{ryde10} and \citet{cunhasmith06} in the near-infrared. 
All NIR data used the H-band, where molecular lines allow
a proper derivation of C, N, and O altogether.
These data tend to show lower oxygen abundances 
relative to other work, and should be more accurate. 
In Fig. \ref{oo} the results in grey correspond to
oxygen abundance derivations that did not take into account 
C and N abundances in red giants,
or that did not employ high-resolution data. Other issues with data on oxygen
are detailed in \citet{friaca17}.
The dwarfs by \citet{bensby13} (we plot those with ages $>$11Gyr) 
have higher temperatures and are not
affected by CNO dissociative equilibrium, but the results 
might instead be affected by
NLTE effects.

Overplotted in Fig. \ref{oo} are chemodynamical evolution models
 by \cite{friaca17}, where yields from hypernovae (Kobayashi et al.
 2006) are taken into account for metallicities [Fe/H]$<$$-$2.0.  The data and models 
 are in good agreement, indicating a specific SFR
of  $\nu_{\rm SF}=$ 0.5 Gyr$^{-1}$, where  
 the inverse of the timescale for the system formation $\nu_{\rm SF}$
 (given in Gyr$^{-1}$), is the ratio of the SFR 
 over the gas mass in M$_{\odot}$ available  for star formation, 
or $\nu_{\rm SF}$ = 1/M(M$_{\odot}$) dM(M$_{\odot}$)/dt.
 Other chemical evolution models
in the literature (see \S 4)
 essentially agree with the behavior shown in Fig. \ref{oo}.

\paragraph{Magnesium}
\label{magnesium}

\begin{figure}
\includegraphics[width=12cm]{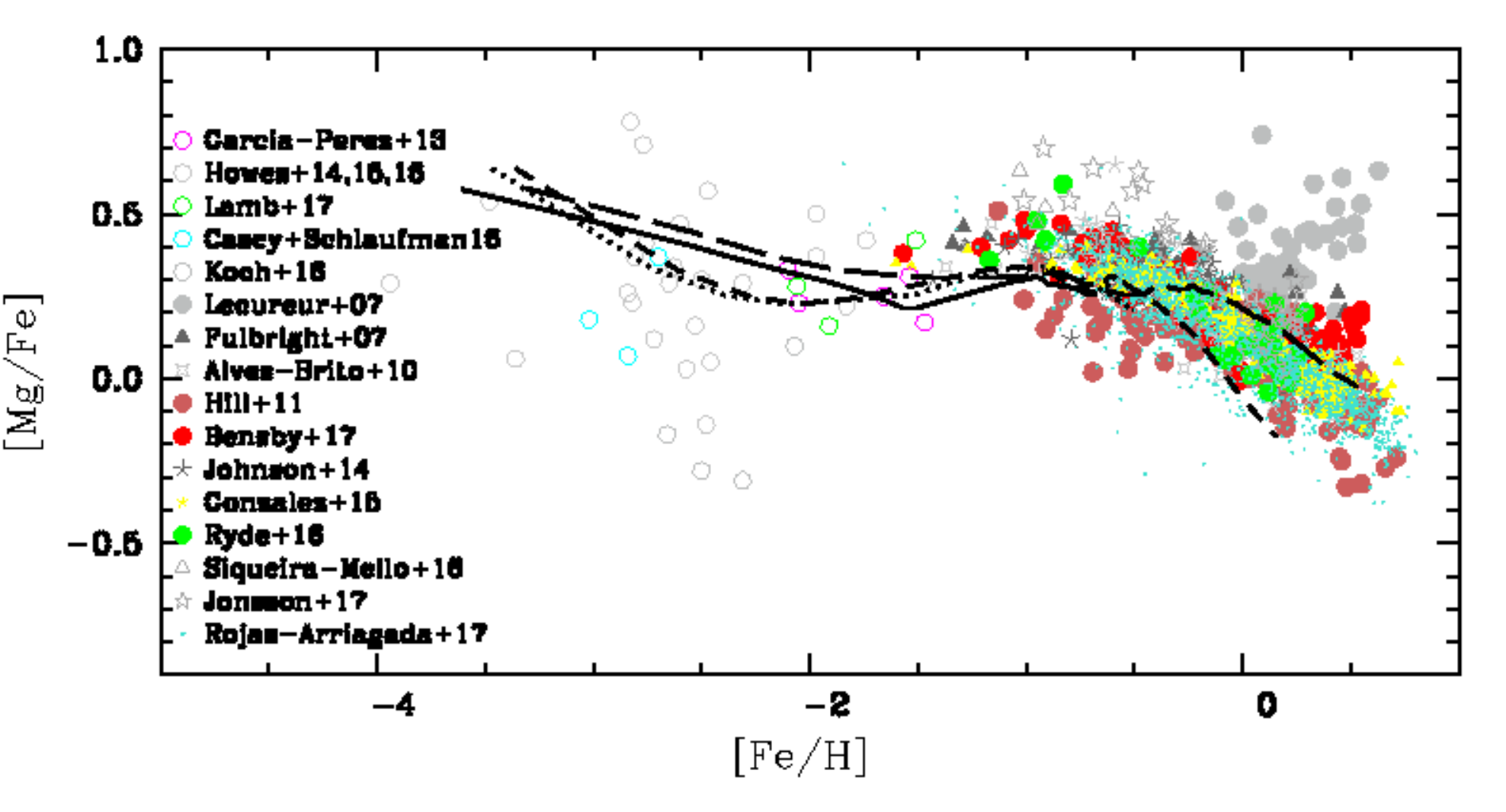}
\caption{ [Mg/Fe] vs. [Fe/H]  with literature abundances for bulge stars. 
magenta open pentagons: \citet{garcia-perez13};
grey open pentagons: \citet{howes16}; 
green open pentagons: \citet{lamb17};
grey open pentagons: \citet{casey15};
grey open pentagons: \citet{koch16};
grey filled circles: \citet{lecureur07};
strong-grey filled triangles: \citet{fulbright07};
grey 4-pointed stars: \citet{alves-brito10};
indianred filled circles: \citet{Hill2011};
red filled circles: \citet{bensby17};
grey stars: \citet{johnson14};
yellow filled circles: \citet{Gonzalez2015a};
red crosses: \citet{ryde10};
grey open triangles: \citet{siqueira-mello16};
grey 5-pointed stars: \citet{jonsson17};
turquoise 5-pointed stars: \citet{Rojas-Arriagada2017}.
Color-coded choices follow explanation in the text.
Chemodynamical evolution models with formation timescale of 0.3 Gyr, 
 are overplotted (same as in Fig. \ref{oo}).}
\label{mgfe} 
\end{figure}

Fig. \ref{mgfe} shows  the available [Mg/Fe] vs. [Fe/H] 
for Bulge stars. 
The  abundances of \citet{fulbright07} were 
revised {following \citet{mcwilliam16}.}
 [Mg/Fe] shows a clear downward relation beyond Solar metallicities,
due to data from \citet{johnson14} and \citet{Gonzalez2015a}. 
This is  contrary to previous findings by \citet{lecureur07}, 
\citet{bensby13} and some stars from \citet{fulbright07}, 
that showed an increase in [Mg/Fe] with increasing metallicity.
Chemodynamical evolution models by Fria\c ca \& Barbuy (2017) for
 $\nu_{\rm SF}$ = 3 Gyr$^{-1}$  are overplotted.
Whereas Fig. \ref{oo} indicates  $\nu_{\rm SF}$ = 0.5-1 Gyr$^{-1}$ 
(timescales 2-1 Gyr), the Mg plateau fit better with 
$\nu_{\rm SF}$ = 3 Gyr$^{-1}$ (timescale of 0.3 Gyr), but the location of the
 turnover is  more compatible with  $\nu_{\rm SF}$ =  0.5-1 Gyr$^{-1}$.
The models reproduce particularly 
well the data by \citet{Hill2011}, \citet{Gonzalez2015a}, \citet{ryde16}, 
and \citet{Rojas-Arriagada2017}.
Table \ref{knee} reports metallicities at which the knee
is located in different samples.

\paragraph{$\alpha$-elements Si, Ca}

\citet[][hereafter WW95]{ww95}
 designated Si through Sc as intermediate-mass
elements, produced by a combination of
hydrostatic oxygen shell burning and explosive oxygen burning,
in proportions difficult to unravel, and varying from
one massive-star model to another. Moreover,
the contribution from SNIa to Si and Ca is also non-negligible.
These include the elements observed in Bulge stars: Si, Ca.
 These elements have a similar behavior
in [X/Fe] vs. [Fe/H] in Bulge stars. 
 In general, an enhancement of [Si,Ca]$\sim$+0.3 up to [Fe/H]$\sim$$-$0.5
and decreasing towards higher metallicity  is found.  
Chemodynamical models by \citet{friaca17} indicated
 that  $\nu_{\rm SF}$ = 0.5 Gyr$^{-1}$
best fits the knee corresponding to the start of SNIa enrichment,
and $\nu_{\rm SF}$ = 3 Gyr$^{-1}$ fits-well  the level of the plateau. 
 Given that chemical evolution models critically depend on
the nucleosynthesis prescriptions, their study is more
useful in terms of trends as a function of metallicity, in particular
indicating the location of the knee
and as constraints on nucleosynthesis \citep{friaca17}.

\subsubsection{Odd-Z elements Na, Al}
Both $^{23}$Na and $^{27}$Al are produced, together with Mg, mainly
in carbon 
 and neon burning during hydrostatic phases
of massive stars (WW95, Sukhbold et al. 2016). 
The bulk of  $^{23}$Na is produced during carbon burning,
requiring an excess of neutrons, and is therefore metallicity 
dependent.
 
 Also, about 10\% is produced in the hydrogen envelope,
 in the neon-sodium cycle, and some $^{23}$Na is made from
neutron capture on $^{22}$Ne in He burning, in the AGB
phase of intermediate-mass stars, 
through Hot-Bottom Burning processes. In this phase
Na is produced, but it is also destroyed
\citep[][and references therein]{renzini15}. 
At the high-metallicity end, Na appears to
 increase with metallicity. \citet{cunhasmith06} were the first to point out a
high Na abundance in the most metal-rich Bulge stars.
\citet{johnson14} confirmed this result,
considered as a behavior typical of a secondary-vs.-primary element. 
This can be due
to metallicity-dependent yields from massive stars, or contribution from 
AGBs \citep{ventura09}. Al is produced by neon burning in
massive stars.
In all Bulge samples 
[Al/Fe] vs. [Fe/H] behaves as an $\alpha$-element,
 as already pointed out by \citet{mcwilliam16}. 
Massive AGB stars can also produce some smaller amount{ s} of
 Al through the MgAl cycle \citep{ventura13}.
\citet{kobayashi06} presented the unique available models for the Bulge,
well-reproducing the behavior of Al-to-Fe,
and that of Na-to-Fe but with overenhanced Na
with respect to observations, and with no upturn of Na 
at the very metal-rich end. 

It is interesting that \citet{rich13} and \citet{johnson14}
noted that
in disk samples, [Na/Fe]$\sim$+0.1 is constant
for all metallicities, thus differing from the Bulge samples.

\subsubsection{Iron-peak elements Sc, {\it{\bf Ti,}} V, Cr, Mn, Co, Ni, Cu, Zn}
\label{iron-peak}

The lower iron-peak element group includes Sc, Ti, V, Cr, Mn, Fe
 with
 21 $\leq$ Z $\leq$ 26.
Depending on temperatures and densities they are produced in explosive 
oxygen burning, 
explosive Si burning
 (WW95, Nomoto et al. 2013).
The upper iron-group  elements  Co, Ni, Cu, Zn, Ga and Ge,
with 27 $\leq$ Z $\leq$ 32,
 are produced in mainly two processes, 
namely,  neutron capture on iron-group
nuclei during He burning and later burning stages,
also called weak {\it s}-component
 and the $\alpha$-rich freezeout in the deepest layers
 (WW95, Limongi et al. 2003, hereafter LC03, Woosley et al. 2002). 

\begin{figure}
\includegraphics[width=12cm]{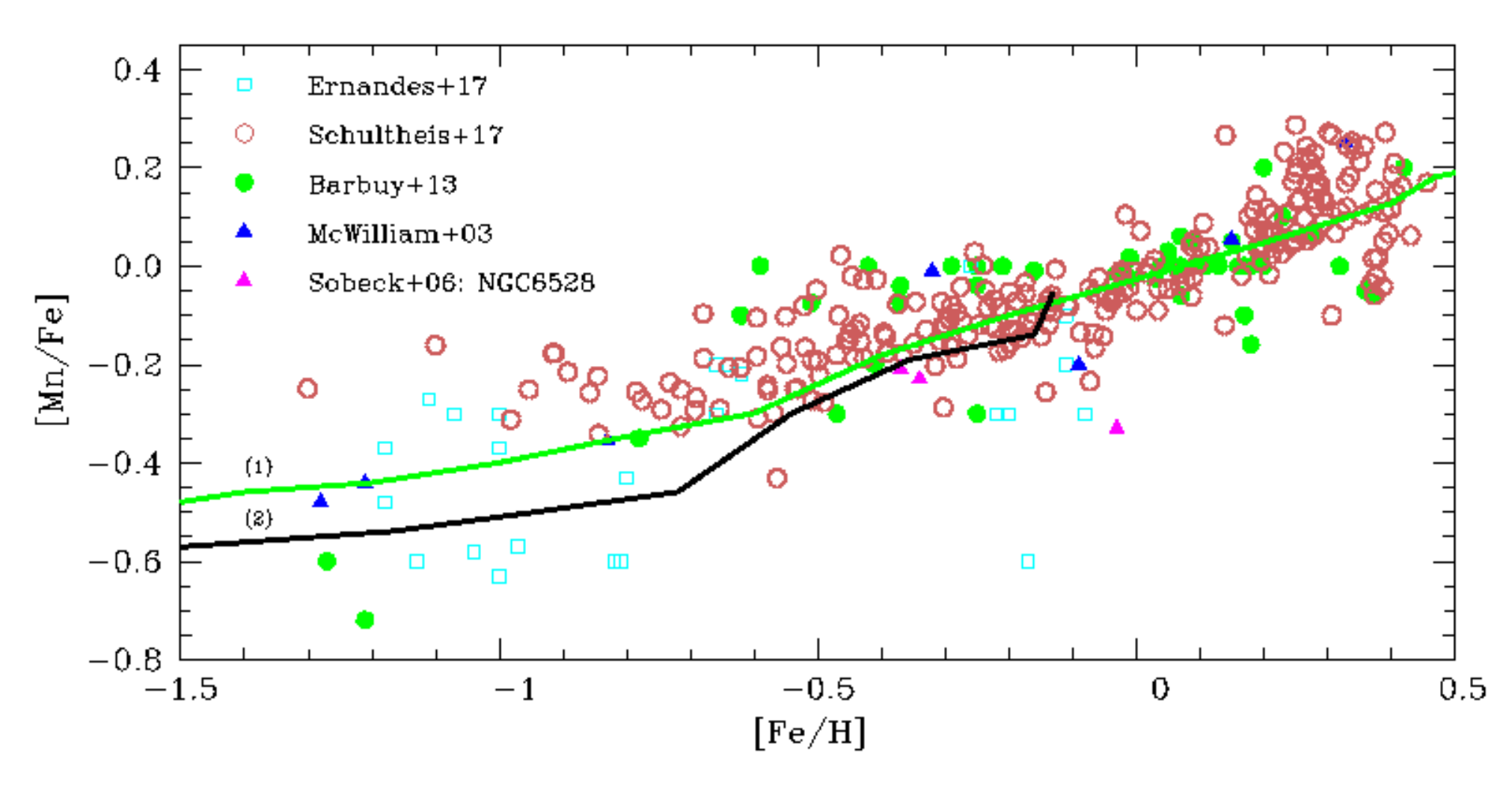}
\caption{[Mn/Fe] vs. [Fe/H] for giant stars from 
\citet{mcwilliam03} (blue filled triangles), 
\citet{barbuy13} (green filled circles), 
\citet{Schultheis2017} (indianred open circles).
Globular clusters data are from
\citet{sobeck06} (red filled triangles)
 and \citet{ernandes17} (open cyan squares).
For Mn, solid lines correspond to chemical evolution models by 
(1) \citet{Cescutti2008} (green), 
and (2) \citet{kobayashi06} (black).}
\label{mn} 
\end{figure}

Observed abundances of Sc, Ti, V, Cr, Mn, Fe, Co, Ni, Cu, Zn
 are available in Bulge stars.
\citet{johnson14} derived abundances of Cr, Co, Ni and Cu
with Cr, Co and Ni  varying in lockstep with Fe.
\citet{Schultheis2017} derived abundances of Cr, Co, Ni and Mn,
with Cr and Ni similar to Fe, except that [Cr/Fe] decreases 
for metal-rich stars.
\citet{bensby13,bensby17} derived Cr, Ni, Zn, where Cr and Ni
scaled with Fe.  \citet{ernandes17} obtained abundances
of Sc, V, Mn, Cu, Zn in bulge globular-cluster stars, with
Sc, and V varying in lockstep with Fe, differently from other work
as concerns Sc.
 In summary, most studies indicate that the
elements Sc, Ti, Mn, Cu, Zn do not scale  with Fe.

\paragraph{Scandium} Sc was found to behave as an $\alpha$-element in thick disk
stars  and with a dual behavior in halo stars (likewise
 $\alpha$-elements) by \citet{nissen00}.  Enhanced [Sc/Fe] values
are derived for metal-poor Bulge field stars by \citet[][and references therein]{howes16}
For Bulge globular clusters instead, \citet{ernandes18} find
a mean value [Sc/Fe]$\sim$0.0, possibly  
 decreasing at the high-metallicity end.

\paragraph{Titanium}
 Ti is an iron-peak element, but it is widely assumed to be an 
$\alpha$-element in 
the literature \citep[e.g.][and references therein]{mcwilliam16,friaca17}.
This is due to most of the Ti  abundance to be under the
form of $^{48}$Ti; it corresponds to 73.73\% in the Sun \citep{asplund09}. 
It is found to be enhanced in metal-poor halo stars, as well as
in Bulge stars for [Fe/H]$\simless$$-$0.8, with [Ti/Fe]$\sim$+0.30  
\citep[e.g.][among others]{alves-brito10,bensby13}.

\paragraph{Manganese}
 Mn is moderately underproduced in massive stars \citep{Sukhbold2016},
and it might produced in SNIa.
\citep{gratton89} suggested a plateau of [Mn/Fe]$\sim$$-$0.4 in metal-poor stars
 increasing for [Fe/H]$\simgreat$$-$1.0. 
A lower plateau with  [Mn/Fe]$\sim$$-$0.5 in the metallicity 
range $-$4.0$\simless$[Fe/H]$\simless$$-$2.0 was later measured
\citep[e.g.][]{cayrel04,ishigaki13}. Fig. \ref{mn} shows that
this behavior is also found in  Bulge stars measured by 
\citet{mcwilliam03}, \citet{barbuy13}, and \citet{Schultheis2017} in field
stars, and \citet{sobeck06} and \citet{ernandes17} in
globular clusters.
 Fig. \ref{mn} compares the observations to
models by \citet{kobayashi06} and \citet{Cescutti2008}.
\citet{Cescutti2008} adopted metallicity-dependent yields
for both SN II and SN Ia,  based on  yields from WW95 for SNII,
 from \citet{iwamoto99} for SNIa. 

 Mn abundances for thin and thick-disk dwarf F-G stars by
\citet{battistini15}, where NLTE corrections for hot
 turn-off stars are non-neglibible
\citep{bergemann08},
indicate Mn vaying in lockstep with Fe. Even if such
corrections are negligible in RGB stars, it would be important
to further analyse these issues
\citep[see detailed discussion in][]{mcwilliam16}.
Finally, an interesting aspect is revealed with
[Mn/O] vs. [O/H]  indicating a different behavior
 between thin-disk, thick-disk and Bulge stars
 \citep{feltzing07,barbuy13}.

{\it Copper}
\citet{johnson14} show a low Cu abundance at low
 metallicities, increasing with increasing metallicity for Bulge
RGB stars. 
For supersolar metallicities, [Cu/Fe] values appear enhanced.
The data from \citet{ernandes18} for Bulge globular clusters fit 
the lower envelope of field stars, and are lower  for moderately 
 metal-poor clusters. Models from \citet{kobayashi06}
 suitably fits the \citet{ernandes18} data, except at the lower metallicities.
\citet{mcwilliam16} showed that [Cu/O] has much less spread than
 [Cu/Fe] data, indicating a production of Cu and O in the same
 massive stars.
 Both  Co and Cu are produced in the alpha-rich freezeout as
 primary elements \citep{Sukhbold2016}.  They are also produced in the weak-s
process in massive stars, and in this case as secondary elements.
It is interesting to note that their abundances can constrain the
relative efficiencies of these two contributions (Woosley, private
commun.). The
data appear to indicate that the weak s-process is dominant for Cu,
and alpha-rich freeze-out for Co.

{\it Zinc}
Available data are gathered in Barbuy et al. (2015), and more 
recent data are given in \citet{bensby17},  and for metal-poor Bulge stars (same references
as in Fig. \ref{oo}).
[Zn/Fe]$\sim$+0.3 at $-$1.3$\simless$[Fe/H]$\simless$$-$0.5, and it decreases
 with increasing metallicity, reaching the Solar ratio at around
[Fe/H]$\sim$0.0. At supersolar metallicities
\citet{bensby13,bensby17} find a constant [Zn/Fe]$\sim$0.0
for dwarf-microlensed stars,
 whereas \citet{barbuy15} find a steadily decreasing [Zn/Fe] with 
 increasing metallicity.
The high Zn abundances at lower metallicities are suitably
explained by explosive nucleosynthesis in hypernovae \citep{nomoto13}.
At the high-metallicity end, a constant [Zn/Fe] would imply 
that the ISM is being enriched in both Zn and Fe
 on similar timescales (either both are coming from SNIa, 
which is unlikely, or the Zn yields from CCSNe vary with 
metallicity such that it mimics the timescales of Fe enrichment by SNIa), 
whereas  a decrease of [Zn/Fe] would correspond
to the enrichment in Fe  (and none in Zn) by SNIa,
likewise for the $\alpha$-elements.

In summary,
 Ti and Zn are primary elements, produced in massive stars,
and  have an $\alpha$-like behavior, similar to O, Mg, Si, Ca.
Similarly to  halo and thick-disk stars
\citep[e.g.][]{sneden91,nissenschuster11,reddy06},
Mn  is deficient in metal-poor stars, confirming that it is
underproduced in massive stars; Mn-over-Fe steadily increases at the higher
metallicities due to a metallicity-dependent enrichment by SN
Ia. Co vary in lockstep with Fe,
compatible with a dominant production in alpha-rich freeze-out.
 Cu behaves as a secondary element, indicating its production in the weak-s
process in massive stars.
V, Ni, Cr, and possibly Sc and Co tend to vary in 
lockstep with iron. The models well-reproduce their behavior in the Bulge,
 except for yields of Sc and V. It is important to stress that
[Mn/O], and maybe also [Zn/Fe] at the high-metallicity end,
 appear to be  discriminators between Bulge and thick disk.
                  
\begin{figure}
\includegraphics[width=15cm]{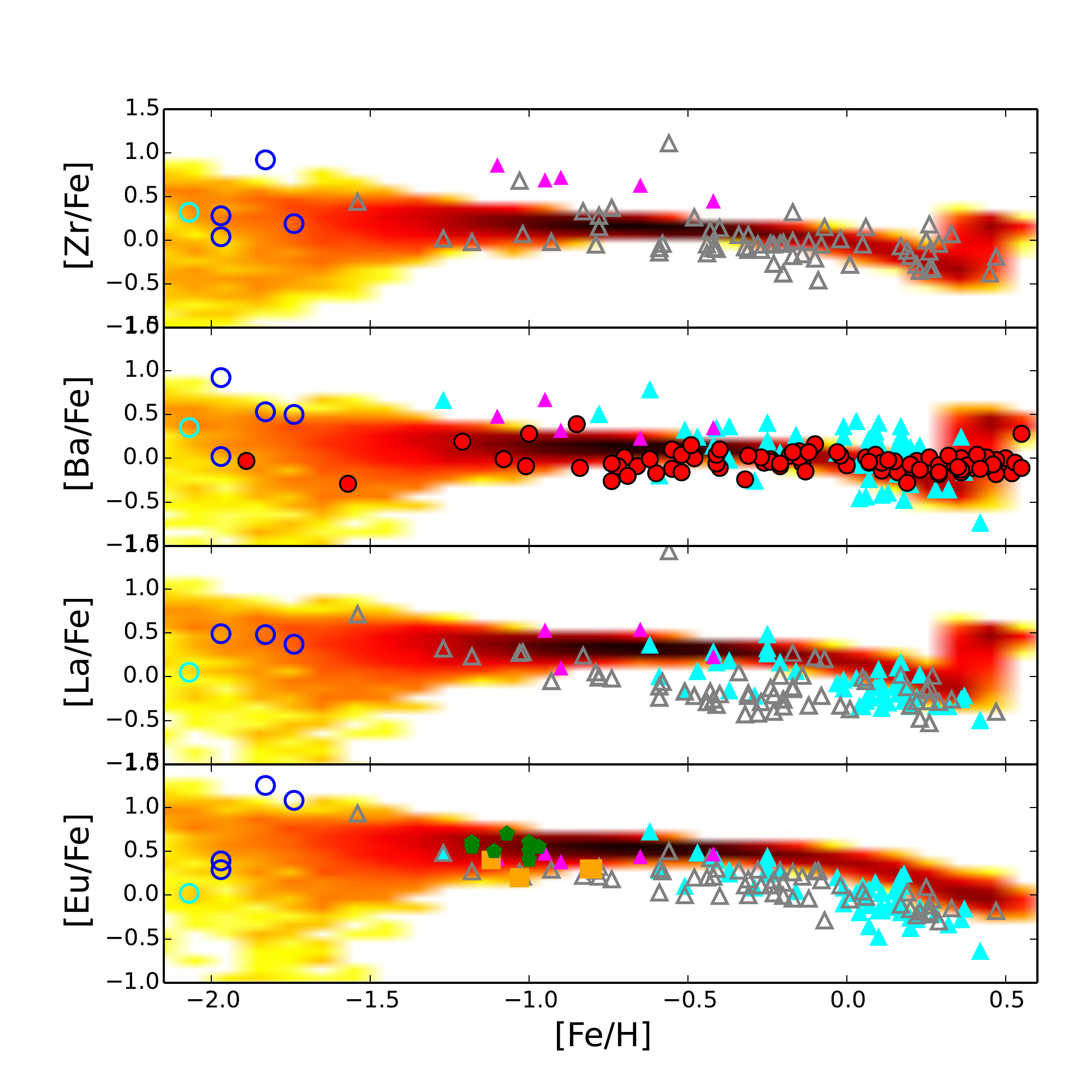}
\caption{[Zr,Ba,La,Eu/Fe] vs. [Fe/H]. Symbols: 
magenta pentagons: M62;
black open pentagons
red filled triangles: Bulge field dwarfs by \citet{bensby13,bensby17};
grey open triangles: Bulge field red giants from \citet{rich12};
cyan filled triangles: \citet{vanderswaelmen16};
light-blue open circles: \citet{howes16};
acquamarine open circles: \citet{lamb17};
magenta filled triangles: \citet{siqueira-mello16};
 green filled pentagons: HP~1: \citet{barbuy16}; 
 orange filled squares: NGC 6522: \citet{barbuy14}.}
\label{heavy} 
\end{figure}

\subsubsection{Heavy elements Y, Sr, Zr, Ba, La, Eu}
 
 Heavy-element abundances have different abundance patterns and scatter
 for different stellar populations, given their complex nucleosynthesis
 (Cescutti et al. 2018, in preparation).
The slow neutron-capture, {\it s}-process, is subdivided in weak,
 main and strong components, according to their production sites. 
The weak {\it s}-process takes place in He core and C-shell-burning phases of
 massive stars, producing heavy elements up to A$\sim$ 90. 
The main {\it s}-process takes place in the He burning layers
 of low-mass AGB stars, during thermal pulses, producing
elements with 90$\leq$A$\leq$208. AGBs of low metallicity
are responsible for a strong {\it s}-component
\citep{bisterzo14,kappeler11}.
The production  site of rapid neutron-capture
 {\it r}-process elements has not yet been
firmly established. The main candidates are: 
a) Neutrino-driven winds arising in CCSNe
\citep{woosley94} were reviewed extensivey by \citet{arcones13}, and
they might account for the
weak {\it r}-process, or the  'lighter element primary process (LEPP)' 
(mainly Sr, Y, Zr) process \citep[][and references therein]{bisterzo17};
b) Neutron star mergers (NSM), where nucleosynthesis
occurring during the merging
and milliseconds afterwards \citep{freiburghaus99},
 might naturally provide the neutron-rich environment needed
for the {\it r}-process. \citet{wanajo14} and \citet{goriely15}
 relativistic models produce a strong {\it r}-process,
including r-elements in the full-mass range (90$<$A$<$240).
The recent detection of {\it r}-process elements in the neutron star merger
GW170817 revealed by LIGO and VIRGO experiments greatly reinforces this 
as a confirmed {\it r}-process site (Pian et al. 2017).
Other possible sites are, among others:
c) neutron-rich material ejection via magnetic turbulence in 
magnetohydrodynamically driven jets (MHDJ) from CCSNe, resulting from
 massive stars characterized by a high rotation rate and a 
large magnetic field necessary for the formation of bipolar jets
 \citep[][and references therein]{Nishimura2015},
d)  Electron-capture supernovae
 could produce elements of the first neutron-capture peak
\citep{wanajo11b}.

A new interpretation of neutron-capture heavy-element enrichment
takes into account rotation in massive stars. 
An efficient {\it s}-process can be triggered
 in rapidly-rotating massive low-metallicity stars
([Fe/H]$<$-1.0) stars \citep{frischknecht12}. This is due to
rotational mixing allowing large amounts of $^{14}$N from the
H-burning shell to migrate to the He-burning core, being
transformed to $^{22}$Ne, and then favoring neutron production
via the $^{22}$Ne($\alpha$,n) reaction \citep{meynet06,frischknecht16}.
These authors are able to show that large amounts of {\it s}-process
elements can be produced. This interpretation is in contrast with
the well-accepted view that very early in the Galaxy, the typical
{\it s}-elements (such as Ba, La) abundances are due to their {\it r}-process
fraction only, as first suggested by \citet{truran81}.
\citet{chiappini11} identified the possibility that
a spread in abundance ratios of neutron-capture elements 
could be explained by nucleosynthesis in spinstars. \citet{bisterzo17}
estimated this contribution for halo stars, showing its importance
in particular in enrichment of first-peak heavy elements. 
\citet{Cescutti2018} developed chemical evolution models
for heavy elements in the Bulge. Their inhomogeneous models aim
 at explaining the spread in abundances of heavy elements,
including yields from spinstars (\S \ref{inhomogeneous}).

Most probably more than one process is needed to explain the
complex observed abundances of neutron-capture elements in the
Bulge, as it seems to be the case in the halo \citep{Spite2018}.
 Reviews on abundances of neutron-capture elements in halo
stars can be found in \citet{beers05}, \citet{Sneden2008}. 
In this review we will describe abundances of heavy elements in
Bulge stars only. 
Fig. \ref{heavy} shows [Zr,La,Ba,Eu/Fe]
from the literature, together with inhomogeneous models by
\citet{Cescutti2018}.

 Europium is an {\it r}-process element at a 94\% level \citep{bisterzo11},
and behaves as an $\alpha$-element.
Similarly to $\alpha$-elements, it is enhanced in
 metal-poor stars with [Eu/Fe]$\sim$+0.4, and starts
to decrease at [Fe/H]$\sim$$-$0.4 to $-$0.7, reaching
[Eu/Fe]$\sim$0.0 at [Fe/H]$\sim$0.0. 

 [Zr/Fe] from the first peak of the neutron-capture elements, 
 [Ba/Fe] and [La/Fe] from the second peak, appear to
 decrease slightly with increasing metallicities. 
This would not be expected if
compared with halo stars, where they increase with increasing
metallicity, reaching the Solar ratio at [Fe/H]$\simgreat$-2,
 as first shown by \citet{spite78}, a result confirmed and
further discussed by \citet{bisterzo17}.
The enhancement in [Zr/Fe] and [Ba/Fe] differ from the
constant Solar ratios in the old inner-disk sample by
Trevisan \& Barbuy (2014).
 The good agreement between the models and the data is remarkable
as concerns [Zr/Fe] and [Ba/Fe]. For [La/Fe] as well, except for
results from \citet{johnson12} where the ratios are about 0.15 dex lower
than the other data and the models.
         
Further indications are given by ratios between different neutron-capture
elements. The ratios [Ba/Eu] and [La/Eu] are indicative of
relative contributions from the {\it s}- and {\it r}-processes, with
 [La/Eu]$_s$=+0.8 to +1.1
and  [Ba/Eu]$_s\sim+1.1$  indicating a pure
s-process, whereas [La/Eu]$_r$=$-$0.4 to $-$0.6, and [Ba/Eu]$_r\sim-0.7$ 
are typical of a pure {\it r}-process   \citep{bisterzo14,vanderswaelmen16}.
 The available samples correspond to 
a large spread in these ratios in between these two values,
 indicating variable contributions of {\it s}- and {\it r}-processes.
 As discussed above, {\it s}-process elements can be due
to transfer from a companion AGB star, or else to spinstars.
A first-to-second neutron-capture peak ratio [Zr/La] has a large
spread at all metallicities, with the spread increasing for 
[Fe/H]$\simless$$-$1.0 that might be explained with the inhomogeneous models.

\subsubsection{Summary}
As a general conclusion, abundances of the $\alpha$-elements are 
powerful indicators of the SFH in the early Galaxy, and of the time of start
of enrichment by SNIa. The enhanced [$\alpha$/Fe] measured in Bulge stars
demands that an early-fast chemical enrichment has occurred.
 The odd-Z elements Na and Al are little studied in
terms of chemical evolution, and these elements could represent the bridge
between the field and globular cluster stars through
 Na-O and Mg-Al anticorrelations. Iron-peak elements can distinguish
between nucleosynthesis processes, and some ratios (e.g. [Mn/O]) can
be discriminants of stellar populations. Nucleosynthesis of heavy elements
is the most complex, given the various paths possible.
Inhomogeneous models might explain their spread in abundances
at lower metallicities.

% Table
\begin{table}
\tabcolsep3.25pt
\caption{Level of abundance ratio plateau, and knee when it starts to drop,
for comparable populations of Bulge and thick-disk stars. }
\label{knee}
\begin{center}
\begin{tabular}{|l|c|c|c|c|c|c|c|c|c|c|}
\hline
Reference            & stars   & [O,Mg/Fe] & [Fe/H] & Reference            & stars   & [Mg/Fe] & [Fe/H]   \\
\hline
                  & B/TD & Plateau & knee &                   & B/TD & Plateau & knee    \\
\hline
1  &  B   &+0.30  & $-$0.55$\pm$0.03 & 2  &  B      &  +0.36  & $-$0.4~to~$-$0.5 \\
3,4  &  B      &+0.41   & $-$0.5~to~$-$0.4 & 4 &  TD      &  +0.36  & $-$0.6 \\
5 &  B     &+0.310$\pm$0.11  & $-$0.37 & 5 &  TD     & +0.304$\pm$0.07 & $-$0.43 \\

\hline
\end{tabular}
\end{center}
\small{1. \citet{friaca17};  2. \citet{Hill2011}; 3. \citet{bensby13}; 4. \citet{bensby17};
5. \citet{Rojas-Arriagada2017}}

\end{table}

\subsection{Comparison of Bulge and thick-disk chemical properties}
\label{thickbulge}
The identification of Bulge stellar populations with thick-disk ones, or else an
older Bulge with its own identity is an issue of intense debate in the
literature.  As discussed in \S 4 several mechanisms are at play, such as
high-redshift clumpy formation of bulges and thick disk, mergers and secular
evolution. The debate now concentrates on which one of these are dominant in the
evolution of spiral galaxies and bulges.  It is difficult to disentangle these
processes by studying external galaxies (see \S 4.4), and the hope is that for
the MW it would be possible to offer new insights by obtaining accurate
abundance ratios in samples for bona-fide Bulge and thick-disk stars.
Expectations are that differences on a) the level of the abundances plateau (in
the case of $\alpha$-elements), and b) the metallicity at which the so-called
knee occurs, which reflects the enrichment by SNIa, could be used to assess
differences in the SFR (and IMF) of the two components, i.e. their SFH.

The comparisons are usually done between Bulge and local thick-disk stars.
The controversy on oxygen abundances of Bulge and thick-disk
populations being similar
has so far no conclusion (see \S \ref{oxygen}) -
 \citep*[see also discussion in][]{friaca17}. 
For  Mg, Ca, Si, Ti, 
\citet{alves-brito10} concluded that they are similar.
As for Mg, in a reanalysis of Mg abundances from \citet{fulbright07}, 
\citet{mcwilliam16} has shown
that the Bulge is more Mg-enhanced than the thick-disk data from
\citet{bensby05}, and the same is found by
\citet{Schultheis2017} from APOGEE data,
confirming previous findings by \citet{anders14},
and \citet{bensby17}, 
suggesting that SFR was faster in the inner Galaxy.
\citet{Rojas-Arriagada2017} found their Bulge stars only slightly more
enhanced than thick disk ones, in terms of plateau level,
but detected a difference of $\Delta$[Fe/H]$\sim$0.06
 in the turnover between the two populations.
McWilliam (2016) concluded that the Bulge
[Mg/Fe] appears shifted by $\sim$+0.16 dex compared
to the solar-neighborhood thick-disk stars by \citet{bensby05}.
Such behavior is also confirmed by \citet{bensby17}.
Table \ref{knee} gathers some of these results from the literature. 
For the odd-Z element Na, \citet{rich13} and \citet{johnson14}
pointed out that [Na/Fe]$\sim$+0.1 at all metallicities
 in disk samples, thus differing from Bulge samples.
The iron-peak element Mn shows different behaviors in [Mn/O] vs. [O/H]
between Bulge, thick-disk, and thin-disk stars \citep{feltzing07,barbuy13}.
For Zn at high metallicities, 
\cite{bensby13} and \citet{bensby17} find [Zn/Fe]$\sim$0.0
 whereas \citet{barbuy15} find a decrease with increasing [Fe/H]. 

\subsection{Initial Mass Function (IMF)}
\label{imf}

The present-day stellar mass function (PDMF)
 is needed to infer the total stellar mass of a stellar population,
including their fainter dwarfs and stellar remnants. 
The  IMF is even more fundamental because
of its implications on the chemical enrichment of subsequent stellar 
generations. 
Several methods have been employed to measure the PDMF and infer
the IMF. In the solar-neighborhood, counting of stars in clusters and the field 
has given fairly robust measurements
except for the lowest mass brown dwarfs 
\citep[see][for recent reviews]{Bastian2010, Krumholz2014}.

In the Bulge, two methods have been used for estimating the PDMF. The first is by counting stars in deep 
HST observations of small Bulge fields and converting the luminosity function to a stellar mass function
using a stellar mass-luminosity relation \citep{Holtzman1998}. \citet{Zoccali2000} used NIR HST data in a field at
$b=-6\dg$ to estimate the PDMF (and for these masses, IMF) in the range $0.15\nto1\Msun$. They found $dN/dM\propto
M^{-\alpha}$ with a best main sequence slope $\alpha_{\rm ms}=-1.33\pm0.07$, and a hint of a steepening for masses
$>$0.5$\Msun$. This agrees with the IMF obtained by \citet{Holtzman1998} in BW over the common mass range, and is
similar to the disk IMF obtained by \citet{Kroupa2001}.  \citet{Calamida2015} derived the IMF in the SWEEPS
field at $(l\narreq1^\circ.35, b\narreq-2^\circ.65)$, using proper motions to separate Bulge stars from foreground disk
stars, and also correcting for unresolved binary stars. They found that for the same mass range their IMF could be
fitted with two power laws with a break at $0.56\Msun$, a slope of $-$2.41$\pm$0.50 for higher masses, and a shallower
slope $-$1.25$\pm$0.19 for the lower masses, for $50\%$ binary fraction. Alternatively, \citet{Calamida2015} obtained an
excellent log-normal fit to their IMF with a central mass $M_c\narreq 0.25\pm0.07\Msun$ and dispersion
$\sigma\narreq0.50\pm0.01\Msun$. These results are in good agreement with the nearby disk IMF as in \citet{Kroupa2001}
and \citet{Chabrier2003}.

A second method makes use of the observed timescale distribution of microlensing events (MLTD).
 The timescale of each
event depends on the square root of the lens mass, as well as on the relative distances and velocities of the lens and
the source. Then, with a dynamical model providing these distances and velocities statistically, the lens mass
distribution and then IMF can be inferred from the MLTD \citep{Han1996,CalchiNovati2008}. Using the MLTD of $\sim$3000
microlensing events from the OGLE-III survey \citep{Wyrzykowski2015} and an accurate dynamical model from
\citet{Portail2017a}, \citet{Wegg2017} measured the distribution of lens masses in the Bulge and inner disk and from
this inferred the IMF down to brown dwarf masses. For a broken power-law parametrisation they obtained
 $\alpha_{\rm  ms}$=1.31$\pm$0.10$|_{\rm stat}\pm0.10|_{\rm sys}$ and 
$\alpha_{\rm bd}=-0.7\pm0.9|_{\rm stat}\pm0.8|_{\rm sys}$ in the
main sequence and brown dwarf regimes, where the systematic errors cover the range $0\nto100\%$ in the binary fraction
for unresolved lenses. These values, and equivalent results obtained for a log-normal IMF, are again similar within
uncertainties to local disk values from \citet{Kroupa2001} and \citet{Chabrier2003}.  In summary, the IMF in the Bulge
and inner disk is indistinguishable from that measured locally, despite these being regions where the stars are
predominantly $10\Gyr$ old and formed on a fast $\alpha$-element enhanced timescale.

\section{Models for the chemodynamical properties of the Bulge}
\label{s:models}

The modeling of the Bulge has seen an enormous progress in recent years, 
mainly due to the large and complex observational data 
(both in the local and high-redshift Universe) combined
 with the evolution in the tools used for N-body and hydrodynamical simulations.
The data and models have suggested six (most probably mixed) scenarios for Bulge formation \citep[see recent reviews][]{somerville15,Babusiaux2016,bournaud16,brooks16,shenli16,naabostriker17,nataf17}, namely:
\begin{itemize}
\item Primordial collapse where Bulge and the thick disk form early on
simultaneously (also referred to as in situ scenario) via strong gas accretion;
\item Bulge formation by hierarchical merging of sub-clumps, prior to the disk
formation;
\item Merging of early thick-disk sub-clumps, migrating to the center and forming the Bulge;
\item Major merger hypothesis, where disk galaxies result from major mergers of gas-rich galaxies;
\item Formation of the Bulge from the disk through a bar instability
 ( "secular evolution");
\item Formation of a Bulge component triggered
 by the accretion of dwarf galaxies. 

\end{itemize}

Ideally one would like to predict the dynamics and chemical abundances of Bulge stars  within all of these scenarios and compare them
 with observations in the Bulge. In practice we still lack such complete models and large unbiased samples of \emph{genuine} Bulge stars. Despite these difficulties, the theoretical descriptions of Bulge formation and evolution have seen advances thanks to the
parallel efforts made by using classes of models that do not treat the full problem but concentrate on specific aspects (just as it happens with the data). 

In this Section  \S4 we will discuss what we have learned from chemical evolution models (\S4.1), from \emph{non-evolutionary} chemodynamical models (\S4.2), 
from inhomogeneous chemical models (\S4.3), from evolutionary chemodynamical models (\S4.4), and from self-consistent chemodynamical models in the cosmological framework (\S4.5).

Historically the above listed scenarios for bulge formation have been  grouped in essentially two main families (see \S1 for a brief description on early theoretical work on bulge formation), namely, 
those forming classical bulges and those involving disk evolution forming pseudobulges (disky and B/P bulges). 
As it will become clear along \S4, early bulge formation scenarios that once were considered to form 
only classical bulges, can also lead to the formation of pseudobulges
 on short timescales. In the first group we find scenarios where the Bulge was formed early,
 on short timescales, and in  the second group those where most of the stars now in the Bulge were formed from gas (and stars) transferred from the disk. The bar-like nature of the MW bulge seems to rule out most of the ClB models for the bulk of the stars, but these models might still apply to a minor component of the present Bulge.
What are the most common expectations in each of the two groups? Expectations are that in the first group the stellar populations would be dominated by old stars (ages $\geq$10 Gyr), thus showing enhanced $\alpha$-elements (with respect to Solar) for most of the metallicity range, and a more isotropic kinematics. Typical expectations in disk evolution scenarios are that the Bulge will have a B/P structure, with stars showing anisotropic kinematics. The expectations of scenarios in this latter group with respect to chemistry are less clear as they can differ from each other depending on the assumptions on when/how was the pseudobulge formed. If very early on, then most of the stars would be thick-disk like (i.e. old), whereas in a more continuous secular scenario, an important part of the Bulge will be made by younger 
(inner thin-disk like) populations. In the former case, the chemistry as well as the dominance of old stars could be similar
 to that expected by early bulge formation scenarios. Therefore, the expected fraction of young stars ($<$5Gyr) and the detailed chemical-patterns expected from the second scenario  will be dependent on the timescales of the secular processes.

The B/P shape of the Bulge (see \S2) is not incompatible with the properties expected from scenarios where most of the Bulge stars formed on short timescales. In addition, the observational evidence of coexisting probably
 merger-built and secular-built bulges reinforces this idea
\citep{prugniel01,mendez-abreu14,fisherdrory16}.
These analyses, based on morphological, photometric, and kinematic properties of bulges have found that many of them have a composite nature, showing pseudobulges, central disks or B/P bulges
\citep{erwin15}, and conclude that the high frequency of composite bulges in barred galaxies points toward a complex formation and evolutionary scenario. 

Finally, a minor contribution of young populations seem to exist in the innermost regions of the Bulge (see \S \ref{NSD}), where the current star formation could be powered by radial gas flows coming from the possible following sources:
 stellar mass-loss (AGBs), gas infall, bar-induced inflow, and bar-spiral arms interactions. This gas is most probably enriched, and expectations are that the metallicity of these young stars would be at least Solar. 
 
Observational constraints such as the Bulge MDF (and its variation along different fields - \S3.2), the fraction of young and intermediate age stars (as a whole, as well as its variations within the Bulge region - \S3.1), and a detailed map of the chemo-kinematic relations of Bulge stars (\S2.2 and \S2.3) are key to constrain the best scenario for Bulge formation. The challenge in the observational side has been to extract these constraints, and at the same time account for selection biases of the studied samples (using different tracers) which potentially affect ages, metallicities and kinematic distributions (see \S 2, \S 3).

\subsection{Chemical evolution models}

Three basic observational early results for the Bulge
 have led to the development of the first Bulge chemical evolution models
 in the early 90's. These were: a) the fact that the bulk of the
 Bulge seemed to be composed of old stars, as indicated by
 the pioneering CMDs for field stars by 
\citet{terndrup88}, b) the broad metallicity distribution of Bulge stars
 reaching oversolar metallicities, as shown by \citet{rich88} 
and c) the [$\alpha$/Fe] decline observed in Bulge stars 
appeared to  occur at higher metallicities with respect to
thick-disk and thin-disk stars of the Solar-vicinity
 (see \S \ref{alphaelements}, and Table \ref{knee}).

Triggered by these early results, \citet{matteuccibrocato90}
 developed one of the first Bulge chemical evolution models, aiming 
  to explain \citet{rich88}'s result which suggested a metallicity peak
 in the Bulge at roughly twice the Solar value 
(a value that was later revised downwards with better data by
 \citealt{mcwilliamrich94}, see \S 3.2),
 and a metallicity distribution that resembled much more to a closed box
 model than the one traced by solar-neighborhood G-dwarfs, where
 a deficiency of metal-poor stars with respect 
to predictions of a closed box model is seen
\citep*[e.g.][]{gilmore89,pagel97}. 

Differently from previous simpler approaches,
where only the global metal content had been predicted by models assuming the instantaneous recycling approximation,
\citet{matteuccibrocato90} traced the chemical evolution of Fe, Si, Mg and O
 individually, including the contribution of SNIa, important for
 Fe yields. After exploring the effects of different assumptions on the star formation efficiencies, IMFs, and collapse timescales, the authors concluded that the Bulge should have formed on short timescales ($<$ 1Gyr), with an efficient SFR
and an IMF with a larger proportion of massive stars \citep[][or a flatter IMF]{Salpeter1955} than the one in the solar-neighborhood \citep*[e.g.][]{scalo86,Kroupa1993,Kroupa2001,Chabrier2003}. 
A solid prediction of these models was that almost all Bulge stars would be $\alpha$-enhanced with respect to the Sun, which is now an observationally established fact  (see \S 3.4). It became clear that the [$\alpha$/Fe] vs. [Fe/H] diagram observed in the Solar Vicinity was not universal, but should vary for different Galactic components and different galaxies (nowadays known as the time-delay model), and proven very useful to studies of Galactic Archaeology.

With the more detailed chemical information for Bulge stars that have started to appear thanks to high-resolution spectrographs installed on 8-m class telescopes, chemical evolution predictions for a large number of individual chemical species appeared in the literature 
\citep*[e.g.][]{matteucci99,ballero07,Cescutti2011}. However, the assumption that in the Bulge the IMF is flatter than locally seems now ruled out by more recent data (see \S \ref{imf}).
By simultaneously predicting different chemical species, important constraints on the nucleosynthetic sites of the different elements can be used a) to test stellar yields from the different stellar models, at different metallicities, and b) help constraining the SFH of different stellar populations.

Confirming earlier findings by Zhao et al. (1994) and Soto et al. (2007), 
in 2010-2014 observations of larger samples towards the Bulge have demonstrated this Galactic component to 
 host at least two main stellar populations
\citep*[e.g.][and more recent ones]{Babusiaux2010,Hill2011,Babusiaux2014}, with different chemistry, ages and kinematics (as discussed in \S2 and \S3).
 As a response to the new data, several groups started modelling the
 chemical evolution of the Bulge as two distinct populations, an older 
metal-poor and $\alpha$-enhanced population, and a more metal-rich with 
disk-like abundance ratios 
\citep*[e.g.][]{Bekki2011,grieco12,tsujimoto12}.
  Interestingly, the first chemodynamical models of a MW-like galaxy had already suggested a double (or even multi-) component Bulge stellar populations (\S4.4). 
  
Models forming the Bulge in a  two-step process like the ones quoted above are able to reproduce the bulge MDF by combining a metal-poor and a metal-rich population, with different star-formation histories. In particular, the work of \citet{grieco15} models a large large variety of chemical species.
 The authors assume that the metal-poor population formed first and on a short timescale, similarly to previous models, and that later on, a more metal-rich population formed from enriched gas whose origin can be either from the inner-disk gas flows or from material left by the first Bulge population. The authors also suggest the oldest Bulge population to present chemical gradients similar to the ones found in \citet{pipino08,pipino10}
 chemodynamical models (\S4.4), adding one more reason for variation of the MDF at different Bulge locations, i.e. those would not only be due to an overlap of the (at least) two different stellar populations co-existing in the Bulge, but also due to an intrinsic abundance gradient of the oldest Bulge populations created by dissipative collapse (see discussion on vertical-gradient in \S \ref{vertical}).

\citet{tsujimoto12}, on the other hand, 
suggested that a significant fraction of the present Bulge is composed of stars initially in the inner part of the thick disc,
introducing a radially varying IMF, which tends to be more top-heavy in the inner parts. However, a radially varying IMF seems not to be supported by the 
recent APOGEE data \citep*[e.g.][]{hayden15} which show a thick-disk
 [$\alpha$/Fe]-[Fe/H] sequence that is independent both from the Galactocentric distance and the height from the Galaxy's mid-plane. On the other hand, the APOGEE data do not rule out that part of the Bulge could be composed of thick-disk stars.
Finally, \citet{cavichia14} presented a chemical evolution model for the Bulge and disk, which includes radial flows. In this case an IMF similar to the local one is assumed, together with a collapse timescale of 2 Gyr (slightly longer than most of the models for the Bulge presented above). The authors have explored different gas flow pattern velocities, also investigating the role of a Galactic bar of different ages.  According to the authors, a radial inflow of gas with velocities lower than 1 km/s already increases the SFR in the Bulge region
 by $\sim$19\%, which helps shifting the MDF to larger values,
 without the need of a flatter IMF or a second infall episode.

\begin{figure}
  \centering
  \includegraphics[width=12.4cm]{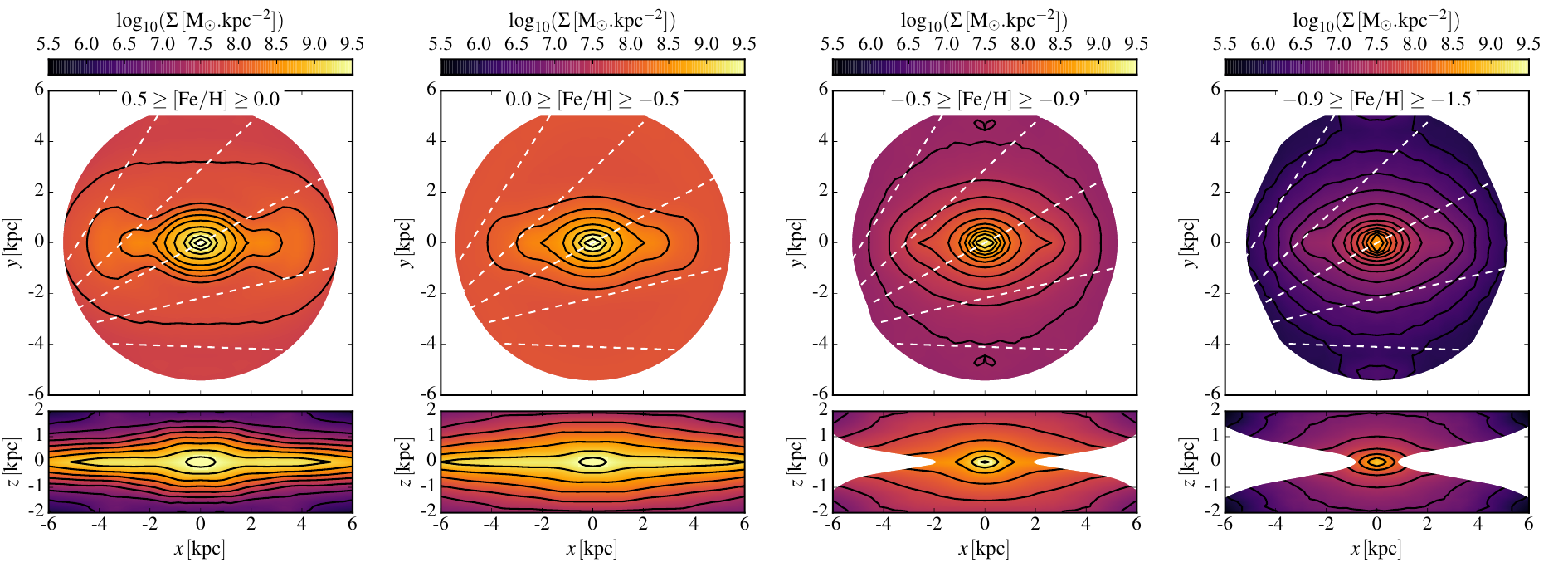}\\
  \caption{Face-on and side-on surface densities of Bulge stars in several metallicity bins, according to the fiducial
    model of \citet{Portail2017b} fitted to the ARGOS and APOGEE chemo-kinematic data. Regions where estimated
    systematic uncertainties are larger than $30\%$ or outside $5.5\kpc$ from the Galactic Center are masked. The white
    dotted lines indicate lines of sight at $l=30\dg$, $15\dg$, $0\dg$, $-15\dg$ and $-30\dg$. The face-on views show
    the barred structure of the populations, with a rounder, more centrally concentrated density distribution seen
    inside $\sim\! 1$ kpc for stars with [Fe/H]$<-0.5$. The edge-on views shows the highly flattened, disk-like shapes
    in all bins except the most metal-poor. Adapted from \citet{Portail2017b}, with permission.}
  \label{fig:SurfaceDensityMbins}
\end{figure}

\subsection{Chemodynamical equilibrium models: the dynamical structure of stellar populations in the Bulge}
\label{s:chemo-dynamics}
Another important piece of information for understanding the formation of the Bulge is provided by dynamical models
addressing how the different stellar populations are distributed in the Bulge. The dynamical models reviewed in
\S~\ref{s:dynamics} constrain the mass distribution and gravitational potential of the inner Galaxy, which combines
contributions from the Bulge and long bar, and the NSD, main disk, and dark matter halo. Different stellar
populations\footnote{A population could be defined in terms of a range in metallicity, age, or abundances, or some
  combination of the physical properties of stars.} will generally have different orbit distributions in this joint
potential, resulting in different spatial distributions and kinematics for each population.  From the observed positions
and velocities of a sufficient number of stars of that population, its specific orbit distribution can be reconstructed
using chemo-dynamical equilibrium (CDE) models.  These models assume that the stars have reached dynamical equilibrium
in the overall gravitational potential.  They do not consider how the chemical properties of the gas and stars have
evolved over time. A notable property of these models, resulting from Jeans' Theorem \citep[e.g.][]{BinneyTremaine2008},
is that they constrain also the stellar density and kinematics in regions where no data were available for the orbit
reconstruction.

\citet{Portail2017b} recently presented the first self-consistent CDE models for the Bulge and entire inner MW.  They
reconstructed the 3D density, kinematics, and orbital structure of stellar populations in different metallicity bins,
using the overall barred gravitational potential from \citet{Portail2017a}, distance-velocity-metallicity data in
multiple fields from the ARGOS and APOGEE (DR12) surveys, and correcting for the survey selection functions.  Figure
\ref{fig:SurfaceDensityMbins} based on their study shows face-on and edge-on surface densities for stars with
$\mathrm{[Fe/H]} \geq 0$ (metal-rich bin A), $-0.5\leq\mathrm{[Fe/H]}\leq 0$ (intermediate bin B),
$-0.9<\mathrm{[Fe/H]}\leq-0.5$ (metal-poor bin C'), and $\mathrm{[Fe/H]} \leq -0.9$ (metal-poorest bin D'). The last two
bins are divided at $\mathrm{[Fe/H]}\narreq-0.9$ rather than at $\mathrm{[Fe/H]}\narreq-1.0$ as in
\citet{Ness2013b,Portail2017b}, such that bin C' includes the metal-poor side of the local thick disk MDF but most of
the Bulge RRL MDF is contained in bin D'. In order of decreasing metallicity, the photometric selection of the
ARGOS sample leads to $(23\%, 43\%, 29\%, 5\%)$ of ARGOS stars in the bins (A, B, C, D), while the
final CDE model has $(52\%, 34\%, 12\%, 2\%)$ in bar-supporting and $(38\%, 47\%, 14\%, 1\%)$ in non-bar supporting
orbits in these bins, respectively.
Since estimated metallicity errors are $\sim\!0.1$ dex \citep{Ness2013a}, much
smaller than the total range, current uncertainties in the MDF (Section~\ref{MetallicityDF}) can only have small effects on
the metallicity ordering.

Fig.~\ref{fig:SurfaceDensityMbins} illustrates that stars in all metallicity bins are significantly barred, but in terms
of mass, most of the support to the bar is provided by metal-rich stars. Bin A stars contribute most to the
Galactic bar and B/P Bulge; they have dynamical properties consistent with a disk origin. Stars in bin B are hotter and
rotate slightly faster than stars in A, they are more extended vertically, and contribute somewhat less to the bar and
B/P shape. They are consistent with a disk origin formed from stars located initially at larger radii
\citep{DiMatteo2014}.  Metal-poor stars in bin C' ($\mathrm{[Fe/H]}$$\leq$$-$0.5) rotate slower and have higher dispersion
than the more metal-rich stars. They indicate
 a weaker support to the bar and do not support the B/P shape.  Outside the
central $\kpc$, these metal-poor stars are found to have the density distribution of a thick disk; in these regions
their vertical profile is exponential with scale-height $\nsim 500$ pc \citep{Portail2017b}. These stars also show
cylindrical rotation \citep{Ness2013b, Portail2017b}, suggesting a thick-disk nature.

In the inner Bulge ($x\lta1\kpc$, $z\lta0.6\kpc$), however, Fig.~\ref{fig:SurfaceDensityMbins} shows evidence for an
extra component of these metal-poor stars with short scale-height rising towards the Galactic center; this component is
seen also for stars in the even more metal-poor bin D'. A central concentration of metal-poor stars was also reported by
\citet{Dekany2013, Pietrukowicz2015} in RRL stars, by \citet{Schultheis2015} in metal-poor APOGEE stars, and by
\citet{Zoccali2017} from the GIBS survey. These could be thick-disk stars on orbits compressed by the deep gravitational
potential of the NSD, stars from the inner halo-Bulge, or stars from a ClB formed by early mergers or strong gas
accretion at early times but whose existence has not yet been established. Finally, the very metal-poor stars in bin D'
constitute a thick, high-dispersion, more slowly and not cylindrically rotating, but still barred population, which is
only weakly constrained by the ARGOS data \citep[see][]{Portail2017b}. This metallicity range includes the RRL
population but probably also the metal-poor tail of the bin C' populations; clarifying the nature of this population
requires further data and analysis.

The combined orbit distributions of all metallicities in the model of \citet{Portail2017b} naturally reproduce the
observed vertex deviations in the Bulge, as seen in Fig.~\ref{vertexdeviation}.  The absence of a significant vertex
deviation for stars in this plot with [Fe/H]$<-0.5$ has been interpreted as a signature of a ClB by \citet{Soto2007}.
However, the stars with [Fe/H]$<-0.5$ in BW shown in Fig.~\ref{vertexdeviation} are predominantly from the thick-disk
component of bin C, which therefore must be the main cause for this decrease in the vertex deviation. The fraction of
stars in a ClB defined structurally as in Section 2, is small according to these models. This implies that the dominant
population of the old Bulge stars is likely to include the barred thick disk and many of the metal-rich barred Bulge
stars.

  In summary, the fraction of stars in a ClB defined structurally as in
\S 1 is small according to these models, less than of order 10 percent
for the whole bulge (see also \S 4.4). This implies that the dominant
population of old Bulge stars is likely to include the barred thick
disk and many of the metal-rich barred Bulge stars (see \S 3.1).

%%%%%%%%%%%%%%%%%%%%%%%%%%%%%%%%%%%%%%%%%%%%%%%%%%%%%%%%%%%%%%%%%%%%%%%%

\subsection{Inhomogenous models and the first stellar generations}
\label{inhomogeneous}

Although the previously discussed dynamical models
cannot constrain the timescales of enrichment of the different
coexisting populations in the Bulge volume, previous results,
combined with those discussed in Sect. 3.1 imply that many of the
old stars in the Bulge are in thick disk-like or even bar-like
configuration (and that the fraction of not very old stars so far
found in the Bulge seems to be larger in the strongly barred
populations). Independently from its present kinematic and structural
configurations, one can still constrain its origin by studying the
detailed chemistry of the old stars. The advantage is that chemistry
is preserved from birth (for most elements), and that abundance ratios
would hopefully be much more sensitive to the different SFH of the 
different components (which could instead overlap both in metallicity
and ages). In this Section we now discuss some constraints added by
the analysis of the detailed chemical patterns and scatter of the oldest
bulge stars.

The nature of the first stellar generations remains a hot debated topic in the literature, as models of their
formation are complex, and have difficulties in predicting from first principles their properties, among which their mass distribution which can
range from top-heavy IMFs, to more normal ones also including low- and intermediate-mass stars \citep[][and references therein]{bromm13,Karlsson2013}. 
Studies of detailed chemistry in the oldest stars found in the MW offers extra constraints to those models
\citep[e.g.][and references therein]{frebel15}.
Of relevance to the topic of the present review is that some of these fossil records should be found not only in the Galactic halo but also in the central regions of the MW (see \S \ref{mostmp}).
Among the \emph{old Bulge} stellar population, the oldest stars would have formed soon after the Big Bang, 
carrying in their atmospheres the fingerprints left by the first stellar generations on the gas from which these fossil objects formed.
There has been
an effort to obtain detailed chemical abundances for very metal-poor stars first in the halo \citep{beers05}, and more recently
 in the Bulge (see \S \ref{mostmp}), where so far, very few stars 
with [Fe/H]$\simless$$-$2.0 have been found (\S \ref{mostmp}).
 This is expected if, in the \emph{old Bulge} the chemical
 enrichment proceeded in a faster pace than in the MW halo. 
This would imply that stars as old as the ones found in the halo at metallicities [Fe/H]$<$$-$2.5 could be found in the Bulge at larger metallicities
 ($-$2.5$<$[Fe/H]$<$$-$1) as proposed by \citet[][see also Wise et al. 2012]{chiappini11}.

Inhomogeneous chemical evolution models are a useful tool to test the above ideas and help
to constrain both the nature and the nucleosynthesis of the first stellar 
generations contributing to the chemical enrichment, while also giving insights on the enrichment timescale of different stellar populations, such as the halo and the \emph{old Bulge}.
These models have been very useful for interpreting the intrinsic scatter observed in several neutron-capture elements in 
 very-metal poor stars ([Fe/H]$<$$-$2.5), while preserving the small scatter in other abundance ratios such as the
 $\alpha$-elements \citep[e.g.][]{argast04,font11,Cescutti2008,Cescutti2013,chiappini13},  
and are the appropriate tool for studying the chemical enrichment of
the earliest phases of the Galaxy assembly, when the interstellar medium was less well-mixed \citep{thornton98}. 
Such models take into account the stochasticity of the IMF and SFR,
 and provide an important test bench for studies of the production sites
 of chemical elements, such as {\it r}-elements, helping one identifying key abundance ratios,
 as well as the most promising stellar yields, before implementation in self-consistent cosmological simulations
\citep{Scannapieco2017}.

Although the first inhomogenous models have been developed for the halo, they can also be useful to interpret the metal-poor data 
in the Bulge \citep{chiappini13,barbuy14}. Via the study of the detailed chemical patterns 
in the most metal poor Bulge stars ($-$2.5$<$[Fe/H]$<$ -1),
one can also address the question of how metal-poor are the oldest Bulge stars, and if the \emph{old Bulge} is just an extrapolation of the halo towards the 
inner regions, or if its enrichment has proceeded in a different pace. The results of \citet{Cescutti2018} are promising as they can potentially be used to chemically disentangle \emph{bonafide old Bulge} from halo stars populating the central regions of the Galaxy. 
Their results suggest that some of the most metal-poor stars found in the Bulge region by e.g. \citet{howes16} could actually be halo stars trapped in the bar potential (see \S \ref{heavy}).

\subsection{Evolutionary chemodynamical models}
\label{evolutionary}

Chemodynamical models not only describe the chemical evolution of the gas and stars, but also follow the physical processes related to the assembly of the Galactic components, which in pure chemical evolution models are introduced by hand (gas accretion, mergers, radial gas flows and Galactic winds). Because of their complexity, various simplified approaches have also been used. Here we describe
these different approaches in the following subsections discussing 
 chemodynamical models that consider \emph{evolutionary} aspects on a) dynamics, but assume a simple "tagging" for the chemical properties (\S 4.4.1), b) dynamical and global metallicity properties, but do not consider aspects related to the evolution of individual chemical elements (\S 4.4.2,  \S 4.4.3) and c) hydrodynamical and chemical (\S 4.4.4). Models in fully cosmological framework are described in \S 4.5.

%%%%%%%%%%%%%%%%%%%%%%%%%%%%%%%%%%%%%%%%%%%%%%%%%%%%%%%%%%%%%%%%%%%%%%%%%%%%%%%%%%%%%%%%%%%%%%%%%%%%%%%%%%%%%%%%%%%%%%%%%
\subsubsection{Dynamical models with chemical tagging}
\label{s:mapping}

B/P bulges originate through the buckling instability of bars in disks 
\citep{Raha1991, Debattista2006, Martinez-Valpuesta2006} or through orbit trapping in the
 rotating barred potential \citep{Combes1990, Quillen2002}. These
systems are characterized by exponential density profiles and cylindrical rotation.  An initial ClB
can be spun up by the rotating bar during the growth of the bar and B/P bulge which can lead to cylindrical rotation for
the ClB stars in some models \citep{Saha2013}.  However, the models predict a gently outwards-increasing rotation
profile which is different from the typical rotation properties of B/P bulges. Moreover, N-body simulations also show
that, if sufficiently massive, pre-existing ClBs lead to final composite bulges with Sersic index n$>$1
\citep{Blana2017} and non-cylindrical rotation \citep{Saha2016}. Therefore, limits to the mass of the ClB can be
obtained both from structural and kinematic data. Comparing N-body simulation models with BRAVA data, \citet{Shen2010}
constrained the mass of a possible pre-bar Bulge in the MW to $<25\%$ of the total Bulge mass \citep[see also][]{DiMatteo2014}.

Because bars have the tendency to mix stars from different radii and thus to erase population gradients
\citep{Friedli1994}, the observed metallicity gradient in the Bulge had long been taken as evidence for a ClB
formed through dissipative collapse early in the history of the MW. 
Dynamical models indicated that this picture needed to be revised. 
There are two key points: (a) The reshuffling of stellar matter during
the bar and buckling instabilities believed to be responsible for creating B/P bulges is 
too weak to erase population gradients present in the prior galaxy disk, 
and (b) depending on the spatial distribution and kinematics of the stars
in the disk, the instabilities may spread out stars in the final system, creating or enhancing vertical gradients.

\citet{Martinez-Valpuesta2013} showed that violent relaxation during the bar and buckling 
instabilities in their N-body
disk-galaxy-evolution model changed the Jacobi energies $E_J$ of the particles only slightly, compared to the range of
energies in the part of the disk participating in the evolution. This implies that population gradients with $E_J$ in
the prior system can largely be preserved even if the stars are scattered widely in angles. They illustrated this by
showing that a radial metallicity gradient in the disk can be mapped into outward and vertical metallicity gradients in
the final B/P Bulge.  \citet{Bekki2011} had earlier concluded that vertical metallicity gradients in a thick disk would
survive through the bar instability, thus explaining the final Bulge vertical metallicity gradient.

\citet{DiMatteo2014} analysed three simulations with different ClB mass fractions. Because the stars migrate in the disk
due to angular momentum transfer around the time of bar formation, stars from up to the radius of the outer Lindblad
resonance of the bar are mapped into the final B/P Bulge. Particle $E_J$ were again approximately conserved, and stars
born at small and large radii in the disk end up preferentially at small and large radii in the B/P Bulge,
respectively. \citet{DiMatteo2014} also found that Bulge stars coming from larger birth radii have greater rotational
support and velocity dispersion. ClB stars extend to larger heights, have lower rotation, and less steeply
decreasing dispersion profiles. Comparing their results to Bulge chemo-kinematics they limit the mass of a ClB in the MW
to $<10\%$ of the disk mass. 

While these thin-disk evolution models can reproduce the global kinematic and chemical properties of the Bulge,
\citet{DiMatteo2015} argued that to reproduce the chemo-kinematic and structural properties at different metallicities
an additional pre-cursor thick disk is needed.  The primary arguments are that for pure thin-disk initial models, the
B/P Bulge should show a split red clump for all stars, and that the metal-poor population ($-1.0<\mathrm{[Fe/H]}<-0.5$)
should be a kinematically hotter and faster-rotating replica of the more metal-rich populations. Both are not observed
in the ARGOS data.  To test this scenario, \citet{DiMatteo2016} investigated N-body models in which
the B/P bulge results from the evolution of a galaxy with thin, intermediate, and thick disks with inital velocity
dispersions adjusted to those measured near the Sun, and \citet{Fragkoudi2017} analysed models with thin and thick
disks. In these models, the B/P Bulge is found to be more pronounced in the kinematically cold populations, and the
peanut shape is visible in the thick-disk component only at heights $>1$ kpc from the Galactic plane. Also, the thicker
and kinematically warmer disks lead to weaker and rounder stellar bars that still rotate cylindrically, as observed in
the Bulge.

\citet{Debattista2017} describe a process termed {\sl kinematic fractionation} whereby co-spatial stellar populations
with different in-plane random motions separate when the bar forms and evolves, in that the radially cooler populations
form a stronger bar and a peanut-shaped Bulge, while the radially hotter populations form a weaker bar and a box-shaped
Bulge. In their experiments an initially radially hot thin disk can become thicker than an initially cool thick disk,
provided the ratio $\sigma_R/\sigma_z$ is about three times greater in the initial thin disk. In this regime it is
mainly $\sigma_R$ that drives the evolution, not $\sigma_z$ (i.e., the vertical thickness). In their star-forming
simulation (see next subsection), the younger populations are indeed similarly anisotropic at early times, although the
physical origin for this remains to be clarified. In conclusion, according 
to these models, chemical gradients in the B/P bulge reflect at least in
part the chemical and dynamical structure of the thin and thick disks, from
which it might have grown.

%%%%%%%%%%%%%%%%%%%%%%%%%%%%%%%%%%%%%%%%%%%%%%%%%%%%%%%%%%%%%%%%%%%%%%%%%%%

 \subsubsection{Chemodynamical models with simplified chemistry}
 
One of the first chemodynamical models to appear in the literature was the model of
\citet{samland97}. This was a two-dimensional chemodynamical  model 
forming a large spiral galaxy,
which the authors suggested to be representative of the MW.  The model starts from an isolated, rotating protogalactic gas cloud in virial equilibrium, which collapses forming a halo, disk and bulge components. 
We here concentrate on what the authors have defined as Bulge, 
particles for which [R$^2$ + Z$^2$] $<$ 2 kpc. As clearly shown in their Fig. 3, the predicted SFH for this inner-bulge shows multi components. Two of them seem to be predominant, resembling the modern data we discussed in \S 3 suggesting at least two components, but details on their chemistry were not given.

This model was then extended to a 3-D model and placed into the cosmological framework by
\citet{samland03} by building in a more realistic cosmological accretion history and angular momentum distribution for the gas. The accretion history was found to be closely linked to the SFH, which 
together with dynamics then determined
the enrichment history of the different components of the Galaxy.  
According to this simulation the halo forms first, followed by the bulge and and a thick disk before finally the thin disk forms as the SFR goes down. Around redshift one, a bar forms and produces a triaxial bulge. The final bulge was found  to be composed of at least two stellar subpopulations,
 namely: an early collapse population and a population that formed later in the bar. The component that the authors call inner bulge is defined as particles
 having [Fe/H]$>$+0.17 which have mostly solar [O/Fe] values.
The second, more metal-poor component has metallicities in the thick disk
metallicity range and with enhanced [O/Fe].
 Stars with metallicities $-$1.9$<$[Fe/H]$<$$-$0.85 are attributed 
to the \emph{inner halo}, and would then include part of what we today 
understand by the metal-poor tail of the Bulge metallicity distribution
 (see \S4.3).

\subsubsection{Classical and pseudobulges formed from early major mergers}
\label{majormergers}

\citet{Athanassoula2017} presented a coupled chemical-kinematical-morphological 
approach, consisting of N-body simulations, where gas and star
formation are included, coupled with a chemical evolution code.
The galaxy formation model is the result of a major merger of two
spherical protogalaxies initially composed of dark matter and gas,
 taking place at 8-10 Gyr ago, with an initial metallicity of
[Fe/H]=$-$1.0. Stars formed before the merging form 
a spheroidal centrally concentrated population, whereas those formed during
the merger form mainly a thick disk and/or an extended stellar halo.
This simulation is able to form a classical bulge with 9-12\% of the mass,
and a bar containing a boxy/peanut inner part. 
 
This formation scenario is in agreement with the star formation
 history prediction by the so-called two-infall model of 
\citet{chiappini97}. The latter authors argue for a break in star formation
 history as the cause for the [$\alpha$/Fe] structure in the MW local disk
 \citep{fuhrmann11}, soon after the first 1-2 Gyrs (by redshift z$\sim$2). 
Interestingly, during the recent IAU Symposium 334 devoted to the Milky Way, 
Wetzel et al. (in preparation) presented FIRE simulations where a bimodal
 [$\alpha$/Fe] distribution appeared to result from a major merger event 
early on in the history of the MW. 
Moreover, as shown by \citet{minchev13}, radial migration alone is unable to 
produce the observed discontinuity
\citep[now seen beyond the local volume][]{anders14,hayden15, mikolaitis14}.
In this scenario, the MW as observed today would have resulted 
from an early collision of two galaxies that gave the origin to the bulge.
Thereafter the merging the evolution is secular, starting the formation of a
thin disk.

Gas-rich two-spiral major mergers are the starting point also for
\citet{Sauvaget2018}. In their simulations, the mergers take place at
high-redshifts, starting 12 Gyr ago, with final coalescence occurring at 1.5 to
3 Gyr from the initial time.  Their motivation is the fact that two thirds of
nearby spiral galaxies have pseudobulges or are bulgeless, whereas it has been
shown that more than 50\% of present-day galaxies have experienced a major
merger in the past \citep{hammer09}.  Their simulations show that most of the
early gas-rich major mergers could lead to the formation of disks and bars,
identified as pseudobulges.

These results indicate that the observed properties of the MW may be
consistent with an early merger. Then the evolution after the merging would
be secular, starting with the formation of a thin disk.

\subsubsection{Chemodynamical models with predictions for individual chemical elements}

Chemodynamical models with more detailed chemical information are discussed next.
\citet{nakasatonomoto03} have built a 3D hydrodynamical N-body model for the
formation of the Galaxy, using the GRAPE
\citep{sugimoto90} smoothed particle hydrodynamics
(SPH). This is combined with chemical
evolution  of the Bulge. In their model, most Bulge stars form during
the subgalactic merger, and another group forms later in the inner-disk
region. This again leads to two chemically different groups, in terms of
[$\alpha$/Fe] ratios, the first characterized by SN II and the second
by SNIa enrichment. In their model, there is a merging of
 sub-clumps occurring at t=0.4-0.5 Gyr, causing a starburst, when 60\% of 
Bulge stars are formed already at 0.5 Gyr. 

\citet{kobayashi11} used similar GRAPE simulations but with different
 prescriptions for feedback and chemical enrichment: the models now 
include not only SNII and SNIa, but also hypernovae with
 the goal to explain the large [Zn/Fe] chemical abundances observed in
 the MW halo most metal-poor stars. The initial conditions are the same as in
 \citet{nakasatonomoto03}. In the \citet{kobayashi11} simulations, the Bulge
 again forms from the assembly of small galaxies
early on (redshifts  z$\simgreat$3) and as a consequence, 
the bulge-particles 
are generally 
older than around 10 Gyr, and around 60\% of them do show an important [$\alpha$/Fe]
 enhancement ([O/Fe]$>$+0.3). As  typical in the chemical evolution
 models discussed in previous Sections, due to the short timescale of
 the Bulge formation, this simulation also predicts an [$\alpha$/Fe] plateau
  extending to above Solar metallicities ([Fe/H]$\sim$+0.3 dex),
creating an excess of very metal-rich stars reaching
 [Fe/H]$\sim$1, compared with observations (\S2 and \S3).
 Predictions for other abundance ratios, including iron-peak and
 odd-Z elements are presented in their models (\S \ref{abundances}).

Chemodynamical models adapted from hydrodynamical models of elliptical galaxies,
not in the cosmological framework, are presently available.
\citet{pipino08,pipino10} developed a 1-D hydrodynamical model of the collapse
of an spherically symmetric gas cloud, following the evolution of the gas
density, momentum and internal energy, metallicity and [$\alpha$/Fe] ratio (in
particular oxygen).  For a MW-like bulge model (the collapse of a gas cloud on a
very short timescale due to fast gas consumption -- below 0.5 Gyr - resulting on
a stellar mass of 2$\times$10$^{10}$ M$_{\odot}$, and an effective radius of
$\sim$1 kpc) the authors find that a stellar abundance gradient can be created
(especially in its central regions), which would have an impact on the final
MDFs at different locations in the Bulge.  The model also predicts integrated
Lick-indices H$\beta$, $Mg_2$, Mgb and $<Fe>$ in good agreement with the
integrated light derived Bulge values of H$\beta$ ($\sim$1.45), $Mg_2$
($\sim$0.25), Mg$_b$ ($\sim$3.0) and $<Fe>$ ($\sim$2.0) given in
\citet{puzia02}.  This model is employed by \citet{grieco12} in order to provide
a comparison of the expected Bulge chemical gradients from pure chemical
evolution models with those of the chemodynamical model (their Table 2).

\citet{friacaterlevich98} presented a  1D chemodynamical evolution model
 for elliptical galaxies. This code was later modified to model
 the Bulge, available in 1D and 2D 
\citep{friaca17}, where a single massive 
dark halo hosts baryonic gas which is turned into stars, by assuming 
different star formation rates,
a baryonic mass of 2$\times$10$^9$ M$_{\odot}$, corresponding to a small classical spheroid, 
and a dark halo mass $M_{H}$= 1.3$\times$10$^{10}$ M$_{\odot}$.
The cooling function of the gas $\Lambda(T)$ is calculated
consistently with the chemical abundances in the ISM,
taking into account separately the abundances of the three main coolant 
elements,
oxygen, carbon and iron. Feedback from heating, ionisation,
 mechanical pressure and chemical enrichment is taken into account.

\subsection{Recent bulge formation models in the cosmological framework}
\label{cosmological}

In  galaxy formation models within the $\Lambda$CDM scenario bulges would 
form via mergers.
 In the last years, high-redshift observations have contributed to better
understand bulge formation.

From the observations discussed in \S 3.1, the conclusion that most of the Bulge stars are $\sim$10 Gyr old (still currently debated, but essentially confirmed by the most robust age indicators) would imply that at redshift z$\sim$2 galaxies would be building their bulges. Indeed, these galaxies have shown to contain around 50\% of gas \citep{Daddi2010,Genzel2015,Tacconi2017}, are mostly clumpy \citep[e.g.][]{ForsterSchreiber2009} and show bulges in formation even 
in absence of any merging \citep{tacchella15}.
The latter authors have shown evidence of mature bulges at redshifts of z$\sim$2.2 by measurements of stellar mass and SFR surface-density
 distributions in massive star-forming galaxies. This study has shown that in most massive galaxies (with stellar masses $\geq$ 10$^{11}$ M$\odot$), although a high star-formation activity is sustained at large radii, mature bulges can reside in their central parts due to a quenching of the star formation already on timescales shorter than $\sim$1 Gyr. This important observational result provides insights on the mechanisms leading to bulge formation, and show that high central stellar densities are the result of gas-rich dissipative processes acting at very early epochs (and not by dissipationless mergers, nor by slow secular evolution). These observations go in the direction of models such as the one presented by \citet{dekel14}, where violent disk instability in high-redshift gas rich galaxies also lead to bulges fully formed by gas dissipation.

In situ fast formation of bulges has also been suggested by several simulations of the so-called clumpy disks, where the main formation process relies on long lived clumps which migrate into the central parts of galaxies, early on  \citep*[e.g.][]{noguchi99,immeli04a}. Note that the formation of clumps in simulations of disks with typical masses and gas fractions as observed at $z$$>$1 does not lead to clump formation, unless turbulence is also powered by high rates of cosmological infall and stellar feedback
\citep[][and references therein]{bournaud16}. 
\citet{elmegreen08} and \citet{bournaud09}
 have presented simulations of clump migration resulting in the formation of both classical and pseudobulges. The authors show that long-lived clumps migrate and coalesce into the central bulge, but the final fate of the bulge formed by this channel will be very dependent on assumptions on feedback. Weak feedback simulations cannot prevent the clumps from becoming very massive, resulting in the formation of large
bulge-to-disk B/D ratios  and in the formation of  ClBs (B/D$\sim$0.2-0.3). On the other hand, when 
strong radiative feedback is assumed, the clump mass is regulated
 and they remain gas-rich, which can lead to rotating pseudobulges 
\citep{bournaud12,bournaud16}.
This means that the results are still strongly dependent on sub-grid physics 
such as supernova feedback, thresholds in star formation, AGN feedback, among
 others \citep*[see][]{scannapieco09,scannapieco12,genel12},
 but bulge growth through high-redshift disk instabilities leading to
 migrating dense gas clumps could have been a common process in the early
 Universe.

An encouraging result comes from the hydrodynamical cosmological simulations of two MW
 mass galaxies in which bars could form naturally within the $\Lambda$CDM scenario 
\citep{scannapiecoathanassoula12,martig12}. Ideally, one would like to study the bar formation 
process in simulations where the galaxy itself is growing and accreting mass both by infall of 
pristine gas and mergers. In the fully cosmological simulations of
\citet{scannapiecoathanassoula12} for two MW-like galaxies one of them resulted in a system 
with a well-defined disk, bulge and halo components and a strong and long bar, whereas in the second case a smaller and weaker bar formed in an almost bulgeless galaxy. Also very important, these simulations show that bars can start forming early on, without {\emph waiting for the galaxy to be fully formed before starting their own formation}, as in idealised dynamical simulations discussed in 
\S 4.2 whose focus is the study of dynamical mechanisms, the role of resonances, and the formation of buckled bars. 

Furthermore, bulges and elliptical galaxies, although similar in several 
aspects, do differ because in bulges the contribution of secular evolution
 during disk instability is important
\citep*[][and references therein]{Kormendy2004}.
 Smaller bulges of late-type galaxies should also form on longer timescales
 from bar-driven flows of disk gas towards the galaxy center.  Secular
 evolution through disk instability can produce a pseudobulge via bar
 formation and buckling
\citep*[e.g.][]{samland03,Athanassoula2005,obreja13}.
 These processes have also been modeled before
 (see \S 4.2) but have mainly focussed on observations in the
 nearby Universe, and on the more recent Bulge data (discussed in \S 3).
  The latter weak non-axisymmetric instabilities of nearby galaxies are 
substantially different from the violent clump instability process taking
 place in high-redshift galaxies, as discussed above.  Interestingly,  
wide and deep IR surveys of galaxies in the 1-3 redshift range do not show
 barred disks but massive young clumps
\citep*[e.g.][]{sheth08,simmons14}. The latter data confirm
 the lack of regular barred disks, giving support to the idea that these star
 forming clumps are bound to their host galaxies, as suggested already in 
O'Neil et al. (2000), Immeli et al. (2004b),
\citet{elmegreen05,elmegreen07}, and are a common property of the progenitors 
of nowadays massive spirals such as the MW and one of the main channels for an
 early (mostly in situ) bulge formation.

Higher resolution fully cosmological simulations forming
bulges were presented by \citet{guedes11}, \citet{obreja13},
and \citet{Hopkins2014}
 \citep*[see][]{brooks16}, and these usually provide a fraction  
of old-to-total components in the bulge of 30-50\%. More recent chemodynamical models, in the cosmological framework, including chemical predictions have
 been presented by 
\citet{grand17,grand18}, although in this case the focus is more on
the formation mechanisms of disk galaxies.
These very complex simulations are based on the
 magneto-hydrodynamical simulation
code AREPO \citep{springel10}, that solves fluid equations on a moving mesh.
 Each particle represents an single stellar population (SSP), 
where chemical evolution and
stellar evolution are included, and gas enriched feedback from SNI, SNII and AGBs are considered (although here too, sub-grid physics related to feedback still prevent a more robust prediction 
for the mixing/distribution of the chemical elements along the several stellar generations). These models produce a red spheroidal bulge, but subdominant in many of the haloes. Moreover, in these simulations
massive bulges are produced mainly as a result of major mergers.

In conclusion, the theoretical picture emerging is one in which 
different processes are at work during Bulge formation, including disk
 instabilities due to strong gas accretion, major and minor mergers, and secular processes.
This complexity is confirmed by the multi-stellar population observed in the Bulge (as discussed in \S 3).  Although the last few years have brought an enormous development in cosmological simulations of the formation and evolution of MW-like galaxies, a successful self-consistent fully cosmological chemodynamical evolutionary model of the MW-like bulges is still missing.  

\section{Conclusions}
 The Galactic Bulge hosts one of the oldest stellar populations in the
Galaxy, carrying the imprints of the earliest Galaxy formation phases.
Because in the
Galactic Bulge we can measure kinematic and chemical properties for individual
stars, unique constraints on bulge formation models can be obtained. 
In this
review we summarize recent results on the structure, kinematics, and dynamics
of the Bulge, and on the physical properties of its stellar populations, 
including their
ages, metallicities and chemical abundances, considering elements produced
through different nucleosynthetic channels. These results have established a
composite nature of the Bulge where the spatial distributions, kinematics, and
dynamics of Bulge stellar populations depend on their population parameters.

In order to provide a link between these observations and the formation
of the Bulge, we also review possible bulge formation scenarios and models.
At present, a  successful fully self-consistent model for the Bulge in the
cosmological framework is not yet available (see \S ~\ref{s:models}).
 Different modeling approaches have, so far, focused on reproducing 
different aspects of the data,
investigating several formation scenarios in parallel. Therefore, 
we first provide a list of the main observational facts that have 
emerged regarding kinematics and dynamics (\S~\ref{s:dynamics}) and stellar populations 
 (\S~\ref{s:populations}), and then proceed to their implications
 for the several interpretations of how the Bulge was formed. 
We also indicate which of these resulsts are still debated:

\begin{itemize}

\item 
  Most of the stellar mass of the Bulge is barred (overall $\sim 90\%$ based on RCG star samples).
  Bimodal magnitude distributions at high-latitudes and reconstruction of the stellar density from
  RCG show a B/P Bulge (X-shape), similar to other barred galaxies. Vertical density profiles are
  exponential above $|z|>$500 pc where they are most reliably measured.  At $2-3\kpc$ along the long
  axis, the B/P Bulge transits into a planar long bar, similar to other barred galaxies.  The RRL
  population, tracing $\sim1\%$ of the stellar mass, is mostly unbarred but a fraction of orbits may
  have been trapped by the bar.

\item
  Cylindrical rotation across all latitudes ($|b|<10\dg$) is seen in the BRAVA, ARGOS and APOGEE surveys, 
   characteristic of stellar bars. 
The cylindrical rotation is observed for most Bulge stars over a broad metallicity
range. However, metal-poor and metal-rich stars differ in their vertical velocity 
dispersion profiles, and only the metal-rich stellar component shows
a kinematic vertex deviation.

\item Spectroscopic measurements show a broad MDF ranging from below 1/10 solar to supersolar
  metallicities.  Metal-rich stars are relatively less numerous at high $|b|$.  Although the
  majority of stars in the Bulge down to [Fe/H]$\sim-$1 (at least $-$0.8) are barred, only the more
  metal-rich populations ([Fe/H]$>-$0.5 in all ARGOS fields, and [Fe/H]$>$0 in GIBS/GES samples for
  inner Bulge) participate in the B/P structure.  Detailed structure still varies between MDFs from
  different surveys.  A larger coverage of Bulge stars, bringing the different surveys to a common
  metallicity scale is needed.

\item The Bulge is predominantly old, older than 10 Gyr. Proper-motion cleaned CMDs leave little
  room for younger populations (below 3.5\%). Bona-fide old Bulge populations (such as RR Lyrae and
  Bulge Globular Clusters) peak at a metallicity [Fe/H] $\sim -$1. Microlensed dwarfs with
  [Fe/H]$<-$0.5 towards the Bulge are old (10 Gyr or older), whereas more metal-rich ones (around
  15\%) appear to be younger than 8 Gyr, in contradiction with the CMD findings. It is unclear
  whether selection effects or foreground/far-side contamination could be at the origin of this
  discrepancy.  A fraction of around 15\% of stars younger than 5-8 Gyrs (as suggested by
  microlensed dwarfs) should show a counterpart in more advanced stages of stellar evolution.  At
  present this has not been conclusively observed.

\item The [$\alpha/Fe$], [Al/Fe], and [Eu/Fe] abundance ratios are
 generally oversolar, 
 indicative of an interstellar
  medium enriched mainly with products of Type II SNe at the time when the old Bulge stars were born.  [Zn/Fe]
  vs. [Fe/H] behaves similarly as [$\alpha/Fe$], suggesting an early production of Zn in massive stars.
Both [Na/Fe] and [Zn/Fe] are Solar in
  disk metal-rich stars, and they are respectively enhanced, and deficient
 in Bulge metal-rich stars. 
 Bulge stars show an abundance spread in neutron-capture elements (Zr, Ba, La) with respect to iron, 
which increases
  towards lower metallicities (similar to what is observed in very-metal-poor stars). 

\item While in many aspects Bulge and local thick-disk stars
abundances are very similar, they indicate  some differences in the
  abundance ratios of [O/Fe], [Mg/Fe], [Al/Fe], [Zn/Fe], and  [Mn/O]. 
This result has to be further confirmed, 
and studies of inner-thick-disk samples would be needed for a
 more conclusive comparison.
 The $\alpha$-element enhancements with 
respect to Fe extend to higher metallicities, therefore the turnover or 
 knee in the Bulge occurs at higher metallicities than in the thick disk. 
This result is however dependent on  sample selections details. 
These differences are likely not due to IMF variations, because 
current data have shown that
the  IMF in the Bulge is similar to that in the solar-vicinity.

\end{itemize}

 This above list of observational facts constitutes an impressive achievement of the Galactic community in the past
 decade.  These data come together with the information obtained from high-redshift (z$\sim$2)
  external bulges and thick disks (briefly summarised in \S 4). It is this body of data that must
 eventually be accounted for by galaxy formation models in the cosmological framework, which then should be able to
 predict chrono-chemical-kinematic maps of the MW Bulge/inner-disk region. Below we now list some of the more
 straightforward implications of the observational facts, based on the various modelling approaches reviewed in
 Section~\ref{s:models}.

\begin{itemize}

\item Detailed dynamical models of RCG starcounts and kinematics in multiple fields in the inner Galaxy confirm the
  barred structure of the Bulge and place the corotation radius of the Bar at $6.1\pm0.5\kpc$. These models find that
  the Bulge and Bar region out to $\sim 5.3\kpc$ radius contains about 65\% of the MW's total stellar mass, and the
  Bulge alone about 25\%. The relative fraction of dark matter in the Bulge is found to be small, $\sim$15-20\%,
  implying a low dark matter density in the Bulge and a core or mild cusp in the density of the dark matter halo.

\item Best-fit chemo-dynamical equilibrium models of distance-kinematics-metallicity data find that the most metal-rich
  populations in the Bulge and Bar are in a strong bar concentrated towards the Galactic plane, while the more
  metal-poor stars are in a weaker, thicker bar. The superposition of the different populations causes the apparent
  vertical metallicity gradient in the Bulge. Stars in the thick disk metallicity range are mostly found in a thick disk
  bar of vertical scale-length $\sim500\pc$, but a fraction of these stars is in a centrally concentrated component
  which could be a small classical bulge but may also have several other possible origins.

\item Bars and B/P Bulges in N-body dynamical models of unstable, pre-assembled
 disks qualitatively explain many of the
  structural and dynamical properties of the Galactic B/P Bulge and long bar.  Differences between metal-rich and
  metal-poor populations can largely be explained if the metal-poor stars are in a thicker and/or hotter
  disk when the bar and buckling instabilities set in. Chemodynamical collapse models leading to bar-unstable disks
  can quantitatively reproduce much of the chemo-kinematic data in the Bulge.

\item The oversolar abundance ratios with respect to iron
 in most Bulge stars more metal-poor than [Fe/H]$\simless$$-$0.5
 can be explained by chemical evolution
  models where most of the Bulge was assembled on short timescales of
around $\sim$2 Gyr,
 where the enrichment was dominated by SNII and SNIa
  did not have time to contribute much to the interstellar medium from which these stars were born.  The oversolar
  [Zn/Fe] ratios in metal-poor Bulge stars suggest that hypernovae contributed to the early chemical enrichment of the Bulge.
  The
  presence of rapidly rotating massive stars might contribute to explain the intrinsic scatter observed in the
  neutron-capture abundance ratios of the oldest Bulge stars.

\item The chemical differences observed in the comparison of stars in the \emph{local} thick disk with Bulge stars suggest that that thick disk and Bulge could be two genuinely different components, and the Bulge is not
 just a superposition of halo, thick disk and bar components. The caveat here is that still very few data are available for thick disk stars towards the inner regions of the MW.

\item Some chemical differences between Bulge and thin-disk stars at higher metallicities are useful 
constraints on stellar yields of intermediate mass-stars and SNIa, as well as on their metallicity 
dependency (also in massive stars).

\item Although the star formation timescale seems constrained by the data, the efficiency of the star formation is still poorly constrained. Pure chemical evolution models seem to require larger star formation efficiencies than those assumed in chemodynamical (hydrodynamical) models; this parameter affects not only
 the predictions
 for the \emph{knee} (Table \ref{knee}), but also for the MDFs 
(\S \ref{MetallicityDF}).

\item The dominant old ages of Bulge stars is compatible with models where the Bulge is
 assembled on short timescales, which includes chemical evolution models, as well as simulations where 
the Bulge forms from strong gas accretion, early mergers, or gas-rich merging clumps from the thick disk 
migrating towards the inner regions. 
The fact that most of the Bulge mass is barred 
would be accounted for by models where the Bulge stars
 formed at high redshifts, and are able to 
form a bar sufficiently early on depending on the assumptions on 
feedback in the simulations. 

\end{itemize}

Ongoing and future survey data will allow us to test and refine our current understanding. On the dynamical side,
proper motions from VVV and Gaia are expected to greatly improve mass models of the Bulge. Stellar ages obtained
by combining spectroscopic and asteroseismological data will be revolutionary in allowing us to dissect the
Bulge populations as a function of their age. Combined with the metallicity structure and abundances this will
give very strong constraints on formation models, which themselves are expected to improve greatly in the
coming years.

% Acknowledgement
\section*{ACKNOWLEDGMENTS}
{ We thank Bruce Elmegreen, Ken Freeman, Alvio Renzini, and Stan Woosley,
for a detailed reading and important comments on an earlier version
of this review.
We are also very grateful for interesting discussions and contributions from
 Gabriele Cescutti, Amancio Fria\c ca, 
 Fran\c cois Hammer, Vanessa Hill, 
 Noriyuki Matsunaga, Dante Minniti, Thorsten Naab, Matthieu Portail,  
Joseph Silk, Fran\c cois Spite, Volker Springel,
 Grazyna Stasinska, Elena Valenti,
 Chris Wegg and Manuela Zoccali, and to
Thomas Bensby, Melissa Ness, Alvaro Rojas-Arriagada,
Matthias Schultheis, for sending
unpublished data, and/or results in advance of publication.}
  BB acknowledges support
from the Brazilian agencies FAPESP, CNPq and CAPES.
CC acknowledges support from DFG grant CH1188/2-1, and from CheTEC COST
Action (CA16117). OG is grateful for the support of the Max Planck Institute
 for Extraterrestrial Physics.

% References

\end{document}